\documentclass[a4paper,fleqn,usenatbib]{mnras}
\usepackage{mathptmx}
\usepackage[T1]{fontenc}
\usepackage{ae,aecompl}
\usepackage{times}
\usepackage{amssymb}
\usepackage{amsmath}
\usepackage{upgreek}
\usepackage{color}
\usepackage{graphicx}
\usepackage{pdflscape}

\title[ATOMS I]{ATOMS: ALMA Three-millimeter Observations of Massive Star-forming regions -- I. Survey description and a first look at G9.62+0.19}

\author[T. Liu et al.]{
Tie Liu,\thanks{E-mail: liutie@shao.ac.cn (TL)}$^{1,2}$
Neal J. Evans,$^{3,2}$
Kee-Tae Kim,$^{2,4}$
Paul F. Goldsmith,$^{5}$
Sheng-Yuan Liu,$^{6}$
\newauthor
Qizhou Zhang,$^{7}$
Ken'ichi Tatematsu,$^{8}$
Ke Wang,$^{9}$
Mika Juvela,$^{10}$
Leonardo Bronfman,$^{11}$
\newauthor
Maria. R. Cunningham,$^{12}$
Guido Garay,$^{11}$
Tomoya Hirota,$^{8}$
Jeong-Eun Lee,$^{13}$
Sung-Ju Kang,$^{2}$
\newauthor
Di Li,$^{14,15,27}$
Pak-Shing Li,$^{16}$
Diego Mardones,$^{11}$
Sheng-Li Qin,$^{17}$
Isabelle Ristorcelli,$^{18}$
\newauthor
Anandmayee Tej,$^{19}$
L. Viktor Toth,$^{20}$
Jing-Wen Wu,$^{14}$
Yue-Fang Wu,$^{21}$
Hee-weon Yi,$^{13}$
\newauthor
Hyeong-Sik Yun,$^{13}$
Hong-Li Liu,$^{22}$
Ya-Ping Peng,$^{23}$
Juan Li,$^{24,25}$
Shang-Huo Li,$^{24}$
\newauthor
Chang-Won Lee,$^{2,4}$
Zhi-Qiang Shen,$^{24,25}$
Tapas Baug,$^{9}$
Jun-Zhi Wang,$^{24,25}$
Yong Zhang,$^{26}$
\newauthor
Namitha Issac,$^{19}$
Feng-Yao Zhu,$^{1}$
Qiu-Yi Luo,$^{1}$
Archana Soam,$^{28}$
Xun-Chuan Liu,$^{21}$
\newauthor
Feng-Wei Xu,$^{21}$
Yu Wang,$^{21}$
Chao Zhang,$^{17}$
Zhiyuan Ren,$^{14}$
Chao Zhang$^{14}$
\\
Affiliations are listed at the end of the paper}

\date{Accepted XXX. Received YYY; in original form ZZZ}

\pubyear{2020}

\begin{document}
\label{firstpage}
\pagerange{\pageref{firstpage}--\pageref{lastpage}}
\maketitle

\begin{abstract}
The ``ATOMS," standing for {\it ALMA Three-millimeter Observations of Massive Star-forming regions}, survey has observed 146 active star forming regions with ALMA Band 3, aiming to systematically investigate the spatial distribution of various dense gas tracers in a large sample of Galactic massive clumps, to study the roles of stellar feedback in star formation, and to characterize filamentary structures inside massive clumps. In this work, the observations, data analysis, and example science of the ``ATOMS" survey are presented, using a case study for the G9.62+0.19 complex. Toward this source, some transitions, commonly assumed to trace dense gas, including CS $J = 2-1$, HCO$^+$ $J = 1-0$ and HCN $J = 1-0$, are found to show extended gas emission in low density regions within the clump; less than 25\% of their emission is from dense cores. SO, CH$_3$OH, H$^{13}$CN and HC$_3$N show similar morphologies in their spatial distributions and reveal well the dense cores. Widespread narrow SiO emission is present (over $\sim$1 pc), which may be caused by slow shocks from large--scale colliding flows or H{\sc ii} regions. Stellar feedback from an expanding H{\sc ii} region has greatly reshaped the natal clump, significantly changed the spatial distribution of gas, and may also account for the sequential high-mass star formation in the G9.62+0.19 complex. The ATOMS survey data can be jointly analyzed with other survey data, e.g., "MALT90", "Orion B", "EMPIRE", "ALMA\_IMF", and "ALMAGAL", to deepen our understandings of "dense gas" star formation scaling relations and massive proto-cluster formation.
\end{abstract}




\section{Introduction}

Knowledge of the physical factors that control the conversion of interstellar gas into stars is of fundamental importance for understanding the star formation process and the evolution of galaxies. \citet{Schmidt1959} proposed a relation between star formation rate (SFR) and gas density. \citet{Kennicutt1998a,Kennicutt1998b} found a strong observational relation between the star formation rate (SFR) surface density ($\Sigma_{SFR}$) and the surface density of cold gas ($\Sigma_{gas}$), which suggests that the SFR surface density is primarily regulated by the gas mass surface density. This so-called Kennicutt-Schmidt (K-S) law, is a crucial input into theoretical models of galaxy evolution.  The K-S law between $\Sigma_{\rm SFR}$ and total gas surface density ($\Sigma_{\rm gas}$=$\Sigma_{\rm HI}$+$\Sigma_{\rm H_2}$) across entire galaxies and galactic nuclei, has a typical power-law index of $\sim$1.4-1.6 \citep{Kennicutt1998a,Kennicutt1998b,Kennicutt2012}. Resolved studies of nearby galaxies instead show a linear relation of $\Sigma_{SFR}$ with molecular gas, implying a depletion time for gas that is constant with surface density at about 1-2 Gyr \citep[e.g.,][]{Bigiel2008,Tacconi2020}. While significant progress has been made in recent years, there are still fundamental aspects of the origin of star formation relations that remain unaddressed.

\subsection{The role of dense molecular gas in star formation}

\citet{Gao2004} showed that the far--infrared luminosities ($L_{\rm IR}$) in starburst galaxies correlated tightly with the luminosities of HCN line emission, i.e., the dense gas star formation law. A nearly linear relation between SFR, indicated by $L_{\rm IR}$, and dense molecular gas mass, indicated by molecular line luminosities ($L'_{\rm mol}$), have been revealed in a number of entire galaxies \citep{Gao2004,Greve2014,Zhang14,Liu2015} or spatially resolved nearby galaxies \citep{Chen2015,Tan2018} with various dense gas tracers, e.g., HCN, HCO$^{+}$, CS and high-J CO lines. The tightness of this correlation suggests that the denser parts of  molecular clouds, rather than the total clouds, provide the direct fuel for star formation.

\citet{Wu2005,Wu2010} showed that massive, dense clumps (with typical sizes of $\sim$1 pc) in the Milky Way followed a similar relation and suggested that the fundamental units of massive clustered star formation are such massive dense clumps. In addition, the vast majority of dense cores and young stellar objects (YSOs) in nearby clouds are found above a "threshold" surface density of about 120 M$_{\odot}$~pc$^{-2}$ (or $A_{\rm V}\sim$8 mag)  \citep{Heiderman2010,Lada2010,Lada2012}. By comparing various models of star formation to observations of nearby clouds, \citet{Evans2014} found that the mass of dense gas above such a "threshold" surface density of $A_{\rm V}\sim$8 mag was the best predictor of the star formation rate.  \citet{Vutisalchavakul2016} found a similar result for more distant and massive clouds in the Galactic Plane, using millimeter continuum emission from the BGPS survey \citep{Ginsburg2013} to measure the mass of dense gas. The average star formation rate per unit mass of dense gas (star formation efficiency; SFE) is remarkably constant from the scales of nearby molecular clouds to Galactic Plane clouds to nearby galaxies to distant (ultra)luminous IR galaxies \citep{Vutisalchavakul2016}. In contrast, the star formation rate per unit mass of total molecular gas shows much more dispersion \citep{Vutisalchavakul2016}.

However, the dense gas star formation relation and the nature of the gas probed by dense gas tracers such as HCN, HCO$^+$ and CS emission is still far from being well-understood. One attractive explanation for the low star formation efficiency in molecular clouds is that most clouds are not gravitationally bound, but only relatively dense regions (clumps) within them are bound \citep[e.g.,][]{Dobbs2011,Barnes2016}. This may also explain the tight relation between far-infrared luminosities and the luminosities of dense molecular tracers in molecular clumps, i.e., dense gas star formation law \citep[e.g.,][]{Wu2005,Wu2010}. \citet{Liu2016} studied the dense gas star formation law for 146 Galactic clumps, most of which are gravitationally bound. They found that the slopes of the $L_{\rm IR}$-$L'_{\rm mol}$ correlations vary for clumps with different dust temperatures and luminosity-to-mass ratios. Such behavior seems to be a result of different evolutionary stages of the Galactic clumps \citep{Liu2016,Stephens2016}. Recent observations of nearby clouds have also shown that some traditional high-density tracers, such as HCN (1-0) and HCO$^+$ (1-0), are also easily detected in extended translucent regions at typical densities of 500 cm$^{-3}$ and are poor tracers of dense structures such as filaments or cores \citep{Kauffmann2017,Pety2017,Shimajiri2017}.  Most recently, \citet{Evans2020} mapped six distant ($d\sim$3.5-10.4 kpc) clouds in HCN (1-0) and HCO$^+$ (1-0) line emission with high sensitivity (rms$\sim$0.1 K per 0.2 km~s$^{-1}$ channel). They found that in all cases a substantial fraction (in most cases, the majority) of the total line luminosity arises in gas below the $A_{\rm V}\sim$8 mag threshold. These studies have challenged the definition of dense gas associated with the observation of some commonly used tracers, e.g., $J$=1-0 of HCN and HCO$^+$, but are limited in scope.

Therefore, in order to understand the origin of the dense gas star formation law, we need to address the following questions: (1) What is the spatial distribution of various molecular transitions, such as CS, HCN, and HCO$^{+}$, in molecular clouds? (2) How much dense gas in the gravitationally bound clumps is actually participating in star formation or is concentrated in the smallest star formation units, i.e., dense cores that have typical sizes of $\sim$0.1 pc? The project introduced in this paper will systematically study how these "dense gas" tracers behave on small scales of clumps to cores, aiming for a thorough understanding of the dense gas star formation law.

For the remainder of this paper we refer to CS, HCN and HCO$^{+}$ without specifying the transitions. However, we emphasize that the sensitivity to density varies strongly with energy level; the transitions from levels higher than those in this study, e.g., $J=2$ for CS, $J=1$ for HCN and HCO$^{+}$, are increasingly sensitive to higher densities. Multiple transitions of a single molecular species or possibly single transitions of different species can be indicative of the characteristic densities of a region.

\subsection{The role of stellar feedback in star formation}

Stellar feedback from massive stars can strongly influence their surrounding interstellar medium and regulate star formation through proto-stellar outflows, thermal feedback, photoionizing radiation, radiation pressure, main sequence winds from hot stars or supernova explosions \citep[e.g.,][]{Krumholz2014}. The combined effects of multiple feedback mechanisms can significantly reduce star formation rates and may play a major role in determining the star formation efficiency and the stellar initial mass function  in molecular clouds \citep[e.g.,][]{Krumholz2014}. Although stellar feedback usually has a negative effect on star formation, in the sense of restraining or terminating star formation, it is also possible for feedback to be positive by triggering new star formation \citep[e.g.,][]{Elmegreen1977,Whitworth1994a,Whitworth1994b,Krumholz2014,Wall2020}. By studying a large sample of infrared bubbles, \citet{Thompson2012} estimated that the fraction of massive stars in the Milky Way formed by a triggering process could be between 14 and 30 percent.

With ALMA, \citet{Liu2017} recently studied sequential high-mass star formation in a proto-cluster, the G9.62+0.19 complex. The ALMA observations resolved G9.62+0.19 into a massive filament that fragments into dozens of continuum sources that are at different evolutionary stages (from high-mass starless core candidates, high-mass proto-stellar objects, hot molecular cores, to UC H{\sc ii} regions), as also suggested in other works \citep{gar93,hof01,hof94,hof96,tes00,liu11,Dall'Olio2019}. \citet{Liu2017} also found no evidence for widespread low-mass proto-stellar core population in the G9.62+0.19 complex. They suggested that the sequential high-mass star formation in this region is the outcome of the stellar feedback from evolved H{\sc ii} regions formed in the same natal clump. The core fragmentation may be suppressed due to feedback from young OB proto-stars by heating the cores up and injecting turbulence through outflows, leading to an increase of their Jeans masses.

In addition, thermal feedback (from outflow heating, accretion luminosity or radiative heating from OB protostars) enhanced by stellar radiation heating in conjunction with magnetic field can be important for reducing the level of fragmentation and producing massive stars \citep[e.g.,][]{Offner2009,Krumholz2011,Myers2013}. In contrast to the magnetic field, which is most effective in more diffuse regions, radiation feedback is more efficient in suppressing fragmentation in the dense, central region of proto-clusters \citep{Myers2013}. \citet{Moscadelli2018} studied the feedback of a Hyper Compact (HC) H{\sc ii} region on its parental molecular core in the star-forming region G24.78+0.08. They found that the ionized gas of the HC H{\sc ii} region is expanding into the surrounding molecular gas. The shocks produced by the fast expansion of the ionized gas significantly influence the temperature distribution throughout its parental molecular core. In addition, outflows can also heat gas at parsec scale within clumps \citep[e.g.,][]{Wang2012}.

Studies of stellar feedback in protoclusters are still rare in cases. More systematic study is needed, however, to evaluate how formed OB proto-stars in proto-clusters influence the dense gas distribution and star formation efficiency in their natal clumps. With its unprecedented sensitivity, ALMA enables such a systemic study, through observing a large sample of proto-clusters in a very reasonable integration time.

\subsection{The role of filaments in star formation}

Thermal dust emission imaging surveys with the Herschel Space Observatory have discovered ubiquitous filamentary structures in nearby clouds and distant Galactic Plane giant molecular clouds (GMCs) \citep[e.g.,][]{Andre2010,Andre2014,Andre2019,Molinari2010,Arzoumanian2011,Juvela2012,Konyves2015,Wang2016,Arzoumanian2019,Schisano2020}. High resolution images from the space telescope Herschel and ground-based telescopes reveal complex internal structures inside filamentary clouds, including dense cores along or at the intersections of some of the filamentary substructures \citep{Wang2011,Wang2014,Andre2014,Konyves2015,Zhang15,Liu2018a,Liu2018b,Kainulainen2017}. Velocity coherent fiber-like substructures in filaments are also identified in nearby clouds \citep{Hacar2013,Hacar2016,Hacar2018,Gonzlez2019} and distant infrared dark clouds \citep{Chen2019}. Observations by Herschel revealed that more than 70\% of prestellar cores are embedded in larger, parsec-scale filamentary structures in nearby molecular clouds \citep{Andre2014,Konyves2015}. The fact that the cores reside mostly within the densest filaments with column densities exceeding $\sim7\times10^{21}$ cm$^{-2}$ strongly suggests a column density threshold for core formation \citep[e.g.,][and references therein]{Enoch2007,Andre2014}.

In addition, filaments also play an important role in cluster formation \citep[e.g.,][]{Stutz2016,Stutz2018}. Hub-filament systems have been revealed in high-mass star forming regions, implying a global hierarchical collapse scenario for proto-cluster formation \citep{Liu2012,Liu2015b,Chen2019}. Gas accreted along filaments continuously feeds the hub region, where massive stars may form \citep{Liu2012,Kirk2013,Peretto2013,Lu2018,Chen2019}. To date, however, only a few observational studies have been performed to trace the hypothesized accretion flows along filaments in proto-clusters.

Despite the importance of filaments in star formation, we still have too many missing links from observations to reconstruct a complete picture of star formation inside filaments. For example, do filaments in widely different environments, e.g., Galactic Plane GMCs or nearby clouds, show similar properties? Are hub-filament systems common in high-mass star-forming regions? To deepen our understandings of the roles of filaments in star formation, we need to systematically investigate a sample of filaments in both far and near clouds across the Galaxy. To separate cores from filaments in distant massive clumps, we need resolution of $<0.1$ pc, which requires angular resolution of $<2\arcsec$ at distances of $<$10 kpc. Only with ALMA, such a systemic high-resolution observational study is possible.

\section{ATOMS: ALMA Three-millimeter Observations of 146 Massive Star-forming regions}

To statistically investigate the star formation process in the Galaxy, we initiated the "ALMA Three-millimeter Observations of Massive Star-forming regions (ATOMS)" survey program at ALMA.  The main science goals of the ATOMS project are:\\

\noindent $\bullet$ To deepen the understandings of the dense gas star formation law by studying the spatial distributions of various dense gas tracers in a large sample of Galactic clumps and evaluating how much of molecular gas is participating in star formation;\\
\noindent $\bullet$ To investigate how stellar feedback from formed OB (proto)stars influences the surrounding gas distributions and the next generation of star formation in their natal clumps;\\
\noindent $\bullet$ To resolve filaments and to study their roles in proto-cluster formation.\\

\subsection{The sample of ATOMS}

\begin{figure*}
\centering
\includegraphics[scale=0.5]{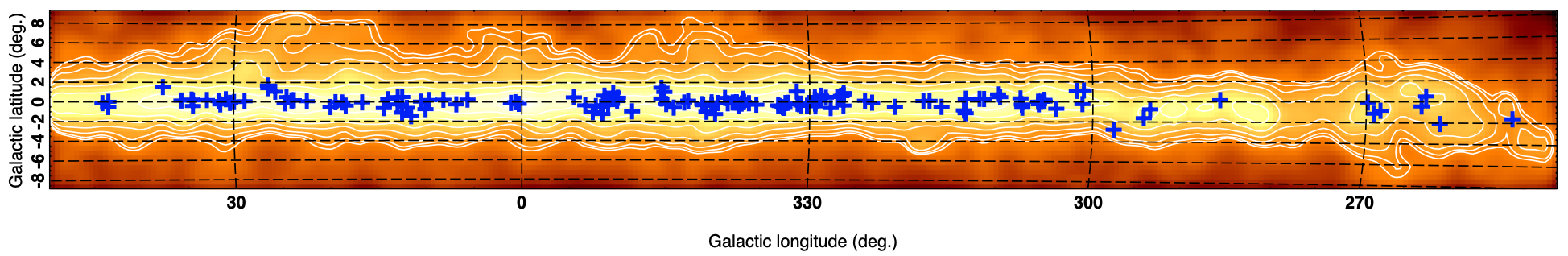}
\caption{Galactic distribution of the ATOMS targets ("+"). The background image and contours show Planck 857 GHz emission. The contours are from 34.6 to 874.5 MJy/sr with 10 steps on a logarithmic scale. \label{Galdist}}
\end{figure*}

\begin{figure}
\centering
\includegraphics[scale=0.6]{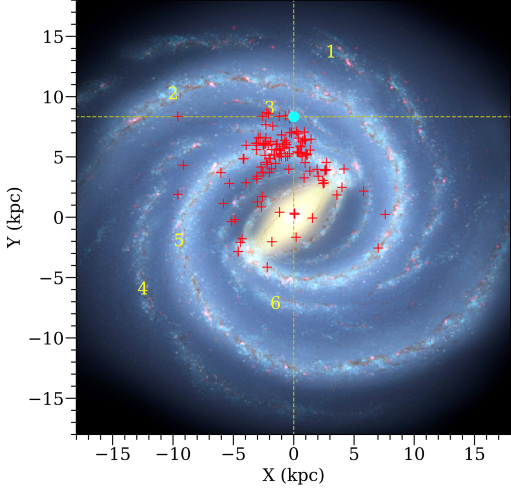}
\caption{Spatial distribution of the ATOMS targets ("+") projected onto a top-down schematic view of the Milky Way (artist's concept, R. Hurt: NASA/JPLCaltech/ SSC). The spiral arms are indicated using numbers from 1 to 6, referring to the Outer, Perseus, Local, Carina-Sagittarius, Scutum-Centaurus, and Norma arms. (Codes used for this plot is from \citet{Yuan2017}). \label{Milkway}}
\end{figure}

\begin{figure*}
\centering
\includegraphics[scale=0.6]{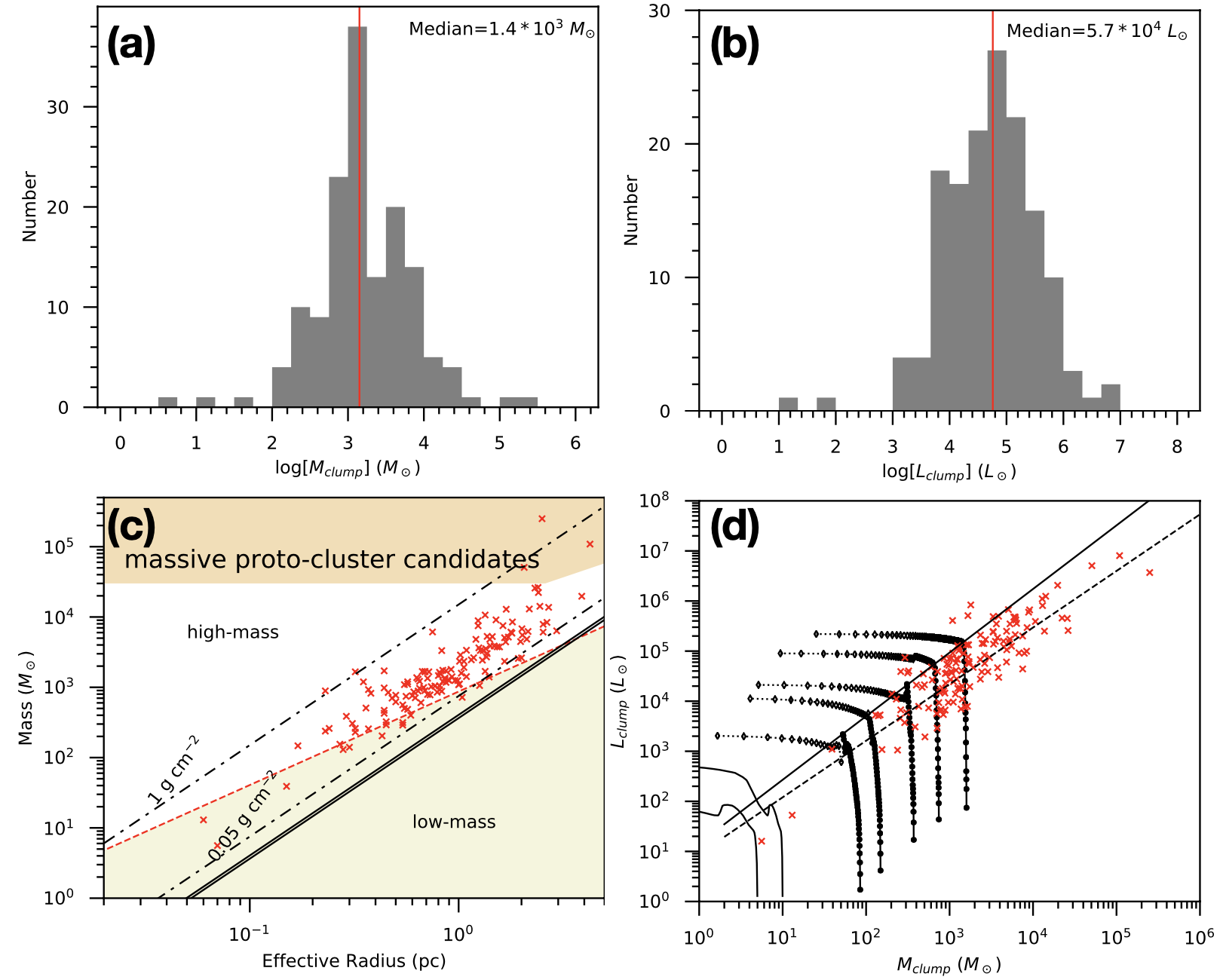}
\caption{(a) Distribution of clump masses of the ATOMS targets; the red line shows the median value; (b) Distribution of clump luminosity; (c) $M_{\rm clump}$ as a function of effective radii for ATOMS targets (red crosses). The unshaded area (above the red dashed line) delimits the region of high-mass star formation regions above the threshold $M>870M_{\sun}(r/pc)^{1.33}$ defined by \citet{Kauffmann2010}. The black solid lines show the mass surface density thresholds for ``efficient star formation" of 116 $M_{\sun}$~pc$^{-2}$ ($\sim$0.024 g~cm$^{-2}$) from \citet{Lada2010} and 129 $M_{\sun}$~pc$^{-2}$ ($\sim$0.027 g~cm$^{-2}$) from \citet{Heiderman2010}. The upper and lower dot-dashed lines show two mass surface density thresholds of 0.05 and 1 g~cm$^{-2}$ for high-mass star formation, which are given by \citet{Urquhart2014} and \citet{Krumholz2008}, respectively. The upper shaded region indicates the parameter space for massive protoclusters, defined in \citet{Bressert2012}. (d) Luminosity-mass diagram for ATOMS targets (red "+"). Evolutionary tracks for stars with final masses of 2.0, 4.0, 6.5, 8.0, 13.5, 18.0, and 35.0 $M_{\sun}$ are from \citet[solid tracks]{Saraceno1996} and \citet[tracks with symbols]{Molinari2008}. The solid and dashed lines are the best log-log fit for Class I and Class 0 sources, respectively, extrapolated in the high-mass regime by \citet{Molinari2008}. (Codes used for panels c, d are from \citet{Yuan2017}).\label{stathist}}
\end{figure*}

The ATOMS sample of 146 sources were selected from the CS $J=2-1$ survey of \citet{Bronfman1996}, a complete and homogenous molecular line survey of ultra compact (UC) H{\sc ii} region candidates in the Galactic plane. The sample of our 146 targets is \textit{\textbf{complete}} for proto-clusters with bright CS $J=2-1$ emission ($T_{\rm b}>$2 K), indicative of reasonably dense gas. It is also complete for the longitude range observable by SEST as well as ALMA \citep{Faundez2004}. Table \ref{atomsample} lists the basic parameters of this sample, including source IDs in observations (column 1), IRAS names (column 2), coordinates (columns 3-4), systemic velocities ($V_{\rm lsr}$, column 5), distances from the sun (column 6), Galactocentric distances ($R_{\rm GC}$, column 7), effective radii ($R_{eff}$, column 8), dust temperature ($T_{\rm dust}$, column 9), bolometric luminosity ($L_{\rm bol}$, column 10) and clump masses ($M_{\rm clump}$, column 11). There are 122 clumps with ATLASGAL and HiGAL counterparts, which have been analyzed in detail by \citet{Urquhart2018}. The basic parameters for these 122 clumps are adopted from \citet{Urquhart2018}. The basic parameters for the remaining clumps are adopted from \citet{Faundez2004}.

Figure \ref{Galdist} shows the Galactic distribution of the ATOMS targets on the Planck 857 GHz emission map. Figure \ref{Milkway} presents the spatial distribution of the ATOMS targets on a top-down schematic view of the Milky Way. From them, we can see that most (139) of the ATOMS targets are located in the first and fourth Galactic Quadrants of the inner Galactic Plane ($|b|<2\degr$). The distances of the sample clouds range from 0.4 kpc to 13.0 kpc with a mean value of 4.5 kpc. About 67\% of the sources are located closer than $\sim$5 kpc (see Figure \ref{Milkway}). We include 27 distant ($d>$7 kpc) sources that  are either close to the Galactic center or mini-starbursts (like W49N/I19078+0901), representing extreme environments for star formation. The sources have Galactocentric distances ranging from 0.5 kpc to 12.7 kpc, and are mostly located on spiral arms (see Figure \ref{Milkway}).

Figure \ref{stathist}a presents the distribution of the clump masses of the sample. The clump masses range from 5.6 to 2.5$\times10^5$ $M_{\sun}$, with a median value of 1.4$\times10^3$ M$_{\sun}$. The distribution of the bolometric luminosity of targets is shown in Figure \ref{stathist}b. The bolometric luminosity ranges from 16 to 8.1$\times10^6$ L$_{\sun}$, with a median value of 5.7$\times10^4$ L$_{\sun}$. All the targets in the sample have bolometric luminosity larger than 1$\times10^3$ L$_{\sun}$ except for I08076-3556 and I11590-6452. I08076-3556, also known as BHR 12 or CG 30, is a bright-rimmed cometary globule located in the Gum Nebula region at a distance of $\sim$400 pc, in which a low-mass binary system is forming \citep{Chen2008}.  I11590-6452 (also known as BHR 71) is also a bright-rimmed cometary globule that is forming a wide binary system \citep{Tobin2019}.

The radii of the sample range from 0.06 to 4.26 pc, with a median value of 0.86 pc, suggesting that the sample contains both nearby dense cores ($R_{\rm eff}\sim$0.1 pc) and distant clouds ($R_{\rm eff}$ of several pc). The dust temperature ranges from 18 to 46 K, with a median value of 29 K. \citet{Liu2016} observed this sample in HCN (4-3) and CS (7-6) lines. The line widths of HCN (4-3) range from 1.5 to 21 km~s$^{-1}$, with a median value of 6.4 km~s$^{-1}$. The line widths of CS (7-6) range from 1.6 to 23 km~s$^{-1}$, with a median value of 4.5 km~s$^{-1}$. Most of the targets tend to be gravitationally bound with an average virial parameter of $\sim$0.8 \citep{Liu2016}.

In a word, this is a very diverse sample that is suitable for statistical studies of the star formation process with different conditions, e.g., different density, temperature, turbulence level and luminosity of dominating OB protostars, in widely different Galactic environments covering a large range of Galactocentric distances.

Figure \ref{stathist}c shows the mass-radius diagram for the ATOMS sample. All the targets are above the mass surface density thresholds for "efficient star formation" given in \citet{Lada2010} and \citet{Heiderman2010}, implying that they are active star-forming regions. We notice that only three of the ATOMS targets are above the surface density threshold of 1 g~cm$^{-2}$, which is a stringent threshold for high-mass star formation given by \citet{Krumholz2008}. However, there are 134 clumps having mass surface density above the threshold for high-mass
star formation regions, $M>870~M_{\sun}(r/pc)^{1.33}$, defined by \citet{Kauffmann2010}. Based on studies of ATLASGAL clumps, \citet{Urquhart2014} also suggested a less stringent empirical threshold of 0.05 g~cm$^{-2}$ for high-mass star formation. There are 132 clumps fulfilling the thresholds given by \citet{Kauffmann2010} and \citet{Urquhart2014}, suggesting that more than 90\% of the ATOMS targets have the ability to form high-mass stars. Figure \ref{stathist}d shows the luminosity-mass diagram for the ATOMS sample. About 143 ATOMS targets are above the evolution tracks for stars with final masses larger than 8 M$_{\sun}$, indicating that the majority of the clouds are forming high-mass stars. The majority of the clumps are in the accelerated accretion phase \citep{Molinari2008}.

In the below, we describe the observations, data analysis, and illustrate the science goals of the ATOMS survey, using a case study for the G9.62+0.19 complex (known as I18032-2032 in Table \ref{atomsample}), which is an active high-mass star forming region located at a distance of 5.2$_{-0.6}^{+0.6}$ kpc \citep[from trigonometric parallax of 12 GHz methanol masers;][]{san09}. The G9.62+0.19 clump is unstable against gravitational collapse even if thermal, turbulent, and magnetic field support are taken into account together \citep{Liu2018c}. We should note that
throughout this paper the analysis of G9.62+0.19 and comparison with previous works, e.g., \citet{Liu2017}, is only applicable to this particular source. The ATOMS will allow to extend the analysis to a much larger sample of sources.

\subsection{ALMA Observations}

\begin{table*}
	\centering
	\caption{Observation logs of the ATOMS survey.}
	\label{obslog}
	\begin{tabular}{ccccccc} 
		\hline
 Source IDs & Obs date  & Min BL &	Max BL & Angular resolution & Maximum recovering scale &  rms per 0.122 MHz channel    \\
	     &            & (m)    &    (m)    & ($\arcsec$)        & ($\arcsec$)              &  (mJy~beam$^{-1}$)     \\
		\hline
        \multicolumn{7}{c}{ACA Observations}\\
        \hline
 1       &2019-10-14             & 8.9 & 48.0 &13.3 &76.2 & 49\\
 2-7     &2019-10-14             & 8.9 & 48.0 &13.3 &76.2 & 63\\
 8       &2019-10-19             & 8.9 & 48.0 &13.3 &72.9 & 46\\
 9-11    &2019-11-08             & 8.9 & 48.0 & 13.3 & 76.2 & 70 \\
 12-23   &2019-11-15             & 8.9 & 48.0 & 13.3 & 76.2 & 70\\
 24-30   &2019-11-13,2019-11-17  & 8.9 & 44.7 & 13.5 & 76.2 & 70\\
 31-36   &2019-11-12             & 9.6 &48.0 & 13.1 & 53.8 & 61 \\
37-62    &2019-11-17,2019-11-18  & 8.9 & 48.0 & 13.3 & 76.2 & 61\\
63-81    &2019-11-12             & 8.9 & 48.0 & 13.1 &76.2 & 93 \\
 82-88   &2019-10-31             &8.9  &45.0 & 13.8 & 76.2 & 58\\
89-101   &2019-11-19             & 8.9 & 48.0 &13.3 &76.2& 83\\
 102-104 &2019-09-30             & 9.6 & 48.0 & 13.1 &53.8 & 51\\
105-110  &2019-10-01,2019-10-13  & 9.1 & 48.0 & 13.1 & 65.4 & 48\\
111-123  &2019-11-02, 2019-11-03 &  8.9 & 44.7 & 13.5 & 76.2 & 80\\
124-131  &2019-10-31,2019-11-01  & 8.9 & 48.0 & 13.3 & 76.2  & 51\\
132-143  &2019-11-01,2019-11-04  &8.9 & 48.0 & 13.3 & 76.2 & 64 \\
144-146  &2019-09-30             &9.6   &48.0 &13.1 &53.8 & 67 \\
        \hline
        \multicolumn{7}{c}{ALMA 12-m Observations}\\
        \hline
  1       &2019-10-19    &15.1 & 783.5 & 1.2 & 14.5 & 3.7 \\
2-7       &2019-10-19    &15.1 &783.5 &1.2 &14.5 & 4.5\\
 8        &2019-11-02    & 15.1 & 500.2 & 1.7 & 19.0 & 6.2 \\
 9-11  &2019-11-11     &15.1 &500.2 &1.8 & 20.3 & 8.1 \\
12-23  &2019-11-13   &15.3  & 500.2 & 1.7 & 19.3 & 9.3\\
 24-30    &2019-11-12    & 15.1 & 500.2 & 1.8 & 19.5 & 9.6\\
 31-36    &2019-11-04    &15.1  & 500.2 &1.8  &20.3 & 5.9\\
 37-62    &2019-11-04    &15.1  & 500.2 & 1.7 & 19.9 & 9.0\\
 63-81    &2019-11-03    &15.1  & 500.2 & 1.7 & 20.1 & 6.4 \\
 82-88    &2019-11-03    &15.1  &500.2 &1.9 & 21.1 & 9.2\\
89-101    &2019-11-03    &15.1  &500.2 & 1.7 & 20.1 & 8.4\\
 102-104  &2019-10-28    & 15.1 & 696.8 & 1.4 & 18.1 & 6.6 \\
 105-110  &2019-10-31    & 15.1 & 696.8 &1.5 &19.9 & 8.1\\
  111-123 &2019-11-01    &15.1 &500.2 &1.7 &18.3 &8.5\\
 124-131  &2019-10-31    & 15.1 & 696.8 &1.5 &19.9 & 7.4 \\
 132-143  &2019-10-31    &15.1 & 696.8 & 1.5 &19.9 & 5.4 \\
 144-146  &2019-10-31    & 15.1 &696.8 &1.5 &19.9 & 7.9 \\
		\hline
	\end{tabular}
\end{table*}

\begin{table*}
	\centering
	\caption{Main targeted molecular lines in the ATOMS survey.}
	\label{spws}
	\begin{tabular}{cccccccccc} 
		\hline
spw & bandwidth & $\delta V$& Species & Transition & Rest frequency &  E$_u$/k & n$_{crit}^a$ (100 K) & n$_{eff}^a$ (100 K) & Note   \\
	&(MHz)	    &  (km~s$^{-1}$)    & &      & (GHz)          &  (K)     & (cm$^{-3}$)&  (cm$^{-3}$)     \\
		\hline
spw1 & 58.59 & 0.424 & H$^{13}$CN	     &1-0       & 86.339918   & 4.14 & 9.7$\times10^4$ & 6.5$\times10^4$  & high-density tracer \\
spw2 & 58.59 & 0.422 & H$^{13}$CO$^+$	 &1-0       & 86.754288   & 4.16 & 2.0$\times10^4$ & 1.1$\times10^4$  & high-density and ionization tracer \\
spw3 & 58.59 & 0.419 & CCH	             &1-0       & 87.316898   & 4.19 & 1.9$\times10^5$ &                  & tracer of photodissociation regions  \\
spw4 & 58.59 & 0.421 & SiO	             &2-1       & 86.84696    & 6.25 & 1.7$\times10^5$ &                  & shock/outflow tracer  \\
spw5 & 58.59 & 0.206 & HCN	             &1-0       & 88.631847   & 4.25 & 1.1$\times10^5$ & 1.7$\times10^3$  & high-density/infall/outflow tracer    \\
spw6 & 58.59 & 0.205 & HCO$^+$	         &1-0       & 89.188526   & 4.28 & 2.3$\times10^4$ & 2.6$\times10^2$  & high-density/infall/outflow tracer \\
spw7 & 1875.00 & 2.973 &CH$_3$OH	     &2(1,1)-1(1,0)A & 97.582798 & 21.56& 4.8$\times10^4$ &                & high-density/hot-cores/shock tracer\\
     &         &       &CS	             &2-1       & 97.980953   & 7.05 & 5.5$\times10^4$ & 4.7$\times10^3$  & high-density/infall/outflow tracer  \\
     &         &       &SO	             &3(2)-2(1) & 99.29987    & 9.23 & 3.0$\times10^5$ &                  & high-density/shock/outflow tracer   \\
     &         &       &H$_{\alpha}$   & H$_{40\alpha}$         & 99.022952&               &   &               & tracer of ionized gas/HII regions              \\
spw8 & 1875.00 &2.914  & HC$_3$N	         &11-10     &100.07639   &28.82  & 9.2$\times10^4$ & 1.1$\times10^4$  & high-density/hot-cores tracer  \\
		\hline
\multicolumn{10}{l}{$^a$ The critical density and effective excitation density are from \citet{Shirley2015}.}\\
	\end{tabular}
\end{table*}

The ALMA observations of the ATOMS survey were conducted from late September to mid November 2019 with both the Atacama Compact 7 m Array (ACA; Morita Array) and the 12-m array (C43-2 or C43-3 configurations) with band 3 (Project ID: 2019.1.00685.S; PI: Tie Liu). The observations including Source IDs, observation dates, minimum baselines, maximum baselines, angular resolution, Maximum recovering scale (MRS) and rms level per 0.122 MHz channel in the ATOMS survey are summarized in Table \ref{obslog}. The ACA observations of G9.62+0.19 (I18032-2032, number 108 in Table \ref{obslog}) were conducted on the 1st and 13th October, 2019, with two executions. The 12-m array observations of G9.62+0.19 were conducted on the 31st October, 2019.

We conducted single-pointing observations with ALMA. The typical 12-m array time on source is $\sim$ 3 minutes, while typical ACA observing time per source is $\sim$ 8 minutes. The angular resolution and maximum recovered angular scale for the ACA observations are $\sim13.1\arcsec-13.8\arcsec$ and $\sim53.8\arcsec-76.2\arcsec$, respectively. The angular resolution and maximum recovering angular scale for the 12-m array observations are $\sim1.2\arcsec-1.9\arcsec$ and $\sim14.5\arcsec-20.3\arcsec$, respectively. These ACA and 12-m array
observations are sensitive to angular scales smaller than $\sim60\arcsec$ and $\sim20\arcsec$, respectively. The achieved angular resolutions of the 12-m array observations are usually better than 2$\arcsec$, enabling us to resolve dense cores ($\sim0.1$ pc in size; or 2$\arcsec$ at 10 kpc distance) in the most distant sources. The obtained sensitivity in 12-m array observations ranges from 3.7 to 9.6 mJy~beam per 0.122 MHz channel.

The observations employed the Band 3 receivers in dual-polarization mode. Eight spectral windows (SPWs) were configured to cover 11 commonly-used lines including the dense gas tracers (e.g., $J$=1-0 of HCO$^{+}$, HCN and their isotopes), hot core tracers (e.g., CH$_3$OH, HC$_{3}$N), shock tracers (e.g., SiO, SO) and ionized gas tracers (H$_{40\alpha}$),  as listed in Table \ref{spws}. The second and third columns of Table \ref{spws} list the total bandwidth and spectral resolution of each spectral window (SPW, column 1). The specie names, transitions, rest frequencies and upper energies are listed in columns 4-7. The critical density (n$_{crit}$), as the density for which the net radiative decay rate from j to k equals the rate of collisional depopulation out of the upper level j for a multi-level system, of these transitions at 100 K are shown in the 8th column. In the 9th column, we also listed the effective excitation density (n$_{eff}$) at 100 K, the density which results in a molecular line with an integrated intensity of 1 K~km~s$^{-1}$ \citep{Evans1999,Shirley2015}.

Two wide SPWs, each one having 1875 MHz bandwidth and $\sim$3 km~s$^{-1}$ spectral resolution are used for continuum emission and line scan survey. The SPWs (1-6) at the lower band have sufficiently high spectral resolution ($\sim$0.2-0.4 km~s$^{-1}$) to resolve line profiles in high-mass star forming regions (typical line width of several km~s$^{-1}$) for studying kinematics.

Calibration and imaging were carried out using the CASA software package version 5.6 \citep{McMullin2007}. The ACA data and 12-m array data were calibrated and imaged separately.

\begin{figure*}
\centering
\includegraphics[angle=0,scale=0.45]{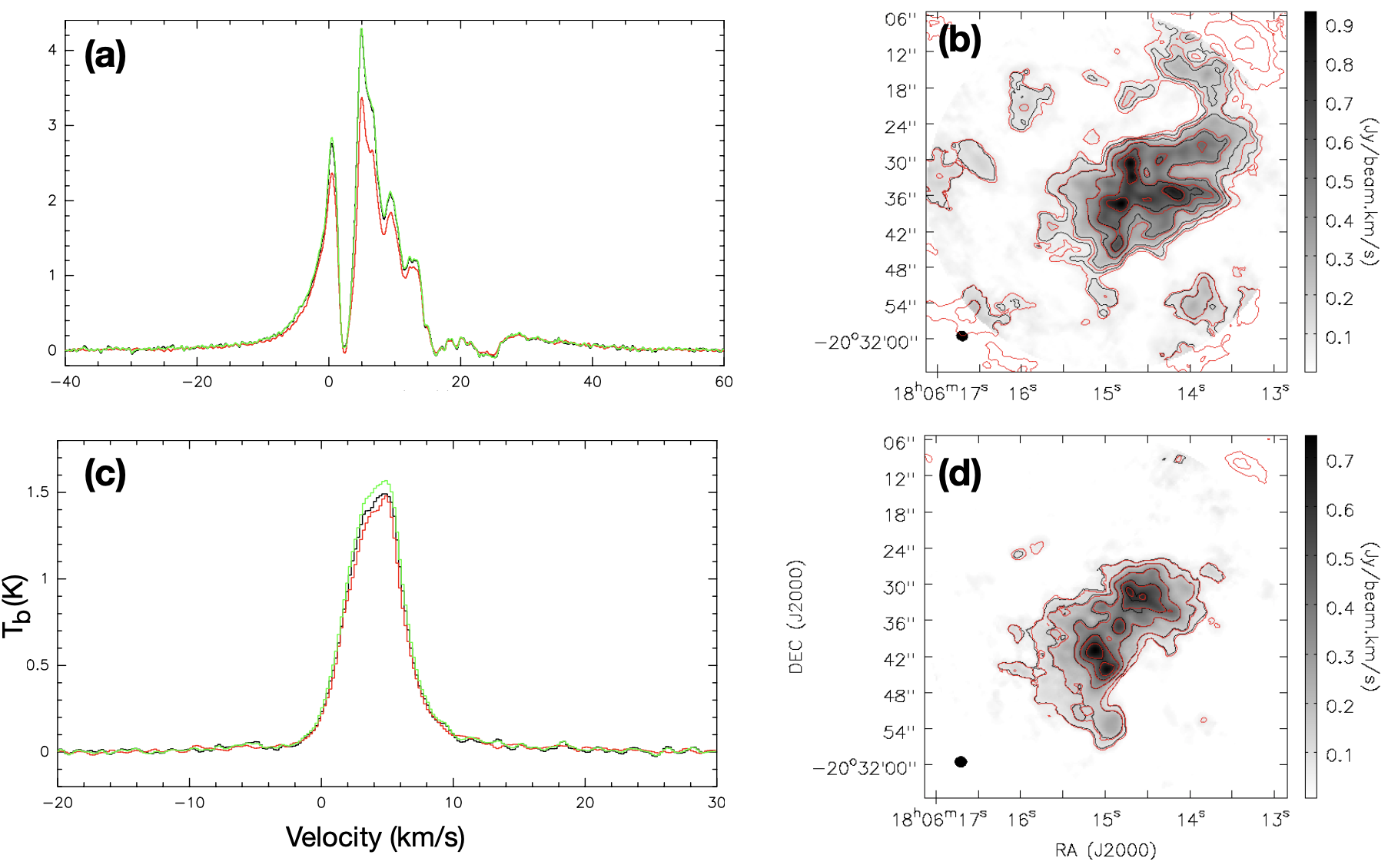}
\caption{(a) HCO$^+$ spectra in G9.62+0.19 averaged over a 30$\arcsec$ region at the central position (R.A.(J2000)=18:06:14.99, decl.(J2000)=-20:31:35.4). The spectra from ACA data, 12-m array data and combined 12-m and ACA data are shown in black, red and green, respectively. (b) Comparison of the integrated intensity maps of HCO$^+$ emission from 12-m array data (gray image and black contours) and combined data (red contours). The contours are [0.1, 0.2,0.4,0.6,0.8]$\times F_{\rm peak}$. The peak integrated intensities ($F_{\rm peak}$) in 12-m array data and in combined data are 0.93 and 0.96 Jy~beam$^{-1}$~km~s$^{-1}$, respectively. (c) Same as in panel a but for H$^{13}$CO$^+$. (d) Same as in panel b but for H$^{13}$CO$^+$. The peak integrated intensities ($F_{\rm peak}$) in 12-m array data and in combined data are the same as 0.75 Jy~beam$^{-1}$~km~s$^{-1}$. \label{combine}}
\end{figure*}

The ACA data and 12-m array data can be combined together to recover very extended emission that is missed in 12-m array observations. In Figure \ref{combine}, we simply combine the 12-m array data and ACA data of G9.62+0.19 in Fourier space with the CASA task "feather" for HCO$^{+}$ and H$^{13}$CO$^{+}$. Panels a and c of Figure \ref{combine} present the spectra of HCO$^{+}$ and H$^{13}$CO$^{+}$, respectively. We found that 12-m array data alone has recovered more than 90\% of the flux in both H$^{13}$CO$^{+}$ and HCO$^{+}$ emission. Panels b and d of Figure \ref{combine} show their integrated intensity maps from 1.5 to 6.5 km~s$^{-1}$. The morphology of integrated intensity maps constructed with 12-m array data alone is nearly identical to that of the maps constructed with combined data (see Figure \ref{combine}b and \ref{combine}d). We will more properly combine the ACA data and 12-m array data in future studies and discuss the data combination issue in a forthcoming paper (Wang et al. 2020, in preparation). The combined data, however, also suffer a non trivial missing flux problem without zero-spacing information from Total-Power (TP) observations. However, this is also advantageous for studying dense structures inside these "ATOMS" clumps. This filters out the large scale emission, making it easier the identification of compact cores or dense filamentary structures (see sections \ref{dustcore} and \ref{gascore}). We should note that dense cores are usually very compact (see Table \ref{continuumpara} for example) and have angular sizes much smaller than the maximum recovering angular scale ($\sim20\arcsec$) in the 12-m array observations, indicating that the missing flux will not be a problem in deriving physical parameters (like fluxes, densities and masses) for dense cores using the 3 mm continuum emission.

In this work, we mainly focus on the spatial distributions of various gas tracers. The 12-m  array data alone are good enough for this purpose. Hence, in this paper, the ACA data and 12-m array data were not combined for the case study of G9.62+0.19..

We construct continuum visibility data using the line-free spectral channels for G9.62+0.19. The continuum image of G9.62+0.19 from 12-m array data reaches a
1$\sigma$ rms noise of 0.2 mJy in a synthesized beam of 1.56$\arcsec\times1.38\arcsec$ (P.A.=87.62$\degr$). The synthesized beams for molecular lines of G9.62+0.19 are slightly different from those for continuum. The typical synthesized beam is 1.9$\arcsec\times1.7\arcsec$ for lines at lower band (SPWs 1-6), and is 1.6$\arcsec\times1.4\arcsec$ for lines at upper band (SPWs 7-8). The 1$\sigma$ rms noise is about 8 mJy~beam$^{-1}$ at SPWs 1-4 per 0.122 MHz channel. SPWs 5-6 have the best spectral resolution of 0.06 MHz and their 1$\sigma$ rms noise is about 10 mJy~beam$^{-1}$ per channel. The 1$\sigma$ rms noise at the two wide SPWs is about 3 mJy~beam$^{-1}$ per channel. The continuum image constructed from ACA data reaches a 1$\sigma$ rms noise of 4 mJy in a synthesized beam of 17.23$\arcsec\times9.46\arcsec$ (P.A.=-82.87$\degr$). The 1$\sigma$ rms noise for the ACA observations are $\sim$40 mJy~beam$^{-1}$, $\sim$80 mJy~beam$^{-1}$, and $\sim$15 mJy~beam$^{-1}$ per channel for lines in SPW 1-4, SPWs 5-6, and SPWs 7-8, respectively. The angular resolution and sensitivity achieved in observations of G9.62+0.19 are comparable to the median values of the whole sample.

\section{Science goals and a first look at G9.62+0.19}

In this section, we present the case study of the proto-cluster complex, G9.62+0.19, and introduce the exemplar sciences cases that can be addressed with the ATOMS survey.

\subsection{Filaments and dense cores}\label{dustcore}

The 3 mm continuum emission will be used to study how the massive clumps fragment into filaments and individual dense cores that will form new stars.

\begin{figure}
\centering
\includegraphics[width=\columnwidth]{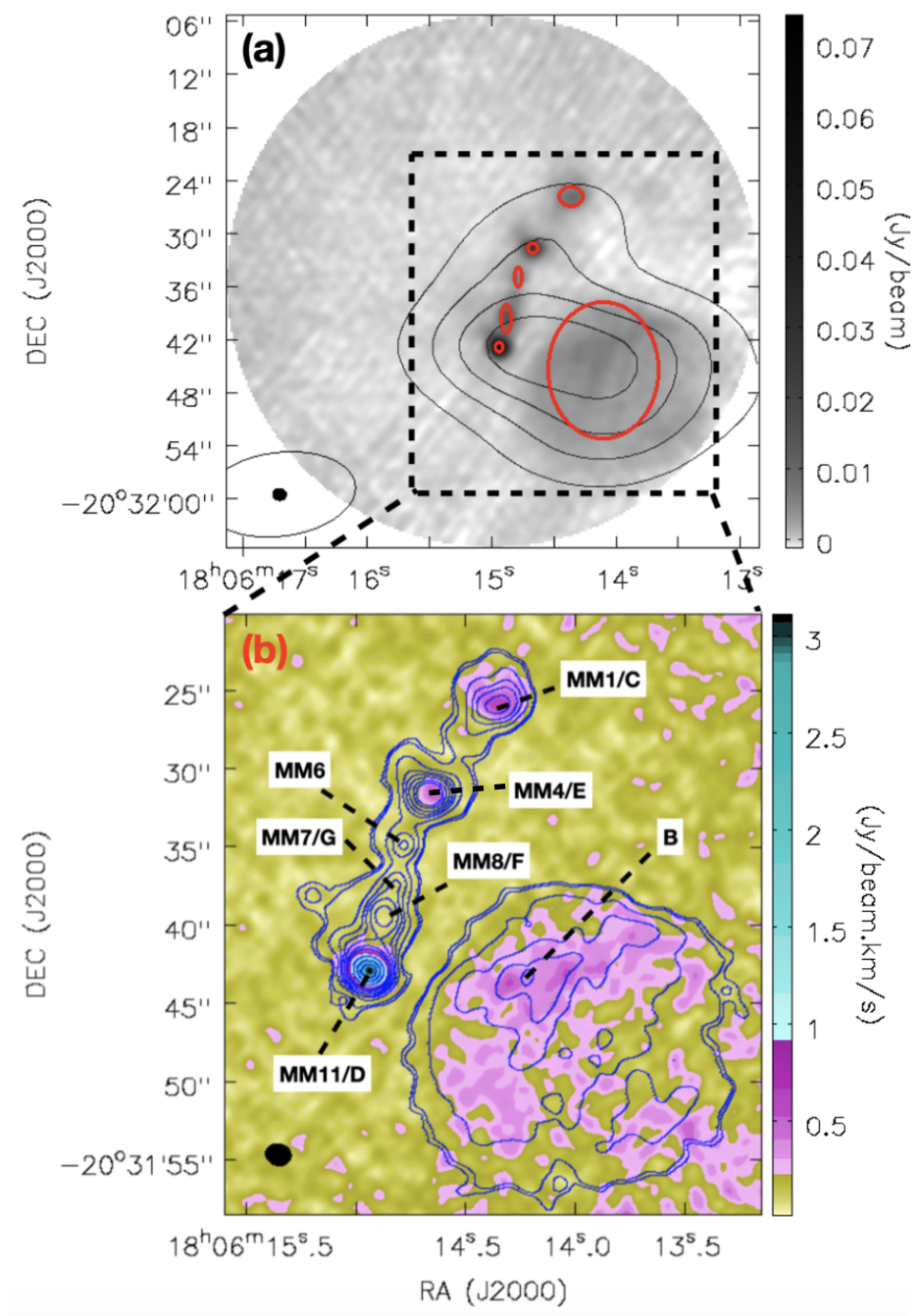}
\caption{The 3 mm continuum emission of G9.62+0.19. (a) The 3 mm continuum emission from ACA observations is shown as contours at [0.2,0.4,0.6,0.8]$\times$0.225 Jy~beam$^{-1}$. The starting (faintest) contour corresponds to $\sim$10$\sigma$ ($\sigma$=4 mJy~beam$^{-1}$). The background image is the 3 mm continuum emission from the 12-m array observations. The red ellipses show the continuum sources identified with FellWalker.  (b) The 3 mm continuum emission from 12-m array observations is shown as blue contours with contour levels of [0.01,0.013,0.03,0.05,0.07,0.1,0.15,0.2,0.4,0.6,0.8]$\times$74.1 mJy~beam$^{-1}$. The starting (faintest) contour corresponds to $\sim$3.5$\sigma$ ($\sigma$=0.2 mJy~beam$^{-1}$). The background (colored) image is the intensity of the H$_{40\alpha}$ line integrated from -50 to 50 km~s$^{-1}$. The names \citep[from][]{Liu2017} of seven main components in the 3 mm continuum emission are indicated in the image.  \label{continuum}}
\end{figure}

Figure \ref{continuum} shows the 3 mm continuum emission of G9.62+0.19. The 3 mm continuum emission from ACA observations is shown as contours in Figure \ref{continuum}a and reveals only one single clump. The 3 mm continuum emission from high-resolution 12-m array data is shown as gray image in Figure \ref{continuum}a and contours in \ref{continuum}b, which reveals the expanding cometary-like H{\sc ii} region, "B", to the west, and a massive fragmented filament to the east.

In the ATOMS survey, we apply the FellWalker algorithm in the
Starlink CUPID package \citep{Berry2007,Berry2015}, for source extraction. The core of the
FellWalker algorithm is a gradient-tracing scheme consisting of
following many different paths of steepest ascent in order to reach
a significant summit, each of which is associated with a clump
\citep{Berry2007}. FellWalker is less dependent on specific
parameter settings than CLUMPFIND \citep{Berry2007}, and has been widely used in continuum surveys \citep[e.g.,][]{Eden2017,Eden2019,Liu2018a}. Source extraction was conducted in regions above a threshold of 3$\sigma$. The minimum number of contiguous pixels (pixel size=0.27$\arcsec$) was set at 12, equaling to those contained in half of the synthesized beam, which was used to define the smallest structures, i.e., "cores" \citep[see][]{Sanhueza2019}. We also tested FellWalker with larger minimum number (16 and 25) of contiguous pixels and found that the results were not changed. The detected sources were fitted with a 2D Gaussian profile and de-convolved with the beam size. With FellWalker, we identified six main components in the 3 mm continuum emission of G9.62+0.19, as shown in Figure \ref{continuum}a.Their de-convolved sizes, peak flux density and total flux density of the 3 mm continuum emission are listed in columns 2-4 of Table \ref{continuumpara}. Another component, MM7/G, is not resolved in 3 mm continuum, but is resolved in other hot core tracers like HC$_3$N (11-10). MM7/G is also marked on Figure \ref{continuum}b and listed in Table \ref{continuumpara}.

\begin{figure}
\centering
\includegraphics[width=\columnwidth]{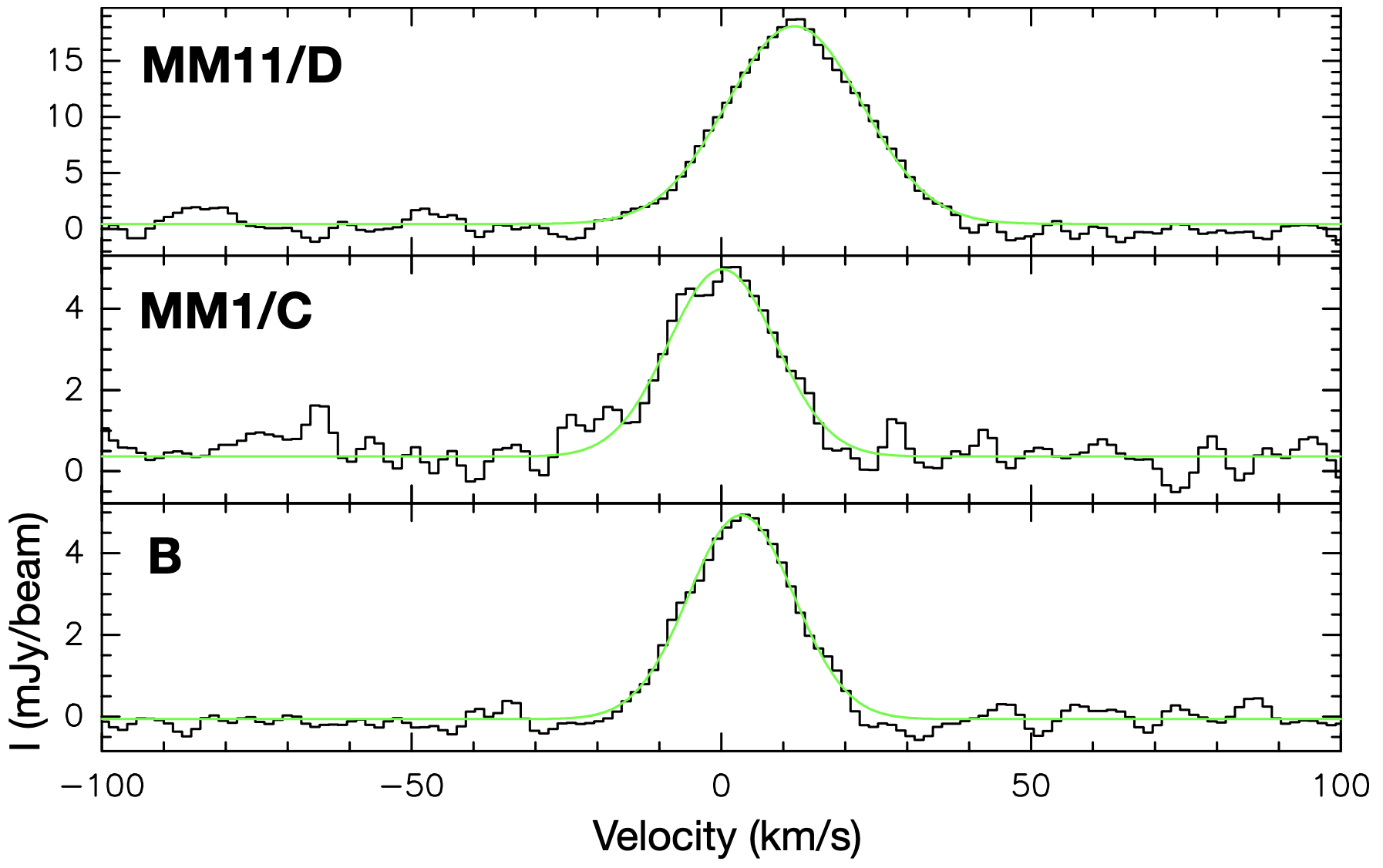}
\caption{Source-averaged H$_{40\alpha}$ RRL in G9.62+0.19. The green lines are gaussian fits. \label{Ha}}
\end{figure}

\begin{table*}
	\centering
	\caption{Parameters of continuum sources in G9.62+0.19.}
	\label{continuumpara}
	\begin{tabular}{ccccccccc} 
		\hline
		Name &  \multicolumn{4}{c}{3 mm continuum} & &\multicolumn{3}{c}{H$_{40\alpha}$}         \\
\cline{2-5}\cline{7-9}
                &  size & $I$       & $S_{\nu}$    & $S_{\rm free-free}$ & & $I_{p}^a$   & $V_{lsr}$ & FWHM   \\
		        &       &(mJy~beam$^{-1}$) & (mJy)          &(mJy)            & & (mJy~beam$^{-1}$)  & (km~s$^{-1}$)  & (km~s$^{-1}$) \\
		\hline
		B                        &      15.30$\arcsec\times13.31\arcsec$(142$\degr$) & 4.6          &  437   &359 & &5.0 & 3.2 & 20.0  \\
	       MM1/C                 &      3.01$\arcsec\times2.10\arcsec$(132$\degr$)   & 8.17         & 32.6   &32  & &6.6 & 0.2 & 20.4  \\
                MM4/E.           &      1.28$\arcsec\times1.02\arcsec$(89$\degr$)    & 25.4         & 40.9   &$<$2& &$<$3&     &$<20$\\
                 MM6             &      1.75$\arcsec\times1.05\arcsec$(173$\degr$)   & 5.4          & 10.4                   \\
                MM7/G            &                                                   & 5.7          & 5.7                    \\
                MM8/F            &      2.53$\arcsec\times1.28\arcsec$(166$\degr$)   &13.2          & 35.5   &$<$3& &$<$3&     &$<20$ \\
                MM11/D           &      1.04$\arcsec\times0.74\arcsec$(72$\degr$)    & 71.6          & 98.8   &73  & &17.8 & 11.8 & 25.8  \\
		\hline
\multicolumn{9}{l}{$^a$ The values for MM4/E and MM8/F are upper limits based on rms level.}\\
	\end{tabular}
\end{table*}

At 3 mm, both thermal dust emission and free-free emission from ionized gas could contribute to the continuum emission. However, since not all the ATOMS targets have been observed at centimeter wavelengths, we cannot easily compile centimeter continuum emission data to help removing any free-free contribution to the 3 mm continuum emission. Instead, we use the H$_{40\alpha}$ radio recombination line (RRL) to estimate the contribution of free-free emission in continuum by deriving the line-to-continuum intensity ratio \citep{Condon2016,Liu2019}. The integrated intensity map of H$_{40\alpha}$ is shown as color image in Figure \ref{continuum}b. H$_{40\alpha}$ is clearly detected towards "B", "MM1/C" and "MM11/D", and is marginally detected toward MM4/E. Figure \ref{Ha} presents the source-averaged H$_{40\alpha}$ spectra. The spectra were fitted with gaussian profiles, and the fit results (peak intensity I$_{\rm p}$, systemic velocity V$_{\rm lsr}$ and line width FWHM) are summarized in columns 6-8 of Table \ref{continuumpara}. The line widths of H$_{40\alpha}$ of  "B" and "MM1/C" are $\sim$20 km~s$^{-1}$. "MM11/D" has the largest line width ($\sim$25 km~s$^{-1}$) and velocity ($\sim$12 km~s$^{-1}$) in H$_{40\alpha}$ line emission. The velocity ($\sim$12 km~s$^{-1}$) of ionized gas in "MM11/D" is significantly redshifted relative to that ($\sim$4.5 km~s$^{-1}$) of the molecular gas, indicating a fast flow in the ionized gas as also seen in the Hyper Compact (HC) H{\sc ii} region G24.78+0.08 \citep{Moscadelli2018}.

We estimated the free-free emission at 3 mm for "B", "MM1/C" and "MM11/D", {which are listed in column 5 of Table \ref{continuumpara}}. An electron temperature of $\sim$6760 K is adopted, which is derived from the electron temperature gradient in the Galactic disk \citep{Quireza2006}. Higher electron temperature will lead to higher free-free emission contributions. The electron temperature could be better constrained by observing multiple RRLs in future \citep{Zhu19}. We found that the 3 mm continuum emission in "B", "MM1/C" and "MM11/D" is dominated by free-free emission, as suggested in previous works \citep{hof96}. The free-free emission in other sources is negligible.  Since the properties of all the continuum sources have been well discussed in \citet{Liu2017}, we will not derive any other parameters from the 3 mm continuum emission in this work.

Fragmented massive filaments are commonly seen in infrared dark clouds \citep[IRDCs;][]{Zhang11,Wang2011,Wang2012,Wang2014,Wang2016,Zhang15,Sanhueza2019}. Those massive filaments in IRDCs also fragment into a chain of dense cores as seen in G9.62+0.19. However, those filaments in IRDCs seem to be undergoing cylindrical fragmentation governed by turbulence \citep{Wang2011,Wang2014}. In contrast, the fragmentation of the massive filament in G9.62+0.19 can be well explained by thermal Jeans instability \citep{Liu2017}.

Within the ATOMS project, we will systematically investigate the fragmentation process of proto-clusters as G9.62+0.19, and compare with IRDC clumps. This will help figure out the dominating mechanisms in dense-core formation in different Galactic environments and conditions.

\subsection{Spatial distribution of molecules}

\begin{figure*}
\centering
\includegraphics[angle=0,scale=0.9]{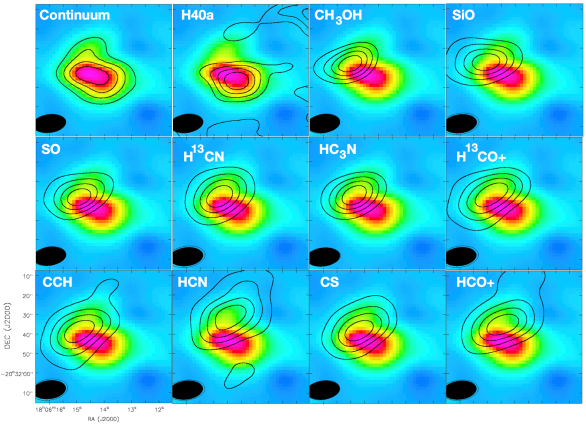}
\caption{The integrated intensity maps from the ACA images of molecular line emission in G9.62+0.19 are shown in contours. The background color image is the 3 mm continuum emission. The contour levels are [0.2,0.4,0.6,0.8]$\times F_{\rm peak}$. The $F_{\rm peak}$ for various tracers are: Continuum: 0.225 Jy~beam$^{-1}$;  H$_{40\alpha}$: 7.8 Jy~beam$^{-1}$~km~s$^{-1}$; CH$_3$OH: 10.3 Jy~beam$^{-1}$~km~s$^{-1}$; SiO: 10.3 Jy~beam$^{-1}$~km~s$^{-1}$;  SO: 28.8 Jy~beam$^{-1}$~km~s$^{-1}$; H$^{13}$CN: 15.3 Jy~beam$^{-1}$~km~s$^{-1}$; HC$_3$N: 34.5 Jy~beam$^{-1}$~km~s$^{-1}$; H$^{13}$CO$^+$: 15.0 Jy~beam$^{-1}$~km~s$^{-1}$; CCH: 13.0 Jy~beam$^{-1}$~km~s$^{-1}$; HCN: 12.1 Jy~beam$^{-1}$~km~s$^{-1}$; CS: 53.9 Jy~beam$^{-1}$~km~s$^{-1}$; HCO$^+$: 23.6 Jy~beam$^{-1}$~km~s$^{-1}$. \label{ACA}}
\end{figure*}

One of the main science goals of the ATOMS project is to investigate the small-scale distributions of various molecular gas tracers within massive star forming regions. In particular, we will be more focused on those dense gas tracers, e.g., $J$=1-0 of HCN and HCO$^{+}$, and $J$=2-1 of CS, which are commonly used in studying star formation scaling relations. By comparing with observational results in nearby clouds \citep{Kauffmann2017,Pety2017,Shimajiri2017}, we will investigate how well those gas tracers unveil the structures and masses of molecular clumps that are located in different Galactic environments and have different physical conditions, e.g., density, temperature, and UV radiation.

\subsubsection{Gas clumps in ACA observations of G9.62+0.19}

The ACA observations have large FOV ($\sim2\arcmin$) and maximum recovering scale (MRS; $\sim1\arcmin$), enabling us to study the overall properties of gas emission at clump scale, such as total line luminosity, emission extent, and virial masses \citep{Liu2020}. Figure \ref{ACA} presents the integrated intensity (from 1.5 to 6.5 km~s$^{-1}$) maps of various gas tracers in ACA observations of G9.62+0.19. All the gas tracers reveal a single gas clump that can be well modeled by a 2D Gaussian profile. We performed 2D Gaussian fitting in the gas emission regions bounded by the 20\% contour of their peak values. From 2D Gaussian fitting (see section \ref{flux ratios}), we found that the spatial distributions of the emission in all lines are smaller than the MRS ($\sim1\arcmin$) in ACA observations. In addition, except HCN and HCO$^{+}$, the other lines have FWHM angular sizes smaller than 20$\arcsec$, suggesting that 12-m array observations with MRS$\sim20\arcsec$ could recover most of the line flux.

In a companion work \citep{Liu2020}, we report studies of the star formation scaling relationships between the total bolometric emission ($L_{\rm bol}$ or $L_{\rm TIR}$) and the molecular line luminosities of J=1-0 transitions of H$^{13}$CN, H$^{13}$CO$^+$, HCN, and HCO$^+$ with data obtained from ACA observations for the whole ``ATOMS" sample.  We found that both main lines and isotopologue lines can well reveal the total masses of dense gas in Galactic molecular clumps.

However, although they are all good tracers of total masses of dense gas, their spatial distributions may be different. The upper panel of Figure \ref{ACA-spec} shows the 3 mm continuum emission in color-scale overlaid with H$^{13}$CO$^+$ integrated intensity map in contours. The emission peaks of various gas tracers as revealed in the ACA observations of G9.62+0.19 are marked on this plot. The emission peaks of most gas tracers (except HCN, HCO$^{+}$ and H$_{40\alpha}$) are close to the H$^{13}$CO$^{+}$ emission peak. The ionized gas traced by H$_{40\alpha}$ is well accociated with the 3 mm continuum emission but is clearly offset from molecular gas and peaks to the south-west. The emission peaks of HCN and HCO$^{+}$ are located to the north-west of the H$^{13}$CO$^{+}$ emission peak, indicating that the main lines may show different spatial distributions from their isotopologue lines. The detailed spatial distributions of various gas tracers can be investigated with the high resolution 12-m array data.

\subsubsection{Moment maps in 12-m array observations}

\begin{figure}
\centering
\includegraphics[width=\columnwidth]{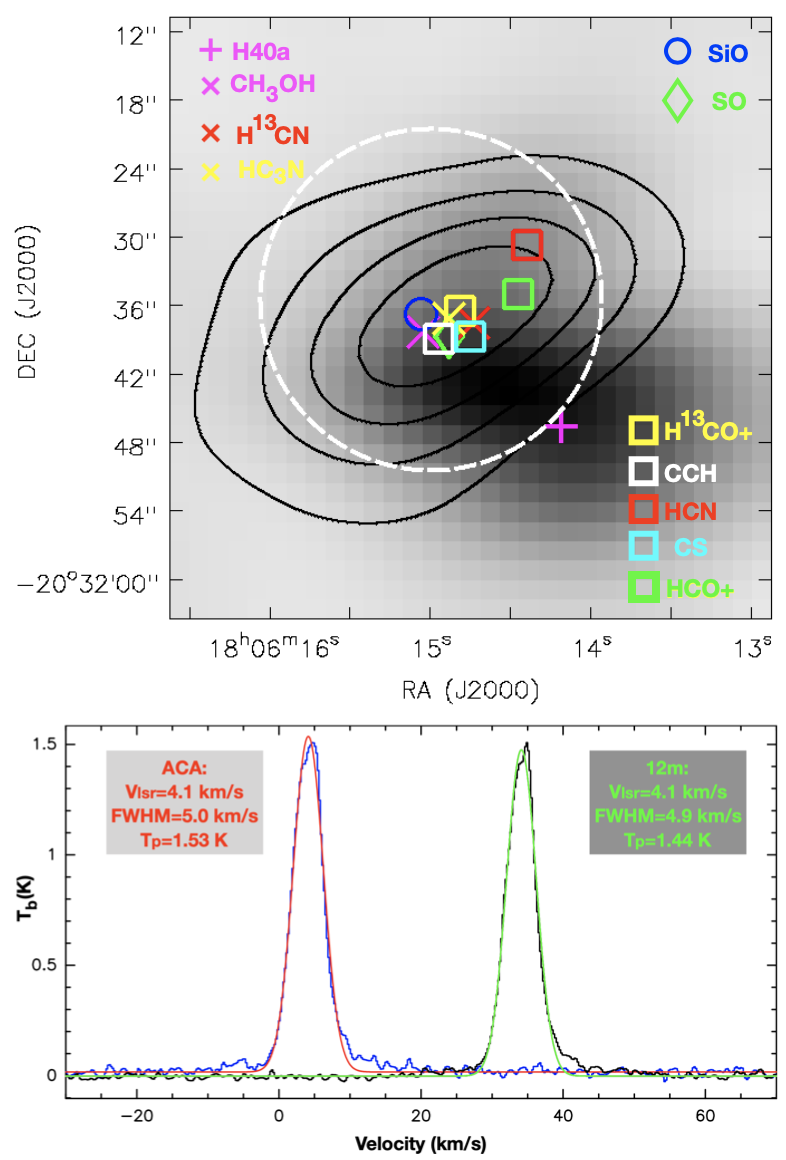}
\caption{(a) The 3 mm continuum emission of G9.62+0.19 in ACA observations is shown as the grey-scale. The integrated intensity from 1.5 to 6.5 km~s$^{-1}$ of H$^{13}$CO$^{+}$ is shown as contours at levels of [0.2,0.4,0.6,0.8]$\times$15 Jy~beam$^{-1}$~km~s$^{-1}$. The peak positions of various gas tracers of G9.62+0.19 derived from ACA observations are marked with different symbols.  (b) The H$^{13}$CO$^{+}$ spectra from the 12-m array and ACA averaged over a 30$\arcsec$ diameter region (white dashed circle in panel a) are displayed in blue and black, respectively. The spectrum from 12-m array data is shifted by 30 km~s$^{-1}$. The red and green curves are gaussian fits. The results of gaussian fits are also displayed on the panel. \label{ACA-spec}}
\end{figure}

In the lower panel of Figure \ref{ACA-spec}, we compare the H$^{13}$CO$^{+}$ spectra from ACA and 12-m array observations. The spectra that were averaged over 30$\arcsec$ region centered at the emission peak are nearly identical in both observations. The 12-m array observation recovers more than 92\% of H$^{13}$CO$^{+}$ flux in ACA observations. From gaussian fits to the spectra, we find a systemic velocity of $\sim$4 km~s$^{-1}$. The line width is about 5 km~s$^{-1}$. To compare the spatial distributions of various gas tracers (see Table \ref{spws}) revealed in 12-m array observations, we integrate their intensity in the same velocity interval that is from 1.5 to 6.5 km~s$^{-1}$ above a threshold of 3$\sigma$. This velocity interval is chosen based on the line width of H$^{13}$CO$^{+}$ to avoid contamination from outflow emission. In Figures \ref{lineimages1}, \ref{lineimages2} and \ref{lineimages3}, their integrated intensity (Moment 0) maps (left panels), intensity-weighted velocity (Moment 1) maps (middle panels) and intensity-weighted velocity dispersion (Moment 2) maps (right panels) are presented as color images. The moment maps were constructed from the data after imposing a cutoff of 3$\sigma$. The 3 mm continuum emission is shown as contours on the moment 0 maps for comparison. The contours on moment 1 and 2 maps are showing the integrated intensity maps of corresponding line emission. Below we will discuss the detailed spatial distributions of various gas tracers based on these maps.

\subsubsection{Extended gas emission}\label{Extended gas}

CS, HCN and HCO$^+$ are commonly said to be dense gas tracers used in determinations of dense gas star formation laws \citep{Gao2004,Wu2005,Wu2010}, but this is not always true \citep{Kauffmann2017,Pety2017,Shimajiri2017}. As shown in Figure \ref{lineimages1}, all the three tracers show very extended emission in the G9.62+0.19 complex. They are poor tracers of the massive filament, where new high-mass stars are forming.  Due to large optical depths or absorption, their moment 1 and moment 2 maps are very complicated and do not show any clear pattern. HCN emission reveals a dense shell-like structure with its emission peak at the position "MM4/E". What we have learned here is that opaque tracers will fail to trace the locations of densest gas on small scales ($\sim$0.1 pc) though they are sort of fine to reveal the overall dense gas distribution on larger scales ($\sim$1 pc; see Figure \ref{ACA}).

Besides these main lines with low effective excitation densities ($n_{\rm eff}<5\times10^3$ cm$^{-3}$), the lines with higher effective excitation densities ($n_{\rm eff}>1\times10^4$ cm$^{-3}$) in Table \ref{spws} including the isotopologue lines (H$^{13}$CO$^{+}$, H$^{13}$CN) also show widespread emission (see Figures \ref{lineimages2} and \ref{lineimages3}). Interestingly, the H{\sc ii} region "B" appears to have produced a cavity in the molecular clump, which is clearly seen in CCH, H$^{13}$CO$^{+}$, H$^{13}$CN, SO and HC$_{3}$N emission. The H{\sc ii} region "B" is bounded by molecular line emission. H$^{13}$CN emission and 3 mm continuum emission together define a nearly circular region that has a radius of $\sim$ 0.5 pc, as shown by the magenta dashed circles in Figures \ref{lineimages1}, \ref{lineimages2} and \ref{lineimages3}. It seems that all high-mass stars in G9.62+0.19 are formed within this 1 pc size natal clump. The Moment 1 maps of these molecular line emission show a clear velocity gradient in the southwest to northeast direction. The gas close to the H{\sc ii} region "B" shows higher redshifted emission. Away from the H{\sc ii} region "B", the gas radial velocity decreases. Line profiles closer to the boundary of the H{\sc ii} region "B" are more redshifted, indicating that the H{\sc ii} region "B" is expanding into its surrounding molecular gas and is reshaping its natal clump.

\subsubsection{Tracers for photodissociation regions}

\begin{figure}
\centering
\includegraphics[width=\columnwidth]{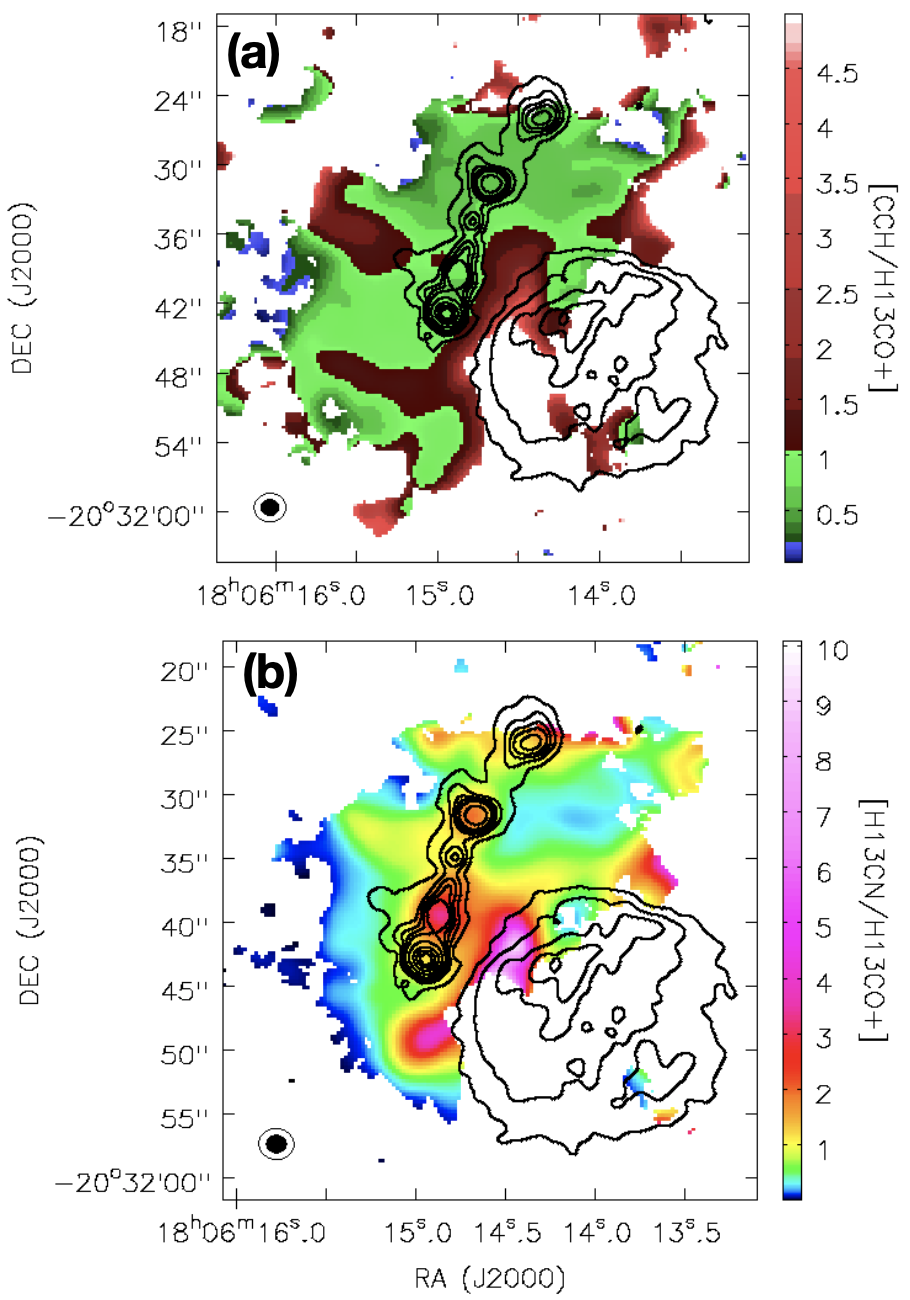}
\caption{Integrated intensity ratio maps of G9.62+0.19. The contours show the 3 mm continuum emission. Contour levels are the same as in Figure \ref{continuum}.  (a) CCH-to-H$^{13}$CO$^+$ ratio; (b) H$^{13}$CN-to-H$^{13}$CO$^+$ ratio. \label{PDR}}
\end{figure}

CCH is a good tracer of photodissociation regions \citep[PDRs;][]{Pety2005,Tiwari2019}. In PDRs, the fragmentation of polycyclic aromatic hydrocarbons (PAHs) due to FUV radiation \citep{Le2003,Pety2005,Montillaud2013} or new gas-phase formation routes \citep{Cuadrado2015} facilitated by heating from high UV flux (G$_0\sim10^4-10^5$ in Habing units) could give rise to a rich hydrocarbon chemistry. However, CCH emission is widespread in the G9.62+0.19. As shown in Figure \ref{lineimages2}, the CCH emission shows a morphology remarkably similar to that of the H$^{13}$CO$^{+}$ emission. It indicates that CCH emission inside massive clumps is not only concentrated in PDRs but is almost omnipresent. CCH has been detected in both nearby dark clouds \citep{Pratap1997} and distant high-mass star forming clumps \citep{Beuther2008,Sanhueza2012}. CCH is produced quickly early on and gets replenished at the clump surfaces by the UV photodissociation of CO \citep{Beuther2008}, leading to its widespread emission in clouds.

HCO$^+$ is more easily recombined with free electrons and its abundance is sensitive to the ionization degree of molecular gas. In far-UV irradiated environments (PDRs), its abundances may decrease. Figure \ref{PDR}a presents the CCH-to-H$^{13}$CO$^+$ integrated intensity ratio map. In the interface between the H{\sc ii} region ``B" and the massive filament, the CCH-to-H$^{13}$CO$^+$ intensity ratio is enhanced by a factor of 2-5 when compared to the massive filament itself and its surrounding regions. The enhanced CCH-to-H$^{13}$CO$^+$ intensity ratio is likely caused by the PDR of the H{\sc ii} region ``B". The H$^{13}$CN-to-H$^{13}$CO$^+$ ratio is also enhanced in this thin layer, as seen in Figure \ref{PDR}b. High HCN-to-HCO$^+$ ratios have also been found in far-UV irradiated environments such as evolved Galactic H{\sc ii} regions \citep{Nguyen-Luong2020}, AGNs \citep{Aladro2015} or luminous infrared galaxies \citep[LIRGs;][]{Papadopoulos2007}. Therefore, it implies that the high CCH-to-H$^{13}$CO$^+$ and H$^{13}$CN-to-H$^{13}$CO$^+$ ratios are good tracers of PDRs.

In the ATOMS survey, we will reveal the PDRs utilizing the CCH-to-H$^{13}$CO$^+$ and H$^{13}$CN-to-H$^{13}$CO$^+$ ratios. We will systematically investigate the chemistry and structure of PDRs \citep[e.g.,][]{Goicoechea2016}. In particular, we will study how PDRs interact with their natal molecular clumps. In the G9.62+0.19 complex, the PDR of the H{\sc ii} region "B" is clearly interacting with its molecular clump as indicated by the velocity gradients in Moment 1 maps of CCH, H$^{13}$CO$^{+}$, H$^{13}$CN, SO and HC$_{3}$N emission (see Figures \ref{lineimages2} and \ref{lineimages3}).

\subsubsection{Widespread shocked gas emission}

\begin{figure*}
\centering
\includegraphics[angle=0,scale=0.5]{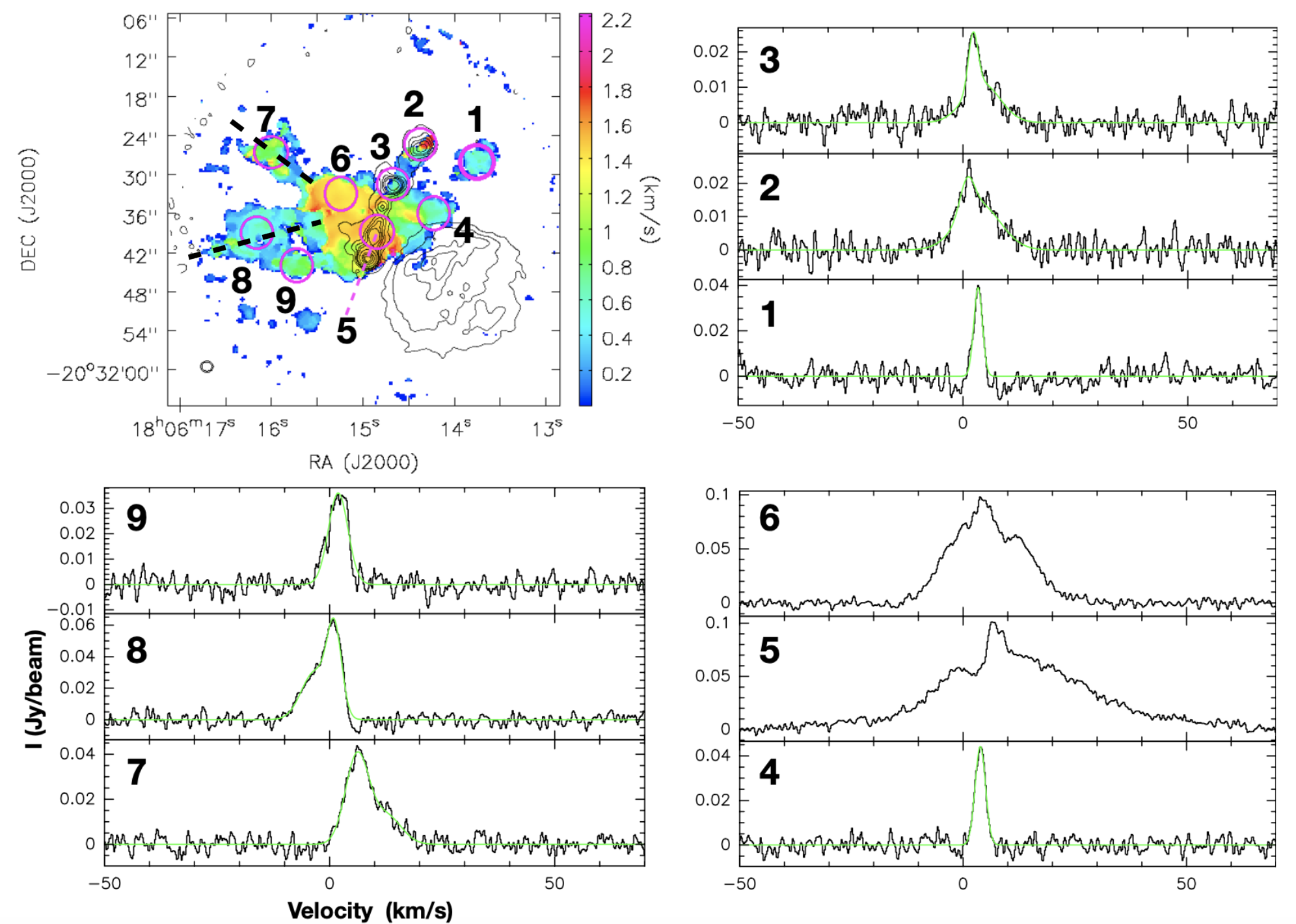}
\caption{SiO spectra averaged over 5$\arcsec$ regions at nine positions across G9.62+0.19. The upper left panel shows the Moment 2 map of SiO overlaid on 3 mm continuum emission contours. The contour levels are the same as in Figure \ref{continuum}. The magenta circles mark the positions where the SiO spectra are extracted. The two black dashed lines mark the elongated structures that may be related to colliding flows. The other panels show the averaged SiO spectra at each position. The green lines are gaussian fits.  \label{sio-spec}}
\end{figure*}

In the ATOMS survey, several shock tracers, e.g., SO, SiO and CH$_3$OH, are observed and will be used to trace the shocked gas emission. As shown in Figure \ref{lineimages2}, SiO emission is widespread in the G9.62+0.19 region. Within the 20\% contour in its integrated intensity map, SiO emission shows a large velocity dispersion ($\sigma>$1.5 km~s$^{-1}$), which is mainly related to outflows. However, in more extended regions, i.e., outside the 20\% contour in integrated intensity map, SiO emission has a much smaller velocity dispersion of $<$1 km~s$^{-1}$. In Figure \ref{sio-spec}, we present the Moment 2 map of SiO emission in the upper-left panel and plot the SiO spectra at nine positions of G9.62+0.19 region in other panels. The SiO spectra were obtained by averaging over 5$\arcsec$ area at each position. At positions 5 and 6, SiO shows broad line emission with high-velocity wings, indicating that SiO  in these regions is affected by outflow shocks. At other positions, however, SiO emission has much narrower line widths. We fit the spectra with gaussian profiles and the results (peak intensity I$_P$, systemic velocity V$_{lsr}$ and line width FWHM) from gaussian fits of the spectra are summarized in Table \ref{SiOpara}. The positions 1-4 are located close to H{\sc ii} regions ("B", "MM1/C", "MM4/E") and SiO could be excited by H{\sc ii} region shocks. Positions 7-9 are away from both outflows and H{\sc ii} regions. SiO emission at these positions may be caused by other mechanisms, e.g., shocks from large-scale colliding flows. Spectra at positions 1, 4 and 9 can be well fitted with a single gaussian profile, while spectra at positions 2, 3, 7, and 8 are better fitted with two gaussians including one narrow component and one broad component. The line widths of the narrow components range from 2 km~s$^{-1}$ to 6 km~s$^{-1}$. Widespread, narrow SiO emission has been detected in infrared dark clouds \citep{Jimenez2010,Cosentino2018,Cosentino2019} and high-mass proto-cluster forming regions \citep{Nguyen2013,Louvet2016}. Such narrow SiO emission can be reproduced by low-velocity shocks in the range 7 km~s$^{-1}$ to 12 km~s$^{-1}$ \citep{Louvet2016}. Such low-velocity shocks are usually attributed to colliding flows or cloud-cloud collision \citep{Jimenez2010,Nguyen2013,Louvet2016,Moscadelli2018}. We noticed that the narrow components of SiO spectra at positions 7 and 8 are clearly redshifted and blueshifted by $\sim$2-3 km~s$^{-1}$ from the systemic velocity (4 km~s$^{-1}$ measured from H$^{13}$CO$^{+}$) with a projected velocity difference of $\sim$5 km~s$^{-1}$, while the wider components of SiO spectra at positions 7 and 8 are redshifted and blueshifted by $\sim$7-9 km~s$^{-1}$ from the systemic velocity with a projected velocity difference of $\sim$16 km~s$^{-1}$. This velocity difference indicates that the two elongated structures associated with positions 7 and 8 could be related to large-scale colliding flows that induce low-velocity shocks.

\begin{table}
	\centering
	\caption{Parameters of SiO lines {\bf toward} G9.62+0.19.}
	\label{SiOpara}
	\begin{tabular}{cccc} 
		\hline
		Position   & $I_{p}$   & $V_{lsr}$ & FWHM   \\
		         & (mJy~beam$^{-1}$)  & (km~s$^{-1}$)  & (km~s$^{-1}$) \\
		\hline
1   & 39  &  3.38(0.05) & 2.25(0.10) \\
2   & 10  &  0.94(0.25) & 3.70(0.93) \\
    & 14  &  3.53(0.49) &11.18(0.64) \\
3   & 16  &  2.29(0.10) & 2.53(0.25) \\
    & 11  &  4.21(0.30) & 9.55(0.77) \\
4   & 44  &  3.87(0.03) & 2.73(0.08) \\
7   & 39  &  6.12(0.19) & 6.12(0.27) \\
    & 13  & 12.62(0.82) & 7.78(1.36) \\
8   & 30  & -3.30(0.22) & 6.97(0.30) \\
    & 54  &  1.07(0.06) & 3.80(0.11) \\
9   & 36  &  1.92(0.06) & 5.13(0.14) \\
		\hline
	\end{tabular}
\end{table}

CH$_3$OH and SO may also trace shocked gas. The upper six panels of Figure \ref{lineimages3} present the moment maps for CH$_3$OH and SO. In outflow regions, CH$_3$OH and SO emission show a large velocity dispersion, indicating that they also trace outflow shocks. Beyond the outflow and filament regions, however, the velocity dispersion of CH$_3$OH and SO is as small ($<$1 km~s$^{-1}$) as that of SiO, suggesting CH$_3$OH and SO may also trace widespread low-velocity shocks generated from either large-scale colliding flows or H{\sc ii} regions. Widespread CH$_3$OH emission that may have originated in a large-scale shock interaction has also been detected in infrared dark clouds \citep{Cosentino2018}. However, CH$_3$OH and SO emission can not only be induced by shocks but also be excited due to pure gas-chemistry in cold cores \citep{Spezzano2017,Nagy2019} or hot cores \citep{Esplugues2013,Qin2015}. Indeed, they reveal well the dense cores in G9.62+0.19 too (see section \ref{gascore}).

In the ATOMS survey, we will for the first time investigate statistically the origin of various velocity shocks in a large sample of massive clumps.

\subsubsection{Gas emission tracing dense cores}\label{gascore}

Figure \ref{lineimages3} presents moment maps of four gas tracers, CH$_3$OH, SO, H$^{13}$CN and HC$_3$N. All of them trace well the dense cores in the massive filament as identified by the 3 mm continuum emission. H$^{13}$CN and HC$_3$N have very similar morphology. Within the 20\% contours in their integrated intensity maps, their morphology is quite similar to that of the 3 mm continuum emission. When compared with other gas tracers in Table \ref{spws}, these four molecular line transitions seem to be better tracers of dense cores as well as tracers of the massive filament. Their emission maps agree with the 3 mm continuum more than the maps of HCN and HCO$^+$.

\subsubsection{Outflow gas emission}

\begin{figure*}
\centering
\includegraphics[angle=0,scale=0.55]{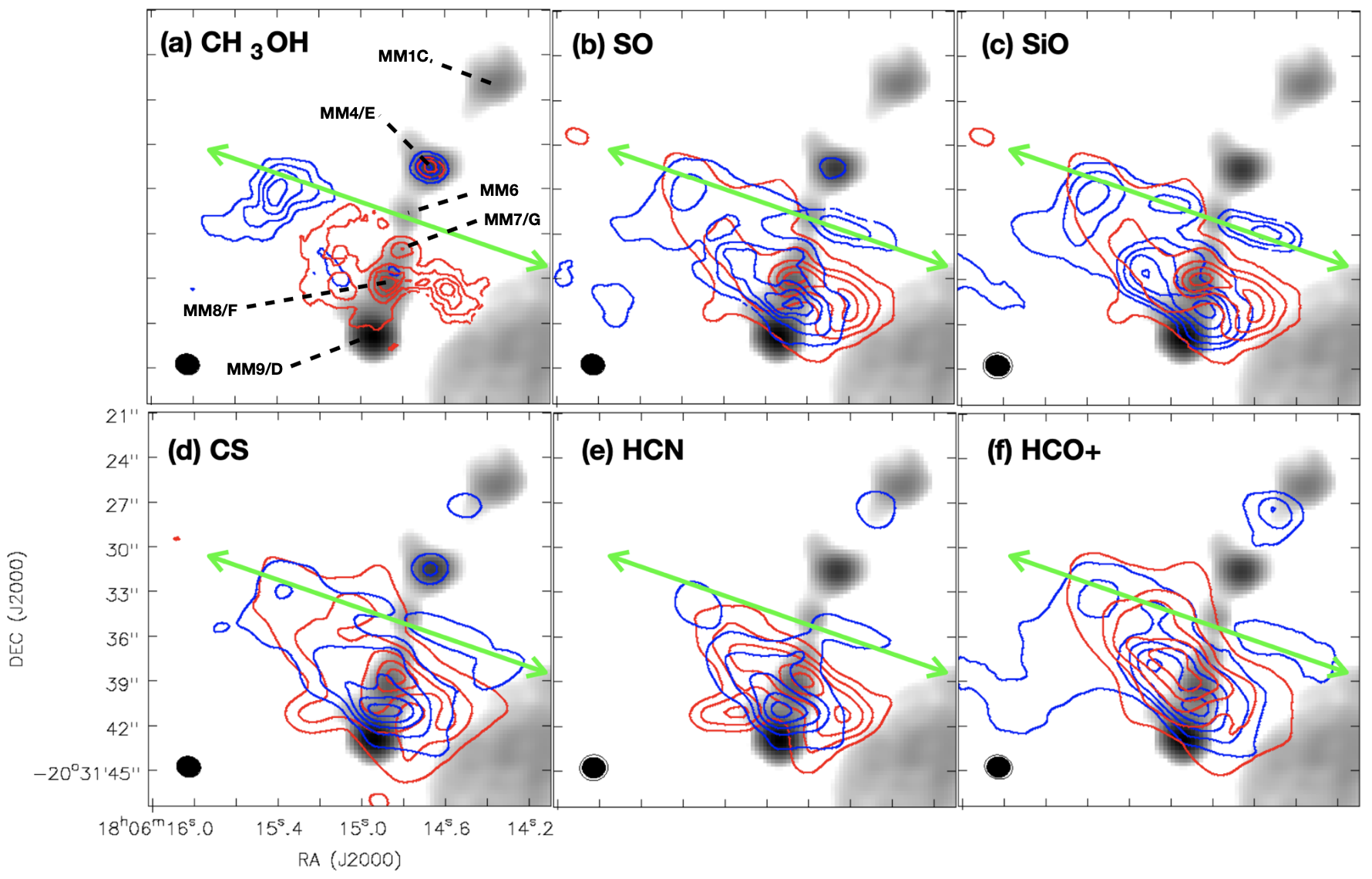}
\caption{Redshifted and blueshifted high-velocity emission of various gas tracers of G9.62+0.19 are shown in contours. The contours are [0.1,0.3,0.5,0.7,0.9]$\times F_{\rm peak}$. The peak integrated intensities ($F_{\rm peak}$) of CH$_3$OH blueshifted and redshifted emission are 0.42 Jy~beam$^{-1}$~km~s$^{-1}$ and 0.32 Jy~beam$^{-1}$~km~s$^{-1}$, respectively; $F_{\rm peak}$ of SO blueshifted and redshifted emission are 1.24 Jy~beam~km~s$^{-1}$ and 2.56 Jy~beam$^{-1}$~km~s$^{-1}$, respectively; $F_{\rm peak}$ of SiO blueshifted and redshifted emission are 1.02 Jy~beam$^{-1}$~km~s$^{-1}$ and 2.99 Jy~beam$^{-1}$~km~s$^{-1}$, respectively; $F_{\rm peak}$ of CS blueshifted and redshifted emission are: 3.89 Jy~beam$^{-1}$~km~s$^{-1}$ and 6.6 Jy~beam$^{-1}$~km~s$^{-1}$, respectively; $F_{\rm peak}$ of HCN blueshifted and redshifted emission are: 5.80 Jy~beam$^{-1}$~km~s$^{-1}$ and 3.94 Jy~beam$^{-1}$~km~s$^{-1}$, respectively; $F_{\rm peak}$ of HCO$^+$ blueshifted and redshifted emission are 1.72 Jy~beam$^{-1}$~km~s$^{-1}$ and 3.41 Jy~beam$^{-1}$~km~s$^{-1}$, respectively. The green arrow shows the direction of the bipolar outflow driven by MM6 \citep{Liu2017}. The outflows to the south are mainly driven by MM8/F \citep{liu11,Liu2017}. The background image shows the 3 mm continuum emission. \label{outflows}}
\end{figure*}

Several gas tracers in the ATOMS survey, including SiO, SO, CS, HCN, HCO$^{+}$ and CH$_3$OH, can trace outflows. Here we demonstrate that the ATOMS data are very useful for identifying protostellar outflows. In Figure \ref{outflows}, we present the redshifted and blueshifted high-velocity emission of these tracers. The HCN line has three hyper-fine components. To avoid contamination of hyperfine component emission from dense cores, the velocity intervals for redshifted and blueshifted high-velocity emission of HCN are [20, 50] km~s$^{-1}$ and [-42, -8] km~s$^{-1}$, respectively. For the other tracers, the velocity intervals for redshifted and blueshifted high-velocity emission are [9, 50] km~s$^{-1}$ and [-42,-1] km~s$^{-1}$, respectively. CH$_3$OH traces part of the high-velocity outflows driven by the hot core MM8/F, but does not trace the high velocity outflows driven by the young high-mass proto-stellar object (HMPO) MM6 \citep{Liu2017}. The other five outflow tracers reveal the high-velocity outflows driven by both MM8/F and MM6. They show similar outflow morphologies. The outflow properties in G9.62+0.19 have been discussed in detail in \citet{Liu2017}. In this paper, we will not further analyze the outflows.

\subsubsection{Chemical differentiation among dense cores}\label{Chemical differentiation}

\begin{figure*}
\centering
\includegraphics[angle=0,scale=0.95]{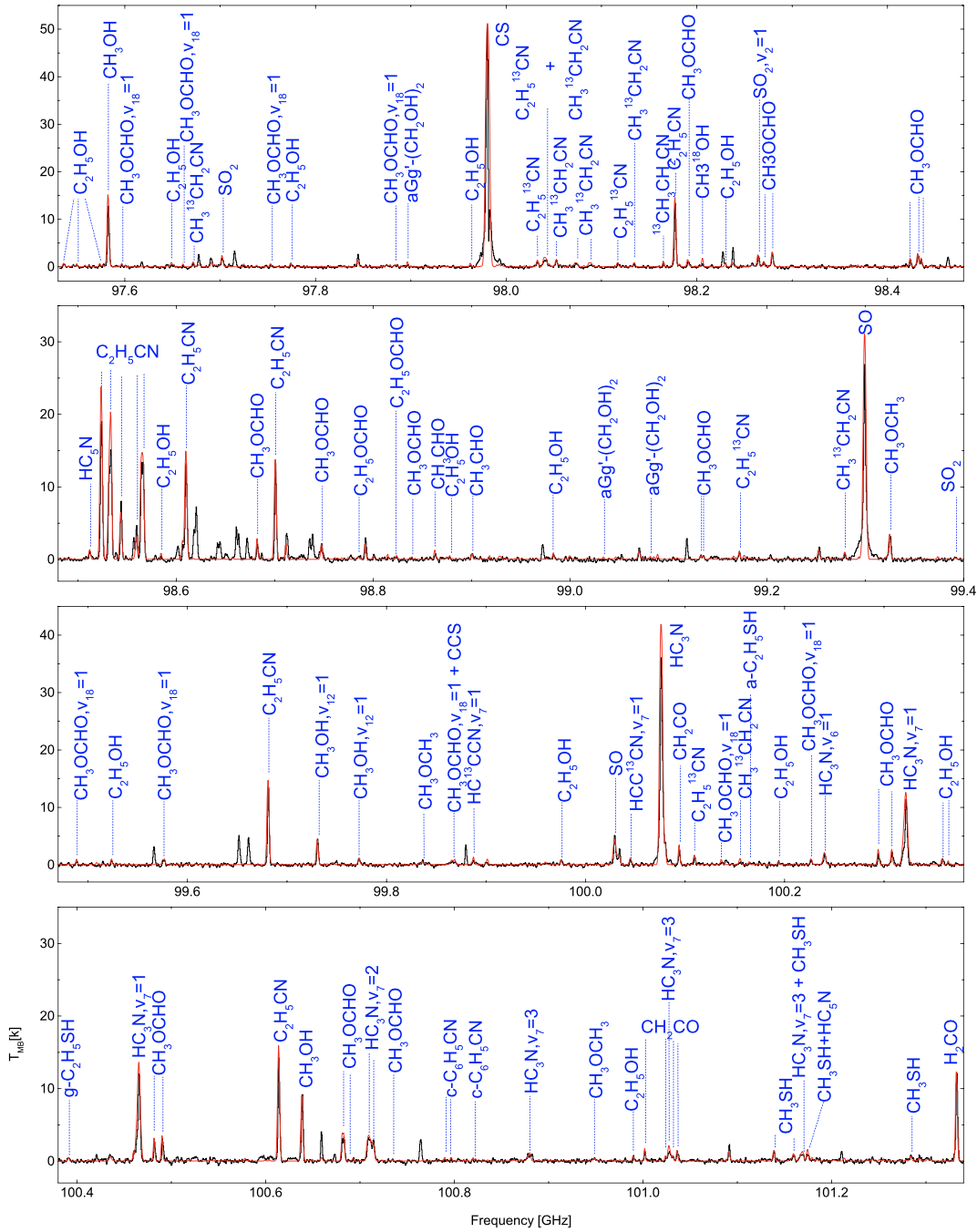}
\caption{Sample spectra at the emission peak of G9.62+0.19 core MM8/F. The black curve is the observed spectra, and red curve indicates the simulated spectra with XCLASS. The identified molecular line transitions are labeled. \label{XCLASS}}
\end{figure*}

The ATOMS survey covers a wide frequency range that includes many molecular transitions in the upper side-band. Figure \ref{XCLASS} presents the observed spectra (in black) at the emission peak of G9.62+0.19 core MM8/F overlaid with simulated spectra (in red) from XCLASS. Several tens of molecular line transitions were identified toward MM8/F. Hence, the ATOMS data are very useful for studying the chemical properties of dense cores. We will identify spectral line transitions using the eXtended CASA Line Analysis Software Suite \citep[XCLASS][]{Moller2017}, which takes the beam dilution, the line opacity, and line blending into account to identify and simulate the observed molecular line transitions under the assumption of local thermodynamical equilibrium (LTE). XCLASS has been widely used in chemistry studies in ALMA era \citep[e.g.,][]{Peng2017,Peng2019,Allen2017,Pols2018}. The detailed chemical analysis in G9.62+0.19 is beyond the scope of this paper and will be presented in a forthcoming paper (Peng, Y.-P. et al., 2020, in preparation).

\begin{figure*}
\centering
\includegraphics[angle=0,scale=0.8]{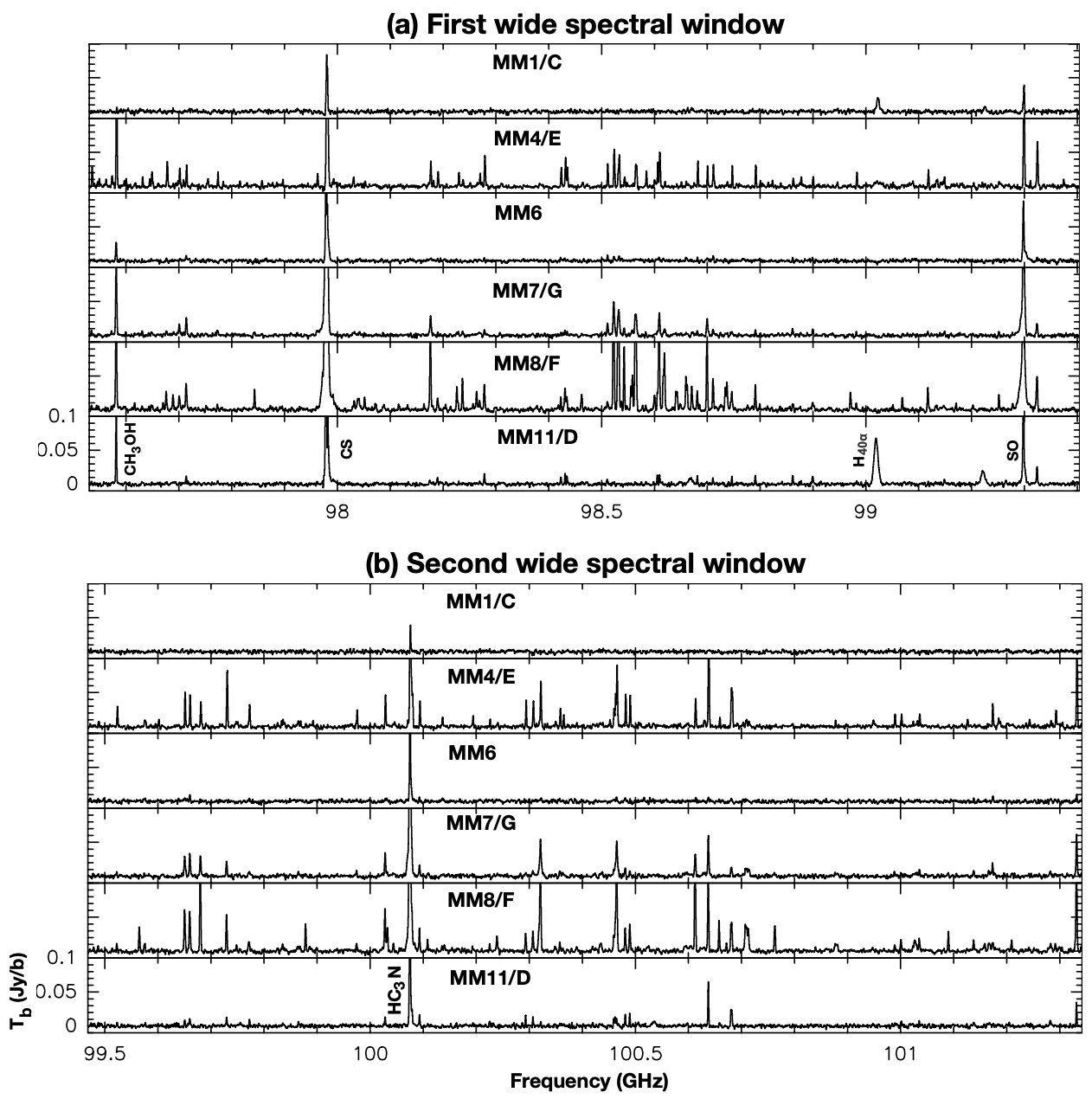}
\caption{The line spectra of the two wide spectral window averaged over 2$\arcsec$ at each core of G9.62+0.19. \label{widespw}}
\end{figure*}

The ATOMS survey data are also very useful for studying chemical differentiation among dense cores that are at different evolutionary stages. Figure \ref{widespw} compares the spectra of the two wide spectral windows at the positions of six cores in G9.62+0.19. Chemical differentiation is clear among these cores. The cometary H{\sc ii} region "B" is the most evolved source in G9.62+0.19 that has ionized its gas and shows no molecular gas emission. MM1/C is also a cometary H{\sc ii} region but is younger than "B". Some strong molecular lines such as HC$_3$N, SO and CS are detected toward it. MM11/D is an UC H{\sc ii} region, which shows strong H$_{40\alpha}$ emission. The molecular line emission toward MM11/D is also much stronger than the corresponding emission toward MM1/C. Cores MM4/E, MM7/G and MM8/F show typical hot core chemistry with line forests in the two wide bands. MM8/F is associated with energetic outflows and has the richest line emission. MM7/G is also an outflow source but may be still at an earlier evolutionary stage than MM8/F \citep{Liu2017}. Its line emission is much weaker than that of MM8/F. MM4/E is a hyper-compact (HC) H{\sc ii} region and is not associated with outflows \citep{Liu2017}. Its molecular line emission is also weaker than MM8/F. MM6 is the youngest high-mass protostellar object (HMPO) in this region that is driving an energetic bipolar outflow, and has not shown hot core chemistry.

Here we simply demonstrate that the ATOMS data are very useful for studying chemical differentiation among dense cores. In future, we will systematically investigate how the chemistry of dense cores changes with other physical properties like temperature, density, luminosity and Galactic environments.

\subsection{Similarities and differences in the spatial distributions of gas tracers}

Due to their different excitation conditions and abundances, different molecular line transitions may trace various physical conditions within molecular clouds, such as density, chemistry, temperature, UV radiation and so on. Through studying the similarities and differences in the spatial distributions of various gas tracers, we could figure out the best tracers for different structures inside molecular clouds, e.g., filaments, dense cores and extended cloud envelopes. In the ATOMS survey, we will mainly apply the principal component analysis \citep[PCA; e.g.,][]{Lo2009,Jones2012,Jones2013} and histogram of oriented gradients \citep[HOGs; e.g.,][]{Soler2019} analysis methods for this purpose. Below we will introduce the use of these two methods from the case study of G9.62+0.19.

\subsubsection{Principal component analysis}

\begin{figure}
\centering
\includegraphics[width=\columnwidth]{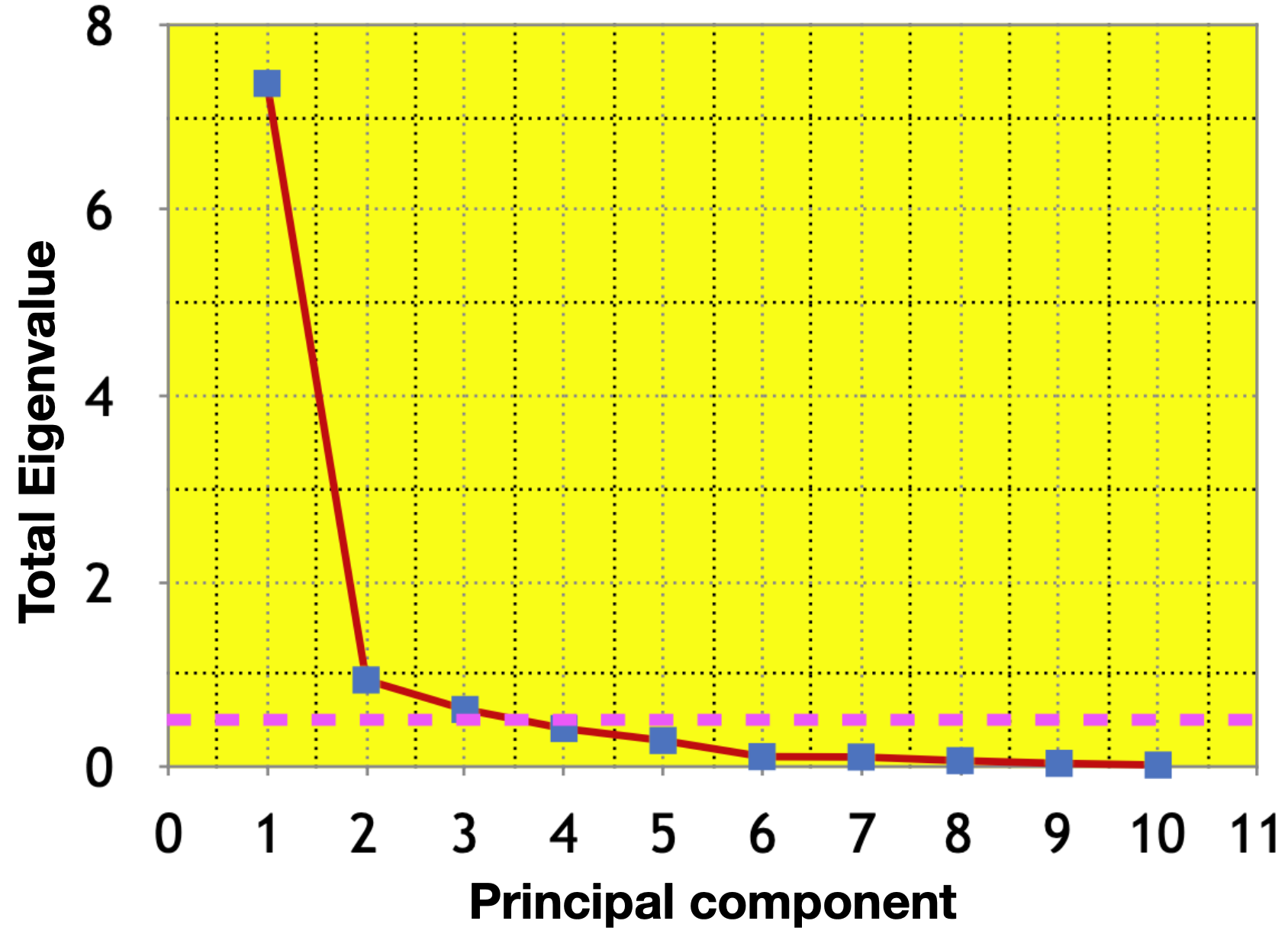}
\caption{The eigenvalue against the component number of the 10 principle components (PCs). The dashed line marks an eigenvalue threshold of 0.5. \label{PCA-plt}}
\end{figure}

The principal component analysis (PCA) is a statistical procedure to convert a set of observations of possibly correlated variables into a set of values of linearly uncorrelated variables called principal components. PCA has been widely used to identify and quantify similarities and differences between various molecular line emission inside molecular clouds \citep{Lo2009,Jones2012,Jones2013,Spezzano2017,Nagy2019}. We performed PCA on the integrated emission maps of various gas tracers to characterize their differences in spatial distribution in the G9.62+0.19 complex following a provider similar to that adopted in these previous works. The integrated intensity maps we used are presented in Figures \ref{lineimages1}, \ref{lineimages2}, and \ref{lineimages3}. These maps are firstly smoothed and re-gridded to ensure that they have the same resolution and pixel size. We then export their pixel values to SPSS software platform\footnote{https://www.ibm.com/analytics/spss-statistics-software} for PCA analysis.  Table \ref{PCApara} lists the correlation matrix of the input molecules, the eigenvectors and the eigenvalues of the Principal Components. From the correlation matrix, one can see that the four dense-gas tracers (CH$_3$OH, SO, H$^{13}$CN and HC$_3$N) are strongly correlated with each other, with correlation coefficients above 0.8. CCH shows strongest correlation with H$^{13}$CO$^+$ with a correlation coefficient of 0.91, indicating that these two molecules have very similar spatial distributions, as also mentioned in section \ref{Extended gas}. CS, HCO$^+$ and HCN are also strongly correlated with each other. This is consistent with the integrated intensity maps (Figure \ref{lineimages1}), showing that CS, HCO$^+$ and HCN have similar
large-scale spatial distribution (also see section \ref{Extended gas}). SiO is weakly correlated with most molecules except SO. The correlation coefficient between SiO and SO is $\sim$0.8, suggesting that the two molecules trace similar shocked gas at large scale.

\begin{figure*}
\centering
\includegraphics[angle=0,scale=0.4]{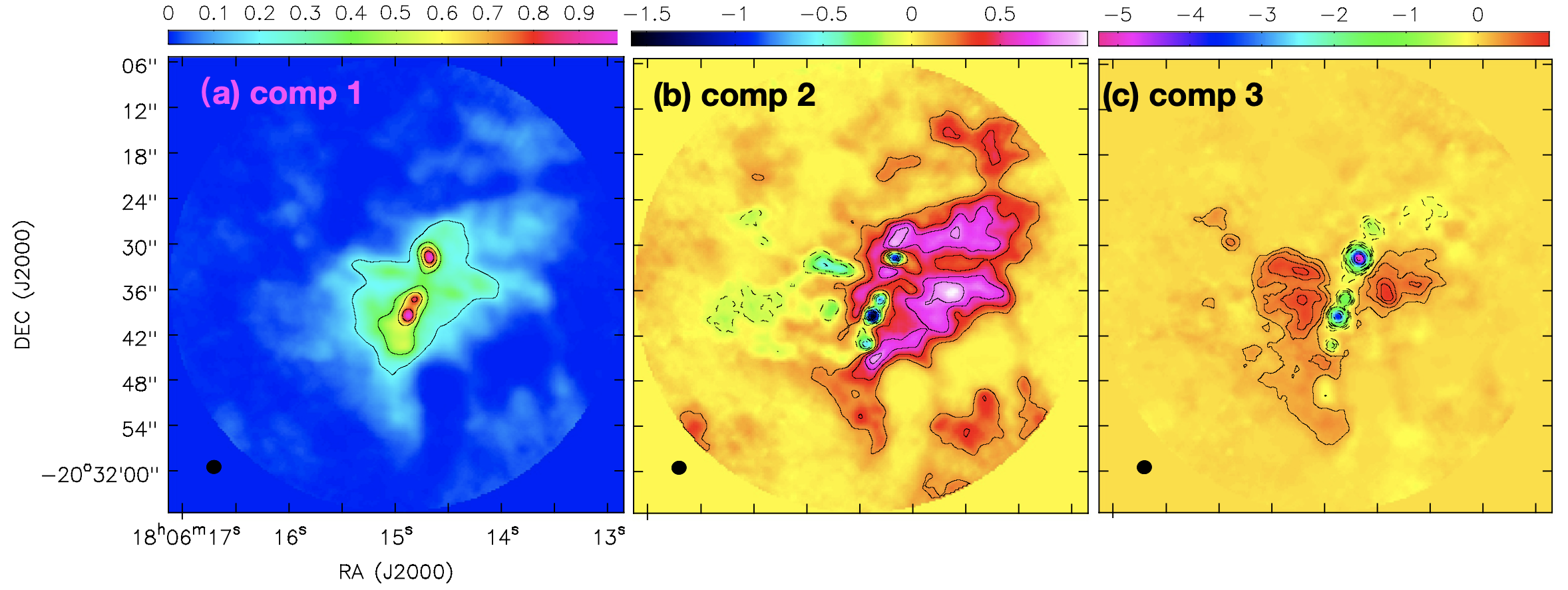}
\caption{Constructed images of the first three PCs. The images are normalized to their maximum values. (a) Principal Component 1. The contours are 0.2,0.4,0.6,0.8  (b)  Principal Component 2. The contours are -1.5,-1.3,-1.1,-0.1,0.2,0.4,0.6,0.8 (c)  Principal Component 3. The contours are -5,-4,-3,-2,-1,-0.8,-0.6,-0.4,-0.2,0.2,0.4,0.6,0.8. \label{PCA-images}}
\end{figure*}

Figure \ref{PCA-plt} presents eigenvalues for each principal components in the PCA analysis. We identified three main principal components (or PCs 1-3) that have eigenvalues larger than 0.5. These three PCs describe 73.62, 9.40 and 6.22 per cent, respectively, of the variance in the data. They together account for nearly 90 per cent of total variation in the data set. The eigenvectors in PCA analysis describe correlations between the variable and the PCs. A (negative) positive eigenvector indicates the molecule is (anti-) correlated with the PC. The larger the value, the stronger the correlation. The eigenvectors of each molecule are listed in the last 10 rows in Table \ref{PCApara}. Constructed images of the first three PCs are presented in Figure \ref{PCA-images}.

For the first PC (PC 1), all the ten molecules have large positive eigenvectors ($>$0.7), suggesting that all ten molecules are positively correlated with each other on large scales and PC 1 resembles the basis of molecular distribution within the G9.62+0.19 clump, as seen in Figure \ref{PCA-images}a. The PC 1 detects well the dense cores as well as the large-scale extended structures.

\begin{figure}
\centering
\includegraphics[width=\columnwidth]{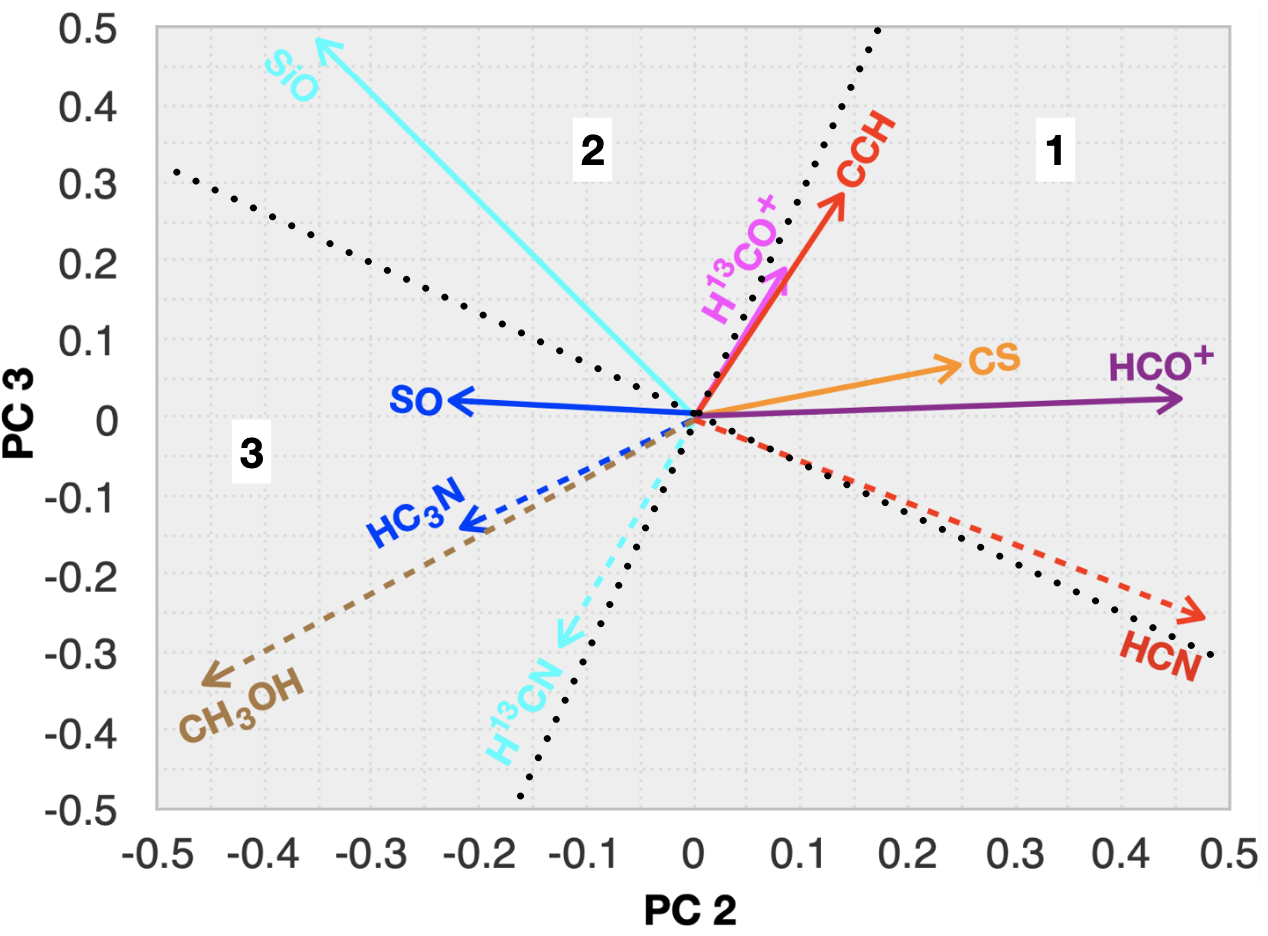}
\caption{Plots of eigenvectors of the second and third PCs of the data set. Each
of the eigenvectors represents the component of that molecule in the relevant
PC. The two mutually perpendicular dashed lines separate the molecules into three groups.  \label{PC2-3}}
\end{figure}

Figure \ref{PC2-3} shows the plot of eigenvectors of each molecule in PC 2 and PC 3. From this plot, we roughly classify the 10 molecules into three groups as separated by the two mutually perpendicular dashed lines. The two dashed lines are carefully chosen to ensure that the angles between a pair of eigenvectors in each group are smaller than 90$\degr$. The first group includes CS, HCO$^+$, HCN, CCH and H$^{13}$CO$^+$, which mainly trace extended gas emission and are poor tracers of dense cores. The third group includes SO, CH$_3$OH, H$^{13}$CN and HC$_3$N, which are dense core tracers. SiO in the second group stands out from other molecules because it mainly traces shocked gas. The first and third groups are clearly anti-correlated in Figure \ref{PC2-3}.  This division of groups, however, is not strict. For example, although SO and SiO are in different groups, the angle between their eigenvectors is only $48\degr$, indicating that they have high similarities somehow. As discussed before, SiO is strongly correlated with SO because both lines are good shock tracers. SO is assigned to the third group because SO is also tracing dense cores as well as other tracers in the same group.

Different groups are also clearly distinguished in the constructed images of the PC 2 and PC 3 (see Figures \ref{PCA-images}b and \ref{PCA-images}c). The dense core emission and shocked gas emission are shown in negative contours in the constructed image of the PC 2, while the extended gas emission is shown in solid contours. From the eigenvector plot of the third PC (PC 3 axis of Figure \ref{PC2-3}), three molecules (HCN, H$^{13}$CN and CH$_3$OH) show negative eigenvector smaller than -0.2. They mainly trace dense cores as shown in negative contours in Figure \ref{PCA-images}c. HCN emission is also strongly anti-correlated to PC 3 because its emission is centered on the dense core ``MM4/E" in the constructed image of PC 3. H$^{13}$CO$^+$ and CCH are strongly correlated with PC 3 with eigenvector larger than $\sim$0.2, while CS and HCO$^+$ are weakly correlated with PC 3 with eigenvector smaller than $\sim$0.1. It suggests that H$^{13}$CO$^+$ and CCH may trace some gas emission structures that are not well traced by CS and HCO$^+$ (Figure \ref{PCA-images}c). Interestingly, we found that H$^{13}$CO$^+$ (also CCH) is clearly anti-correlated with H$^{13}$CN in the PC 2 and PC 3. The angle between their eigenvectors in Figure \ref{PC2-3} is close to 180$\degr$, suggesting that they trace very different structures in these two PCs. It is because H$^{13}$CO$^+$ and CCH do not trace the dense cores at all, while H$^{13}$CN emission coincides with the dense core emission very well.

From PCA analysis, we learned that CS, HCO$^+$, HCN, CCH and H$^{13}$CO$^+$ show high similarities in spatial distribution because they mainly trace extended gas emission and they are poor tracers of dense cores in G9.62+0.19. In contrast, CH$_3$OH, H$^{13}$CN and HC$_3$N trace well dense cores. SiO line is a pure shock tracer. It has high similarity with another shock tracer, SO. However, SO line emission can trace well not only shocked gas but also dense cores. The PCA analysis is very powerful in separating different kinds of gas and will be used in further studies in the ATOMS survey to characterize gas distributions.

\begin{table*}
	\centering
	\caption{PCA analysis of G9.62+0.19.}
	\label{PCApara}
	\begin{tabular}{ccccccccccc} 
		\hline
\multicolumn{11}{c}{The correlation matrix of the input molecular dataset}\\
\hline
      &  CCH	 & CH$_3$OH	& CS	  & H$^{13}$CN	& H$^{13}$CO$^{+}$	&HC$_3$N	  &HCN	  &HCO$^+$	  &SiO	  &SO     \\
\hline
CCH	        & 1.00 	 & 0.56 	& 0.79 	& 0.70 	& \textit{\textbf{0.91}} 	  &0.77 	&0.59 	&0.73 	&0.60 	&0.75   \\
CH$_3$OH	      & 0.56 	 & 1.00 	& 0.57 	& \textit{\textbf{0.83}} 	& 0.62 	  &\textit{\textbf{0.87}} 	&0.48 	&0.47 	&0.56 	&\textit{\textbf{0.84}}   \\
CS	        & 0.79 	 & 0.57 	& 1.00 	& \textit{\textbf{0.81}} 	& 0.76 	  &0.75 	&0.68 	&\textit{\textbf{0.87}} 	&0.57 	&\textit{\textbf{0.81}}   \\
H$^{13}$CN	      & 0.70 	 & \textit{\textbf{0.83}} 	& \textit{\textbf{0.81}} 	& 1.00 	& 0.75 	  &\textit{\textbf{0.94}} 	&0.63 	&0.70 	&0.54 	&\textit{\textbf{0.87}}   \\
H$^{13}$CO$^+$	    & \textit{\textbf{0.91}} 	 & 0.62 	& 0.76 	& 0.75 	& 1.00 	  &\textit{\textbf{0.83}} 	&0.62 	&0.75 	&0.60 	&\textit{\textbf{0.80}}   \\
HC$_3$N	      & 0.77 	 & \textit{\textbf{0.87}} 	& 0.75 	& \textit{\textbf{0.94}} 	& \textit{\textbf{0.83}} 	  &1.00 	&0.62 	&0.69 	&0.66 	&\textit{\textbf{0.92}}   \\
HCN	        & 0.59 	 & 0.48 	& 0.68 	& 0.63 	& 0.62 	  &0.62 	&1.00 	&0.79 	&0.35 	&0.60   \\
HCO$^+$	      & 0.73 	 & 0.47 	& \textit{\textbf{0.87}} 	& 0.70 	& 0.75 	  &0.69 	&0.79 	&1.00 	&0.48 	&0.71   \\
SiO	        & 0.60 	 & 0.56 	& 0.57 	& 0.54 	& 0.60 	  &0.66 	&0.35 	&0.48 	&1.00 	&0.78   \\
SO	        & 0.75 	 & 0.84 	& \textit{\textbf{0.81}} 	& \textit{\textbf{0.87}} 	& \textit{\textbf{0.80}} 	  &\textit{\textbf{0.92}} 	&0.60 	&0.71 	&0.78 	&1.00   \\
  \hline
  \multicolumn{11}{c}{The eigenvectors and eigenvalues of the Principle Components}\\
  \hline
Component					&				1	&2&	3	&4	&5	&6	&7	&8	&9	&10 \\
\hline
Percentage of variance (\%)      &   73.62 & 9.40 & 6.22 & 4.18 & 2.89 & 1.16 & 1.12 & 0.72 & 0.43 & 0.26 \\
  \hline
CCH	    &   0.87 	& 0.13 	&   0.27 &-0.32 &	0.09 	  &  0.13 &	0.10 	&  0.10 	&  -0.06 &  0.02    \\
CH$_3$OH	  &   0.80 	& -0.45 &	-0.34  &0.00 	&  0.07 	& -0.02 &	0.20 	&  0.01 	&  0.07  &  0.01    \\
CS	    &   0.89 	& 0.25 	&   0.06 &0.05 	& -0.33 	&  0.14 &	0.05 	&  -0.08  &	0.06 	 & -0.05    \\
H$^{13}$CN	  &   0.91 	& -0.12 &	-0.28  &-0.07 &	-0.16 	&  0.05 &	-0.18 &	0.07 	  &  0.03 	&  0.08   \\
H$^{13}$CO$^+$	&   0.90 	& 0.08 	&   0.18 &-0.30 &	0.16 	  & -0.11 &	-0.07 &	-0.13 	&  0.08 	&  0.01   \\
HC$_3$N	  &   0.94 	& -0.21 &	-0.14  &-0.08 &	0.04 	  & -0.04 &	-0.10 &	0.06 	  & -0.05 	&  -0.12  \\
HCN	    &   0.74 	& 0.47 	&  -0.25 &0.26 	&  0.31 	&  0.10 &	-0.02 &	-0.02 	&  0.00 	&  0.00   \\
HCO$^+$	  &   0.84 	& 0.44 	&   0.03 &0.13 	& -0.12 	& -0.23 &	0.06 	&  0.09 	& -0.01 	&  0.01   \\
SiO	    &   0.71 	& -0.36 &	 0.48  &0.35 	&  0.07 	&  0.02 &	-0.05 &	0.05 	  &  0.05 	&  0.01   \\
SO	    &   0.95 	& -0.22 &	 0.02  &0.10 	& -0.05 	& -0.03 &	0.03 	& -0.14 	& -0.14 	&  0.04   \\
		\hline
	\end{tabular}
\end{table*}


\subsubsection{Histogram of oriented gradients analysis.}

\begin{table*}
	\centering
	\caption{HOG analysis of G9.62+0.19}
	\label{HOGpara}
	\begin{tabular}{cccccccccccc} 
		\hline
		lines                         &continuum & HC$_3$N  & H$^{13}$CN  &H$^{13}$CO$^{+}$   & CCH  &  CS  &  HCO$^{+}$   & HCN    &  SO  & CH3OH  &SiO\\
		\hline
		\multicolumn{12}{c}{$r$} \\
	        \hline
	       continuum                   &1.00\\
	       HC$_3$N                     & \textit{\textbf{0.32}}            & 1.00            &                    &\\
                H$^{13}$CN             & \textit{\textbf{0.30}}            &  \textit{\textbf{0.43}}           &1.00             &\\
                H$^{13}$CO$^{+}$       & 0.10            & \textit{\textbf{0.33}}            &0.16             &1.00                     \\
                CCH                    & 0.13            &\textit{\textbf{0.28}}            & 0.18            &\textit{\textbf{0.33}}                      &1.00      \\
                CS                     & 0.11            &\textit{\textbf{0.20}}            &\textit{\textbf{0.29}}             &0.08                     &0.16       &1.00 \\
                HCO$^{+}$              & 0.11            &0.17            &\textit{\textbf{0.21}}            &0.18                     &0.12       & 0.19    &1.00 \\
                HCN                    & 0.07            &0.05            &0.11             &0.03                     &0.03       & 0.08    &0.07     &1.00 \\
                SO                     & \textit{\textbf{0.20}}            & \textit{\textbf{0.29}}           &\textit{\textbf{0.29}}             &0.12                      &0.14       &0.19     &0.09      &0.09   &1.00\\
                CH3OH                  & \textit{\textbf{0.24}}            & \textit{\textbf{0.26}}           &\textit{\textbf{0.25}}            &0.02                      &0.03       &0.10     &0.05      &0.08    &\textit{\textbf{0.23}} & 1.00\\
                SiO                    & 0.08            &0.15           &0.06           & 0.12                      &0.07      &0.05     &0.07      &0.04     &\textit{\textbf{0.33}} & \textit{\textbf{0.21}}  & 1.00\\
  	        \hline
		\multicolumn{12}{c}{$V$}\\
	        \hline
	        continuum                & 5278    \\
	       HC$_3$N                  & \textit{\textbf{1138}}         &6972          &                 &\\
                H$^{13}$CN             & \textit{\textbf{1040}}         &\textit{\textbf{1770}}          &4707         &\\
                H$^{13}$CO$^{+}$   & 388           &\textit{\textbf{1961}}          &712           &8011                      \\
                CCH                         &  512           &\textit{\textbf{1714}}         &774           &\textit{\textbf{2051}}                      &9193     \\
                CS                            &  571           &\textit{\textbf{1334}}         &\textit{\textbf{1387}}         &546                       &\textit{\textbf{1267}}     &15850   \\
                HCO$^{+}$               &  485          &\textit{\textbf{1083}}         &964          & \textit{\textbf{1278}}                    & 972      &\textit{\textbf{2373}}     &18196  \\
                HCN                         &  317          & 300           & 503         & -24                        &179      &809       &845     &  16265   \\
                SO                           &  823           &\textit{\textbf{1620}}         &\textit{\textbf{1243}}        &764                         &701      &\textit{\textbf{1514}}    &598      &  554    & 8115\\
                CH3OH                    &  391          &644           &448         &28                           &57        &266      &90        &  179     &661     & 2933\\
                SiO                           &  119          &381           &119         &163                        &138        &169     &169      &  156     &\textit{\textbf{1078}}   & 452 & 3354\\
		\hline
		\multicolumn{12}{c}{$AM$}\\
	        \hline
	        continuum                &1.00\\
	       HC$_3$N                  & \textit{\textbf{0.53}}         &1.00           &                &\\
                H$^{13}$CN             & \textit{\textbf{0.51}}          &\textit{\textbf{0.61}}           &1.00         &\\
                H$^{13}$CO$^{+}$   & 0.35          &\textit{\textbf{0.54}}           &\textit{\textbf{0.41}}         &1.00                      \\
                CCH                         & 0.38          &\textit{\textbf{0.50}}           &\textit{\textbf{0.42}}          &\textit{\textbf{0.54}}                      &1.00    & \\
                CS                            & 0.37          &\textit{\textbf{0.43}}           &\textit{\textbf{0.51}}         &0.33                      & \textit{\textbf{0.40 }}   & 1.00 \\
                HCO$^{+}$               & 0.36          &\textit{\textbf{0.42}}          &\textit{\textbf{0.44}}         & \textit{\textbf{0.42}}                     & 0.37    & \textit{\textbf{0.43}}     &1.00\\
                HCN                         & 0.33          &0.32          &0.37          &0.27                     & 0.29     &0.33     &0.32     & 1.00\\
                SO                            & \textit{\textbf{0.44}}          &\textit{\textbf{0.50}}         & \textit{\textbf{0.50}}         &0.37                      &0.38     &\textit{\textbf{0.43}}      &0.34     & 0.35     & 1.00\\
                CH3OH                    & \textit{\textbf{0.46}}           &\textit{\textbf{0.46}}          &\textit{\textbf{0.46}}          &0.29                     &0.30      &0.35     & 0.31    & 0.34      & \textit{\textbf{0.45}}   &1.00\\
                SiO                           & 0.32          &0.38          &0.33         & 0.33                    &0.31      &0.32      &0.32    &  0.30      & \textit{\textbf{0.54}}   &\textit{\textbf{0.44}}    &1.00\\
		\hline
	\end{tabular}
\end{table*}

\begin{figure}
\centering
\includegraphics[angle=0,scale=0.55]{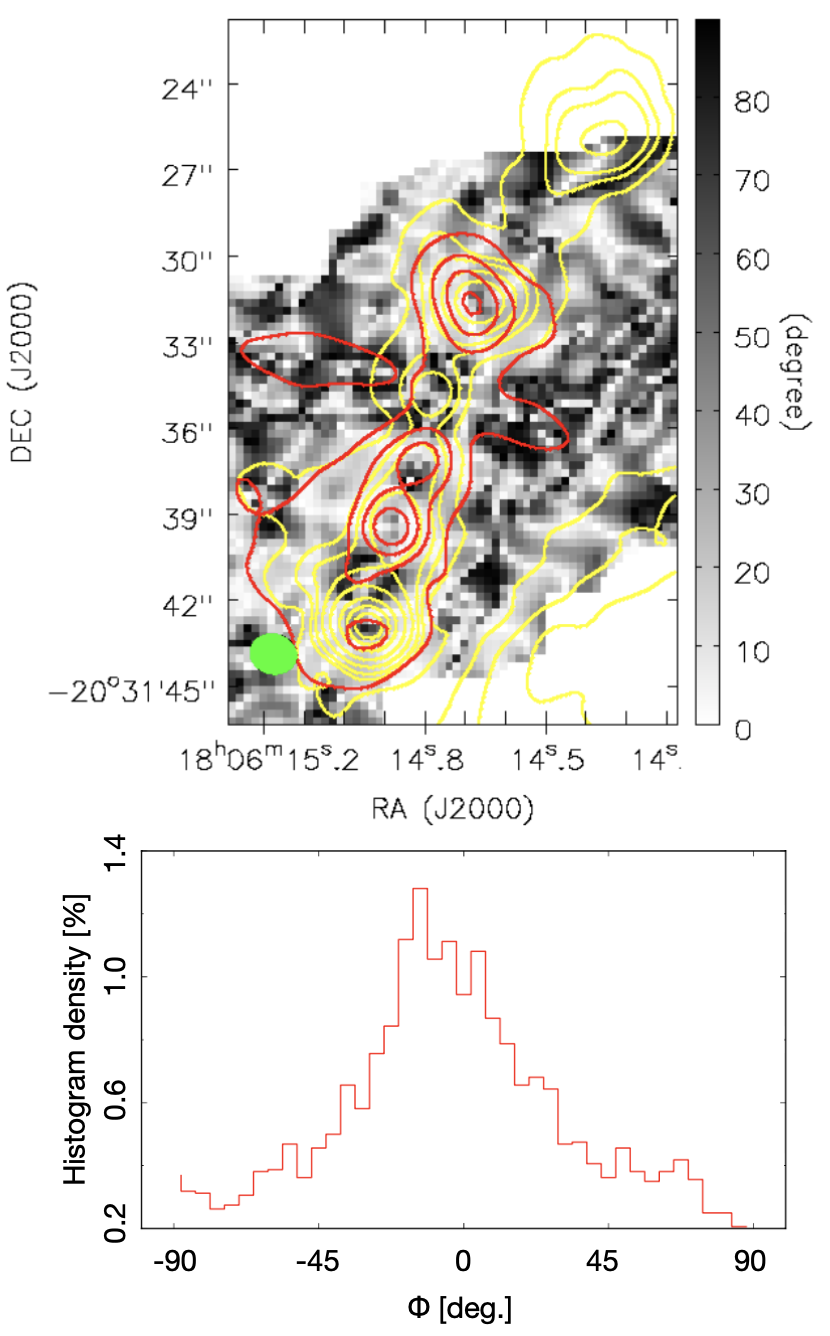}
\caption{Upper panel: The background image shows the absolute values of relative orientation angles ($\phi$) between the 3 mm continuum image and the HC$_3$N image of G9.62+0.19. The yellow contours show the 3 mm continuum emission. The contour levels are [0.01,0.03,0.05,0.07,0.1,0.2,0.4,0.6,0.8]$\times$74.1 mJy/beam. The HC$_3$N integrated intensity map is shown in red contours. The contour levels are [0.2,0.4,0.6,0.8]$\times$3.09 Jy/beam~km/s. Lower panel: Histogram of relative orientation angles. \label{HC3N-ang}}
\end{figure}

An alternative way to study the spatial correlation between different tracers in molecular clouds is the Histogram of Oriented Gradients analysis \citep[HOGs;][]{Soler2019}. In the HOGs method,
it is assumed that the appearance and shape of an object in an image can be well characterized by the distribution of local intensity gradients or edge directions. The HOG method quantifies similarities and differences in two images by studying the occurrences of the relative orientations between their local gradient orientations. \citet{Soler2019} investigated the spatial correlation of the HI and $^{13}$CO, and found a significant spatial correlation between the two tracers in velocity channels where $V_{\rm HI}\approx V_{\rm ^{13}CO}$.

In this work, we performed a HOGs analysis on the integrated intensity maps of various gas tracers following the procedure in \citet{Soler2019}. Three statistical evaluation parameters are derived: mean resultant vector length ($r$), the projected Rayleigh statistic ($V$), and the alignment measurement ($AM$). The definitions of these parameters are given in \citet{Soler2019}. The resultant
vector length, $r$, is a normalized quantity that can be interpreted as the fraction of uniform distribution of angles between gradient vectors. A preferential relative orientation (either parallel or perpendicular) in two images is of statistical significance if $r$ differs from zero significantly. The projected Rayleigh statistic ($V$) is used to determine whether or not gradient vectors have mostly parallel ($V>0$) or perpendicular ($V<0$) relative orientations. The higher the $V$ values, the more similarity in two images. The alignment measurement ($AM$), is an alternative method for estimating the degree of alignment between vectors \citep{Lazarian2007}. High $V$ and $AM$ values indicate mostly parallel relative orientations between the gradient vectors.

Three statistical evaluation parameters from HOGs analysis are summarized in Table \ref{HOGpara}. The three parameters give consistent statistical evaluation results. HC$_3$N is strongly correlated with six molecules (H$^{13}$CN, H$^{13}$CO$^{+}$, HCO$^{+}$, CCH, SO, CS), with $r>$0.2, $V>$1000 and $AM>$0.4. This correlation arises because  HC$_3$N is a good tracer for both dense cores and extended structures (see Figure \ref{lineimages3}). The strongest correlation occurs between HC$_3$N and H$^{13}$CN with $r$ of 0.43, $V$ of 1770 and $AM$ of 0.61. These two molecules have similar geometry in their integrated intensity maps (see Figure \ref{lineimages3}). H$^{13}$CO$^{+}$ is most correlated with CCH, as also mentioned in section \ref{Extended gas}. The $r$, $V$ and $AM$ values for the pair of H$^{13}$CO$^{+}$ and CCH are 0.33, 2051 and 0.54, respectively. CS shows highest correlation with HCO$^{+}$ with $V$ of 2373 and $AM$ of 0.43, suggesting that they have very similar spatial distributions. HCN shows strongest correlations with CS and HCO$^{+}$ with $V>800$ and $AM>$0.3. The SiO image is most similar to SO image with $V$ of 1078 and $AM$ of 0.54, suggesting that SiO and SO may reveal similar shocked gas distribution. CH$_3$OH shows strongest correlations ($V\ga450$ and $AM\ga$0.45) with H$^{13}$CN, HC$_3$N and SO because they trace similar dense gas distribution in the massive filament.

The HOGs analysis is also consistent with the previous PCA analysis. Molecules in the same group defined by the PCA analysis in Figure \ref{PCA-plt} show the strongest correlations in the HOGs analysis.

\begin{figure*}
\centering
\includegraphics[angle=0,scale=0.5]{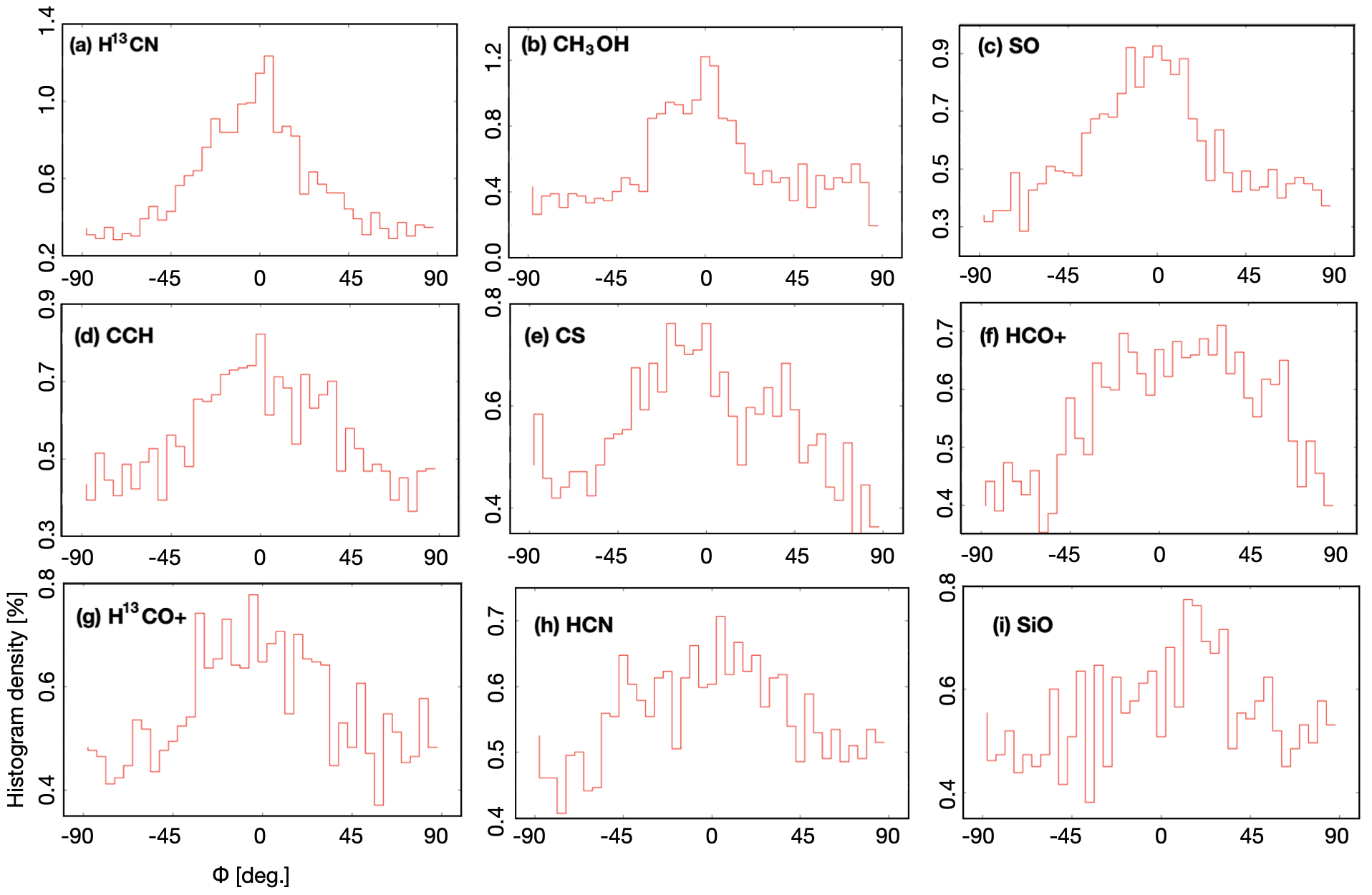}
\caption{Histogram of relative orientation angles between molecular line emission and 3 mm continuum emission. \label{HOG}}
\end{figure*}

We also investigate the similarities and differences between the 3 mm continuum emission and gas emission. We only performed HOGs analysis in the massive filament region. For example, the gray image in the upper panel of Figure \ref{HC3N-ang} shows the absolute values of relative orientation angles ($\phi$; color image) between the 3 mm continuum map (yellow contours) and the HC$_3$N map (red contours) in the filament region.  The histograms of relative orientation angles between 3 mm continuum and molecules are shown in the lower panel of Figure \ref{HC3N-ang} and Figure \ref{HOG}. From the histograms, one can see that the distributions of relative orientation angles for HC$_3$N, H$^{13}$CN, CH$_3$OH and SO are clearly peaked around 0$\degr$ with small dispersions. However, the other molecules show much more flat distributions. It indicates that HC$_3$N, H$^{13}$CN, CH$_3$OH and SO are the best to trace the dense structures (filaments and cores) as the 3 mm continuum emission. The other molecules are poor tracers for those dense structures. As shown in Table \ref{HOGpara}, HC$_3$N shows highest similarity with the 3 mm continuum emission. The $r$, $V$ and $AM$ for the HC$_3$N and 3 mm continuum correlation are 0.32, 1138 and 0.53, respectively. SiO shows largest difference with the 3 mm continuum emission. The $r$, $V$ and $AM$ for the SiO and 3 mm continuum correlation are 0.08, 119 and 0.32, respectively, indicating that SiO cannot trace dense structures at all.

\subsection{Fraction of line fluxes from filaments and cores}\label{flux ratios}

\begin{table*}
	\centering
	\caption{Integrated intensities of molecular lines in ACA and 12-m array observations.}
	\label{lineflux}
	\begin{tabular}{cccccc} 
		\hline
	Molecule  &         \multicolumn{3}{c}{ACA$^a$} & 12-m$^b$ & \\
\cline{2-4}\\
           &Size & peak  &  total   & total & ratio    \\
	               &($\arcsec,\arcsec$)	  &  (Jy~beam$^{-1}$~km~s$^{-1}$)          & (Jy~km~s$^{-1}$)           & (Jy~km~s$^{-1}$)   &    \\
		\hline
H$^{13}$CN	          &17.6$\arcsec\times11.4\arcsec$(149$\degr$)   & 15.3  & 32.3   & 13.5 & 0.42 \\
H$^{13}$CO$^+$	 & 24.0$\arcsec\times12.5\arcsec$(127$\degr$)   & 15.0  & 38.1   & 11.2 & 0.29\\
CCH	                          & 26.2$\arcsec\times13.5\arcsec$(142$\degr$)  & 13.0  & 38.7    & 8.7   &0.23\\
SiO	                          &  18.1$\arcsec\times11.4\arcsec$(93$\degr$)   & 10.3  &20.7    & 4.6 & 0.22 \\
HCN	                          &  34.3$\arcsec\times22.0\arcsec$(131$\degr$)  & 12.1 & 46.4    & 8.5 & 0.18 \\
HCO$^+$	                 & 31.4$\arcsec\times15.7\arcsec$(137$\degr$)   & 23.6  & 89.3   & 15.3 & 0.17  \\
CH$_3$OH	        & 13.7$\arcsec\times7.0\arcsec$(122$\degr$)     & 10.6  & 17.2    & 10.1 & 0.59  \\
CS	                        & 23.5$\arcsec\times13.7\arcsec$(135$\degr$)  & 53.9 & 166.3    & 38.1 & 0.23  \\
SO	                        & 17.2$\arcsec\times10.9\arcsec$(119$\degr$) & 28.8 & 62.2       & 19.4 & 0.31 \\
HC$_3$N	                &15.4$\arcsec\times8.9\arcsec$(140$\degr$)   & 34.5 & 70.2       & 35.9 & 0.51 \\
		\hline
\multicolumn{6}{l}{$^a$ The parameters from ACA data were from 2D Gaussian fits.}\\
\multicolumn{6}{l}{$^b$ The total integrated intensities of molecular lines in 12-m array data are measured}\\
\multicolumn{6}{l}{~~within the massive filament region bounded by the 1\% of peak flux density contour}\\
\multicolumn{6}{l}{~~($\sim3.5\sigma$ level) of the 3 mm continuum emission.}\\
	\end{tabular}
\end{table*}

One of the main science goals of the ATOMS survey is to statistically evaluate what portion of gas in the gravitationally bound clumps is actually participating in star formation or is concentrated in the smallest star formation units, i.e., dense cores. In this section, we explore what fraction of the fluxes of the various lines comes from the massive filament that is forming high-mass stars in G9.62+0.19.

\begin{figure}
\centering
\includegraphics[width=\columnwidth]{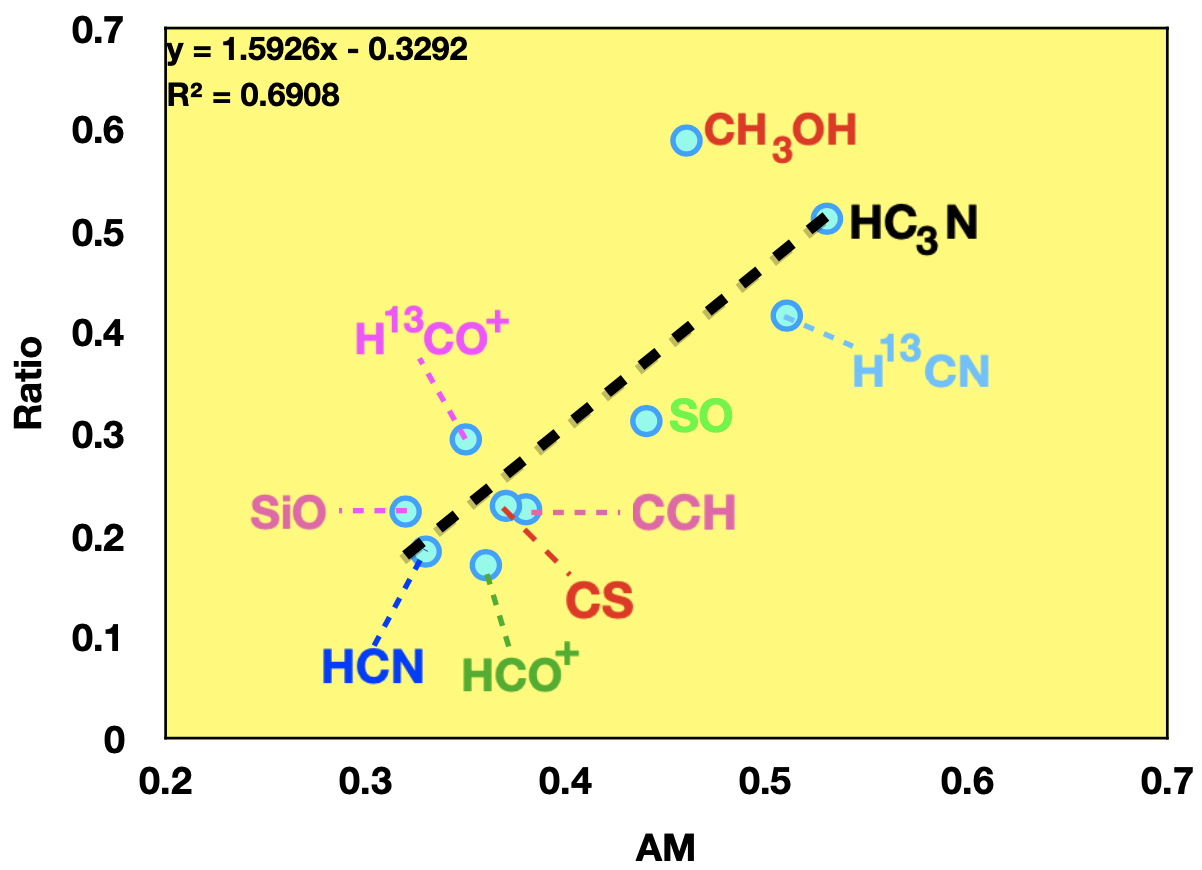}
\caption{The ratio of $S_\nu^{\rm 12m}$ to $S_\nu^{\rm ACA}$ is plotted as a function of the alignment measurement ($AM$) in HOG analysis for 3 mm continuum and molecular line emission.  \label{AM-R}}
\end{figure}

In Table \ref{lineflux}, we compare the total integrated intensities of various molecular lines obtained from ACA measurements with that obtained from 12-m array measurements. The integrated intensities are derived in the velocity interval of 1.5 to 6.5 km~s$^{-1}$. The deconvolved FWHM angular sizes, peak integrated intensities and total integrated intensities ($S_\nu^{\rm ACA}$) of various molecular lines in ACA data are derived from a 2D gaussian fitting, and are listed in columns 2-4 in Table \ref{lineflux}. The total integrated intensities (S$_\nu^{\rm 12m}$) of molecular lines in 12-m array data are measured within the massive filament region that is bounded by the 1\% of peak flux density contour ($\sim3.5\sigma$ level) of the 3 mm continuum emission. We did not perform 2D Gaussian fitting on 12-m array data because the massive filament is far from a 2D Gaussian shape. We also did not measure integrated intensities of molecular lines in the H{\sc ii} region ``B" because there is no molecular line emission associated with it. The 1\% of peak flux density contour of the 3 mm continuum emission defines well the boundary of the massive filament (see Figure \ref{continuum}). The ratios of $S_\nu^{\rm 12m}$ to $S_\nu^{\rm ACA}$ listed in the last column in Table \ref{lineflux} can be used to evaluate how much emission is from the densest filament region. Figure \ref{AM-R} shows the ratios against the $AM$ values that are derived from HOGs analysis. There is a clear trend between the ratios and the $AM$ values. Molecules with higher $AM$ values show higher similarity with the 3 mm continuum emission. Molecules (CH$_3$OH, HC$_3$N and H$^{13}$CN) that have high $AM$ values also show high ratios of $S_\nu^{\rm 12m}$ to $S_\nu^{\rm ACA}$ ($>40\%$), further suggesting that these molecular lines are good tracers of the dense filament. In contrast, only less than 25\% of the fluxes of CCH, HCN, CS and HCO$^+$ are from the dense filament region. These molecules also show largest differences (small AM values) with the spatial distribution of the 3 mm continuum emission.

G9.62+0.19 is a gravitationally bound massive clump \citep{Liu2017,Liu2018c}. Molecular lines HCN J=1-0, CS J=2-1 and HCO$^{+}$ J=1-0 are commonly used in dense gas star formation law studies \citep{Gao2004,Wu2005,Wu2010}. Our studies indicate that even in such a gravitationally bound clump, these commonly used lines are poor tracers of the dense structures, e.g., filament or dense cores. This is consistent with observations in nearby GMCs, where these lines do not well trace the column density structures (cores and filaments) revealed in Herschel images \citep{Pety2017,Shimajiri2017}.  Our studies of G9.62+0.19 indicate that the gas lies within the dense filament as defined by the continuum emission only contributes to less than 25\% flux of these lines. These main lines of dense gas tracers are poor tracers of dense structures because they tend to be optically thick in dense parts of molecular clouds \citep[e.g.,][]{Sanhueza2012,Hoq2013,Shimajiri2017,Liu2020}. The large optical depths of main lines, such as HCN and HCO$^+$, indicate that the density or mass estimates from their line intensity could be misleading. In contrast to the main lines, emission from the isotopologues are optically thinner because of their much lower abundances and hence higher effective excitation densities \citep{Shirley2015}. Therefore, they are potentially better tracers of the column densities and dense structures in molecular clouds. The optical depth issues in dense gas star formation law studies with main lines will be discussed in a companion paper by \citet{Liu2020}.

In dense gas star formation law studies \citep[e.g.,][]{Wu2010}, the line luminosity (L$^{\prime}_{\rm mol}$) is assumed as a tracer for dense gas mass (M$_{\rm dense}$). \citet{Wu2010} found that the median conversion factors from line luminosity to dense gas masses (L$^{\prime}_{\rm mol}$-to-M$_{\rm dense}$) are about 20, 39, 16 and 19 M$_{\sun}$~(K~km$^{-1}$~pc$^2$)$^{-1}$ for CS J=2-1, CS J=7-6, HCN J=1-0 and HCN J=3-2 lines, respectively. In the ATOMS survey, we are more interested in how the total line luminosity of clumps correlates to the total masses (M$_{cores}$) of dense cores, i.e., the smallest star formation units. In G9.62+0.19, the total line flux densities of H$^{13}$CN, H$^{13}$CO$^+$, HCN and HCO$^+$ that are integrated over all velocity channels in ACA observations are 87, 45, 258 and 176 K~km$^{-1}$, respectively. We can convert the total flux densities of these lines into their line luminosity following \cite{Solomon1997}. The line luminosity of the J=1-0 transitions of H$^{13}$CN, H$^{13}$CO$^+$, HCN and HCO$^+$ of the G9.62+0.19 clump are 10, 5, 29 and 19 K~km$^{-1}$~pc$^2$, respectively. \citet{Liu2017} identified 12 dense cores in G9.62+0.19. The total mass of these dense cores is $\sim$400 M$_{\sun}$ assuming an average dust temperature of 35 K \citep{Liu2017}. Therefore, the conversion factors L$^{\prime}_{mol}$-to-M$_{cores}$ in G9.62+0.19 are about 40, 80, 14, and 21 M$_{\sun}$~(K~km$^{-1}$~pc$^2$)$^{-1}$ for the J=1-0 transitions of H$^{13}$CN, H$^{13}$CO$^+$, HCN and HCO$^+$, respectively. In future, we will statistically investigate the conversion factors of line luminosity to dense core masses with the full ATOMS sample, which is very helpful to investigate how star formation efficiency varies across Galactic clumps. The knowledge of these studies can also be applied to star formation in external galaxies.

\section{Discussion}

\subsection{Stellar feedback and sequential high-mass star formation}

The ATOMS aims to investigate how stellar feedback from formed OB (proto-)stars in proto-clusters affect their natal clumps and regulate new star formation. Below we discuss the stellar feedback effect in G9.62+0.19.

As discussed in section \ref{Chemical differentiation} and in \citet{Liu2017}, sequential high-mass star formation is taking place inside the G9.62+0.19 clump, which has a size of $\sim$1 pc. The oldest H{\sc ii} regions in this area seem to have great influence on their natal clump, creating a cavity in molecular line emission. From moment 1 maps (see middle panels of Figure \ref{lineimages2} and \ref{lineimages3}) of several molecular lines (e.g., CCH, H$^{13}$CO$^{+}$, SO and HC$_3$N), one can see clearly a velocity gradient across the clump. Such large-scale velocity gradient and widespread low-velocity SiO shocks surrounding the H{\sc ii} regions are more likely caused by the expansion of the H{\sc ii} region "B". It indicates that stellar feedback from formed OB proto-stars can greatly reshape their natal clumps and change the spatial distribution of gas.

We also noticed that the evolved OB proto-stars in G9.62+0.19 are not located near the center of the clump but are clearly close to the clump edges. Along the massive filament, most evolved sources are located at the ends and the youngest sources are located in the filament center. This phenomenon is similar to the "edge-collapse" process proposed for filaments \citep{Pon2011,Pon2012,Clarke2015}. For long filaments with aspect ratios larger than 2, the
global collapse of the filaments is likely end-dominated \citep{Clarke2015}. The ends experience the highest acceleration, and this leads to the formation of end-clumps which then converge on the centre. In addition, recent ALMA high-resolution observations of massive infrared dark clouds (IRDCs) indicate that there is no significant evidence of primordial mass segregation \citep{Contreras2018,Sanhueza2019}. More massive cores are distributed in the same way as other low-mass cores \citep{Contreras2018,Sanhueza2019}. The observations of those IRDCs are consistent with what we have found here in the G9.62+0.19 clump. It seems that high-mass stars are not necessarily formed at the clump center as expected by the "competitive accretion" model \citep{bonn02,bonn06,bonn08}. Recently, \citet{Padoan2019} proposed a new scenario of massive-star formation, the "Inertial-Inflow Model", in which massive stars are assembled by large-scale, converging, inertial flows that naturally occur in supersonic turbulence.

\begin{figure}
\centering
\includegraphics[width=\columnwidth]{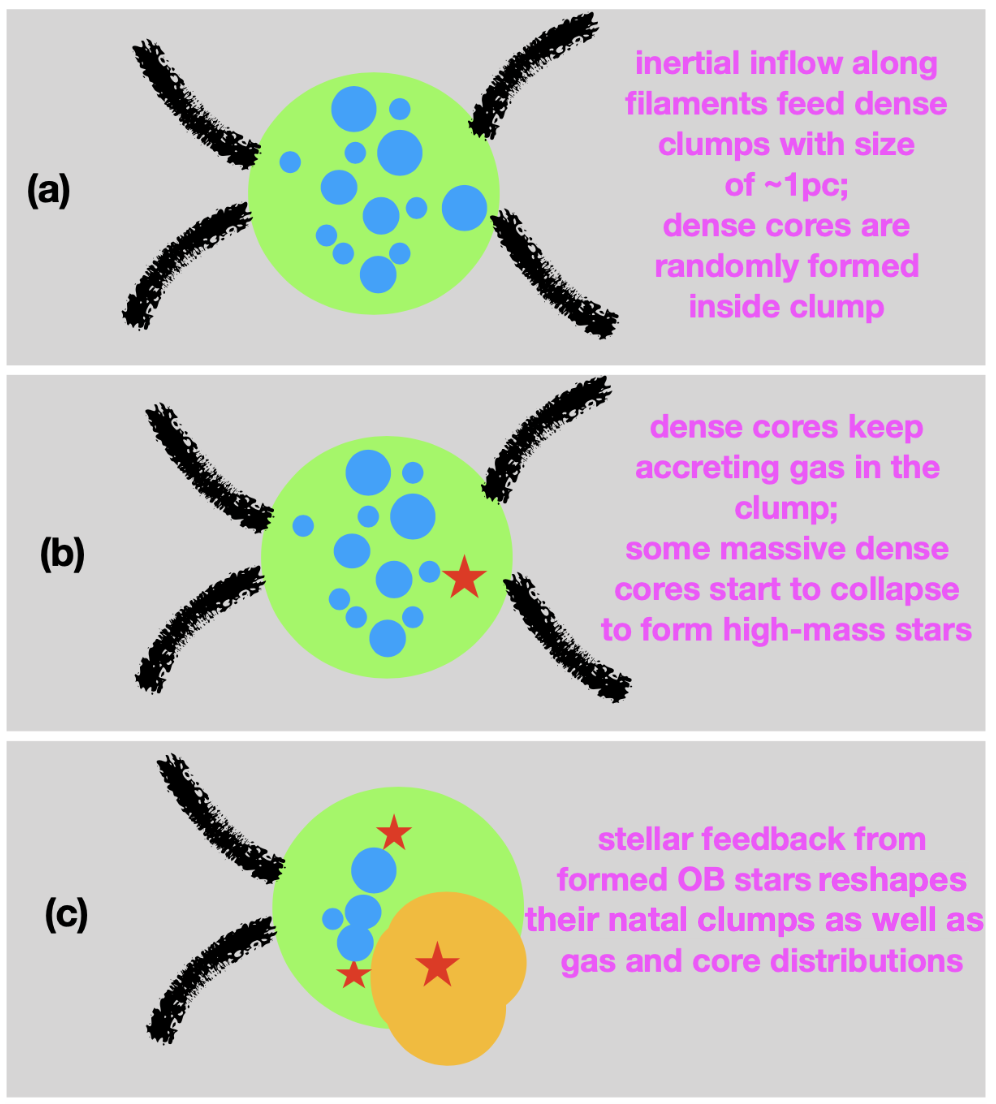}
\caption{Sketch of the scenario to explain the sequential high-mass star formation in G9.62+0.19.  Green filled circle is the massive clump with a typical size of 1 pc. The black lines represent converging flows connected to the clump. Blue filled circles are dense cores with a typical size of 0.1 pc. Red stars represent formed OB stars. The orange area in panel (c) shows an expanding HII region.     \label{model}}
\end{figure}

Based on our knowledge of the G9.62+0.19 clump and recent studies of IRDCs, we propose the following scenario to explain the sequential high-mass star formation in pc-size clumps. The sketch of the scenario is shown in Figure \ref{model}. The three stages involved in proto-cluster formation of this scenario are:

(a) The initial fragmentation of a massive clump does not show significant evidence of primordial mass segregation \citep{Sanhueza2019}.  The clumps are accreting gas from large-scale converging flows \citep{Padoan2019}. These flows generate low-velocity shocks that excite widespread narrow SiO emission.

(b) As proposed in the "edge-collapse" scenario \citep{Pon2011,Pon2012,Clarke2015}, dense cores that are near the clump edges or at the ends of dense filaments inside clumps have higher accretion rates and are more gravitationally unstable. They will be the first to collapse to form proto-stars.  Dense cores and formed proto-stars keep accreting gas from clump gas. High-mass proto-stars could form away from clump center as seen in the G9.62+0.19 complex.

(c) Stellar feedback from formed OB stars reshapes their natal clumps and changes the spatial distribution of gas and dense cores. The expansion of H{\sc ii} regions can also generate low-velocity shocks that excite widespread narrow SiO emission in their surroundings. Some low-mass dense cores could be destroyed as H{\sc ii} regions expand into the clump. Some dense cores in front of H{\sc ii} regions may collide and merge into massive cores or massive filament due to the enhanced density by the H{\sc ii} region shocks. This can explain the lack of widespread low-mass cores in the G9.62+0.19 complex \citep{Liu2017}. New generation of high-mass proto-stars could form in these new massive cores.

Statistical studies of a large number of proto-clusters in the ATOMS survey will help test this scenario.

\subsection{Links to other surveys}

There are several on-going or completed survey programs aiming at studying dense gas in either Galactic Giant molecular clouds or external galaxies. In this section, we will introduce several of these surveys and discuss how they will connect to the ATOMS survey.

The "Millimetre Astronomy Legacy Team 90 GHz (MALT90)" survey (PI: J. M. Jackson) has used the Mopra 22-m single-dish telescope to obtain $3\arcmin\times3\arcmin$ maps in various gas tracers towards $\sim$2000 dense molecular clumps that are at different evolutionary stages \citep{Jackson2013,Sanhueza2012,Stephens2016}. The "MALT90" project aims to characterize the physical and chemical evolution of high-mass star-forming clumps. The "ATOMS" survey observed similar gas tracers as in the "MALT90" project, e.g., HCN, HCO$^+$, and their isotopologue lines, CCH, SiO, and HC$_3$N. The "ATOMS" survey data will be used to study the star formation scaling relations \citep{Liu2020}, and compared with those derived using "MALT90" data \citep{Stephens2016}. The "ATOMS" survey has a much higher angular resolution (2$\arcsec$ vs. 38$\arcsec$) than the "MALT90" project. The "ATOMS" survey will reveal the internal structures and detailed spatial gas distributions within the Galactic clumps that cannot be resolved in the "MALT90" project.

The "Orion B: The Anatomy of a Giant Molecular Cloud" (PI: J. Pety) is a multi-line imaging survey at the IRAM 30-m telescope toward the Orion B molecular cloud \citep{Pety2017}. It covers approximately thirty 3 mm lines with a typical resolution of 26$\arcsec$ (or $\sim$0.05 pc at Orion distance). The "Orion B" project aims to obtain an accurate 3D description of the molecular structure in a GMC and to reveal the detailed anatomy of a molecular emission that is usually hidden behind these galaxy-averaged spectra in nearby galaxy studies. The ATOMS survey has comparable or better spatial resolution ($<$0.05 pc) than the "Orion B" project for the 98 targeted Galactic clumps within 5 kpc. The ATOMS survey aims to reveal the detailed spatial distributions of molecular line emission inside a large sample of Galactic clumps and shares similar science goals as the "Orion B" project.  The comparative studies with the "Orion B" data and the "ATOMS" data will shed light on the spatial distributions of molecular emission across a widely different environment from a nearby GMC to distant massive star forming regions.

The "EMIR Multiline Probe of the ISM Regulating Galaxy Evolution" (EMPIRE; PI: Frank Bigiel) survey is a multi-line mapping survey at the IRAM 30-m telescope that is targeting high critical density molecular lines ($J$=1-0 of HCN, HCO$^+$ and HNC) and CO isotopologues across the disks of 9 nearby, star-forming, disk galaxies \citep{Jimenez2019}. With an angular resolution of 30$\arcsec$ at 90 GHz (or 1-2kpc), The EMPIRE survey aims to constrain dense gas fraction and star formation efficiency across galaxy disks and explore how they depend on local conditions (molecular fraction, hydrostatic pressure, radiation field, stellar potential, etc.). The ATOMS survey also aims to investigate the star formation scaling relations with the same high critical density gas tracers, e.g., $J$=1-0 of HCN, HCO$^+$ and their isotopologue lines, and also to study how star formation efficiency changes across the inner Galaxy. The joint analysis of the ATOMS data and "EMPIRE" data will link "dense gas" from the Milky Way to external galaxies.

The "ALMA\_IMF: ALMA transforms our view of the origin of stellar masses" (PI: Fr\'{e}d\'{e}rique Motte) observed 15 most massive clouds within 6 kpc at 3 mm and 1.3 mm at the $\sim$2000 AU spatial resolution. The "ALMA\_IMF" project aims to constrain the  protocluster core mass function (CMF) in the range 0.5-200 M$_{\sun}$ and to investigate which variables (such as inflows, outflows, or forming filaments) might be correlated with CMF evolution in proto-clusters. In contrast, the desired 3 mm continuum rms level in the ATOMS survey is of $\sim$65 $\mu$Jy, corresponding to a mass sensitivity $\sim$2 M$_{\sun}$ at a distance of 1 kpc and $\sim$50 M$_{\sun}$ at a distance of 5 kpc \citep{Kauffmann2008}. The mass sensitivity is derived with a median dust temperature of 30 K and a dust opacity per unit gas mass of 0.00189 cm$^2$~g$^{-1}$ at 3 mm that is extrapolated from the dust opacity of 0.01 cm$^2$~g$^{-1}$ at 230 GHz assuming a dust opacity index ($\beta$) of 2 \citep{Ossenkopf1994}. Therefore, the ATOMS project will help constrain the high-mass end of the CMFs in protoclusters with a much larger sample but a shallower mass sensitivity than the "ALMA\_IMF" project.

Another ALMA survey project, "ALMAGAL: ALMA Evolutionary study of High Mass Protocluster Formation in the Galaxy" (PI: Sergio Molinari), will observe more than 1000 dense clumps with M$>$500 M$_{\sun}$ and d$<$7.5 kpc with similar linear resolution ($\sim1000$ AU) and a mass sensitivity of 0.3 M$_{\sun}$ at 1.3 mm. The "ALMAGAL" project will systematically investigate what the physical characteristics of core fragments in dense clumps are as a function of evolution and to what extent fragmentation is driven by dynamics in clumps or mass transport in filaments. The angular resolution and mass sensitivity of the "ALMAGAL" project are much better than the ATOMS project. However, the ATOMS survey benefits from the ($\sim$3 times) larger FOV at 3 mm than that of the "ALMAGAL" project at 1.3 mm, indicating that the ATOMS survey are more suitable to study the overall structures and kinematics in proto-clusters. The comparative studies with the "ALMAGAL" data and the "ATOMS" data will help deepen the understandings of hierarchical fragmentation and differential star formation in high-mass proto-clusters.

\section{Conclusions}

We introduce in this paper the observations, data reduction and example science case, G9.62+0.19 (I18032-2032), of the ALMA survey program ATOMS. The ATOMS project, standing for ALMA Three-millimeter Observations of Massive Star-forming regions, has observed 146 active star forming regions at Band 3 in both continuum emission and molecule lines. More than 90\% of the "ATOMS" targeted sources are promising high-mass star forming regions. The ATOMS survey aims to systematically investigate the spatial distribution of various dense gas tracers, the roles of stellar feedback and filaments in star formation.

(1) The 3 mm continuum data of the ATOMS survey will be used to detect massive filaments and dense cores. In G9.62+0.19, the 3 mm continuum emission reveals an expanding H{\sc ii} region "B" to the west and a massive fragmented filament to the east. The 3 mm continuum emission of the expanding H{\sc ii} region "B", compact H{\sc ii} region "MM1/C" and the UC H{\sc ii} region "M11/D" seem to be dominated by free-free emission.

(2) The "ATOMS" survey covers dozens of molecular line transitions, including shocked gas tracers (SiO, SO), a PDR tracer (CCH), hot core tracers (e.g., HC$_3$N and CH$_3$OH) and some lines commonly assumed to trace dense gas (e.g., HCN, HCO$^{+}$ and CS). It also includes a tracer of ionized gas (H$_{40\alpha}$ recombination line). We will investigate the spatial distributions of various gas tracers within a large sample of massive clumps. In G9.62+0.19, the similarities and differences in images of 10 molecular lines in Table \ref{spws} are investigated with the principal component analysis and the histogram of oriented gradients analysis. We found that the 10 lines can be classified into three groups. The first group includes CS, HCO$^+$, HCN, CCH and H$^{13}$CO$^+$, which mainly trace extended gas emission. SiO in the second group stands out from other molecules because it mainly traces shocked gas. The third group including SO, CH$_3$OH, H$^{13}$CN and HC$_3$N traces well dense gas in the massive filaments. Molecules in the same group show strongest correlations in their spatial distribution. SO, CH$_3$OH, H$^{13}$CN and HC$_3$N show morphology similar to that of the 3 mm continuum emission. The other molecular lines including the commonly used dense gas tracers, e.g., CS J=2-1, J=1-0 of HCO$^+$ and HCN, can not reveal well dense structures (dense cores and filament) in G9.62+0.19 on the scale of $\sim$0.1 pc.

(3) In the ATOMS survey, we will for the first time investigate statistically the origin of various velocity shocks using shock tracers like SiO. In G9.62+0.19, SiO shows widespread narrow emission, which is likely caused by slow shocks generated by large-scale colliding flows and H{\sc ii} regions.

(4) In the ATOMS survey, we will identify outflows driven by proto-stars. For G9.62+0.19, SiO, SO, CS, HCO$^+$ and HCN trace well high-velocity outflow gas emission. The outflows, as traced by these molecular lines, show very similar morphology.

(5) The ATOMS survey covers a wide frequency range that includes dozens of molecular transitions that can be used to study chemical differentiation among dense cores. In G9.62+0.19, three dense cores (MM4/E, MM7/G and MM8/F) show hot core chemistry.

(6) The ATOMS survey aims to statistically evaluate what fraction of gas in gravitationally bound clumps is actually participating in star formation or is concentrated in the smallest star formation units, i.e., dense cores. In G9.62+0.19, only less than 25\% of the "dense gas" revealed by HCN, CS and HCO$^+$ is concentrated in the massive filament.

(7) The ATOMS survey aims to investigate how stellar feedback from formed OB (proto-)stars in proto-clusters affect their natal clumps and star formation efficiency. Stellar feedback from formed OB stars (H{\sc ii} region "B") in G9.62+0.19 has reshaped their natal clump and changed the spatial distribution of gas and dense cores. Stellar feedback may be also the cause of the sequential high-mass star formation that is taking place in G9.62+0.19 complex.

(8) Joint analysis with the ATOMS survey data and other survey data, such as "MALT90", "Orion B", "EMPIRE", "ALMA\_IMF", and "ALMAGAL", will help deepen our understandings of "dense gas" star formation scaling relations and massive proto-cluster formation.

\section*{Acknowledgements}

Tie Liu is supported by the initial funding of scientific research for high-level talents at Shanghai Astronomical Observatory. This work was carried out in part at the Jet Propulsion Laboratory, which is operated for NASA by the California Institute of Technology. LB acknowledges support from CONICYT project Basal AFB-170002. Di Li is supported by CAS International Partnership Program No.114A11KYSB20160008. This paper makes use of the following ALMA data: ADS/JAO.ALMA\#2019.1.00685.S. ALMA is a partnership of ESO (representing its member states), NSF (USA) and NINS (Japan), together with NRC (Canada), MOST and ASIAA (Taiwan), and KASI (Republic of Korea), in cooperation with the Republic of Chile. The Joint ALMA Observatory is operated by ESO, AUI/NRAO and NAOJ.








\appendix

\section{}\label{app}

\begin{table*}
	\centering
	\caption{Basic parameters of the targets in the ATOMS survey.}
	\label{atomsample}
	\begin{tabular}{cccccccccccc} 
		\hline
ID & IRAS  & RA &  DEC & V$_{\rm lsr}$  & Distance &  $R_{\rm GC}$ & Radius  & $T_{\rm dust}$ & log($L_{\rm bol}$)  & log($M_{\rm clump}$) & refs$^a$\\
 & &  & &(km~s$^{-1}$) & (kpc)  & (kpc)  & (pc) & (K) & (L$_{\sun}$)  & (M$_{\sun}$) & \\
\hline
1 & I08076-3556	&  08:09:32.39 &	 -36:05:13.2 & 5.9    & 0.4    & 8.5    & 0.07   & 18.0   &  1.2  & 0.7 & 1    \\
2 & I08303-4303	&  08:32:08.34 &	 -43:13:54.0 & 14.3   & 2.3    & 9.0    & 0.32   & 30.0   &  3.8  & 2.4 & 1    \\
3 & I08448-4343	&  08:46:32.90 &	 -43:54:35.9 & 3.7    & 0.7    & 8.4    & 0.15   & 25.0   &  3.0  & 1.6 & 1    \\
4 & I08470-4243	&  08:48:47.07 &	 -42:54:31.0 & 12     & 2.1    & 8.8    & 0.32   & 33.0   &  4.0  & 2.4 & 1    \\
5 & I09002-4732	&  09:01:54.24 &	 -47:44:00.8 & 3.1    & 1.2    & 8.4    & 0.24   & 39.0   &  4.6  & 2.4 & 1    \\
6 & I09018-4816	&  09:03:32.84 &	 -48:28:10.0 & 10.3   & 2.6    & 8.8    & 0.44   & 31.0   &  4.7  & 3.0 & 1    \\
7 & I09094-4803	&  09:11:07.29 &	 -48:15:48.7 & 74.6   & 9.6    & 12.7   & 1.40   & 23.0   &  4.6  & 3.1 & 1    \\
8 & I10365-5803	&  10:38:32.46 &	 -58:19:05.9 & -19    & 2.4    & 8.0    & 0.44   & 30.0   &  4.3  & 2.7 & 1    \\
9 & I11298-6155	&  11:32:05.70 &	 -62:12:24.3 & 32.9   & 10     & 10.1   & 1.36   & 32.0   &  5.2  & 3.4 & 1    \\
10& I11332-6258	&  11:35:31.81 &	 -63:14:44.6 & -15.4  & 1.9    & 7.7    & 0.28   & 30.0   &  3.7  & 2.1 & 1    \\
11& I11590-6452	&  12:01:34.54 &	 -65:08:52.9 & -4.3   & 0.4    & 8.2    & 0.06   & 22.0   &  1.7  & 1.1 & 1    \\
12& I12320-6122	&  12:34:53.38 &	 -61:39:46.9 & -42.5  & 3.43   & 7.2    & 1.00   & 44.6   &  5.6  & 3.0 & 2    \\
13& I12326-6245	&  12:35:34.81 &	 -63:02:32.1 & -39.6  & 4.61   & 7.2    & 0.83   & 34.2   &  5.4  & 3.5 & 2    \\
14& I12383-6128	&  12:41:17.32 &	 -61:44:38.6 & -39.1  & 3.27   & 7.2    & 0.63   & 19.8   &  3.8  & 3.1 & 2    \\
15& I12572-6316	&  13:00:24.43 &	 -63:32:30.4 & 30.9   & 11.57  & 9.8    & 1.63   & 21.5   &  4.6  & 3.9 & 2    \\
16& I13079-6218	&  13:11:14.16 &	 -62:34:42.1 & -42.6  & 3.8    & 6.9    & 1.09   & 28.2   &  5.1  & 3.5 & 2    \\
17& I13080-6229	&  13:11:14.10 &	 -62:45:01.9 & -35.6  & 3.8    & 6.9    & 1.03   & 34.1   &  5.1  & 3.2 & 2    \\
18& I13111-6228	&  13:14:26.43 &	 -62:44:24.1 & -38.8  & 3.8    & 6.9    & 0.96   & 30.0   &  4.8  & 3.1 & 2    \\
19& I13134-6242	&  13:16:43.05 &	 -62:58:30.1 & -31.5  & 3.8    & 6.9    & 0.57   & 29.2   &  4.6  & 3.1 & 2    \\
20& I13140-6226	&  13:17:15.70 &	 -62:42:27.5 & -33.9  & 3.8    & 6.9    & 0.50   & 22.6   &  3.8  & 2.9 & 2    \\
21& I13291-6229	&  13:32:32.38 &	 -62:45:23.5 & -36.5  & 2.9    & 7.0    & 0.55   & 28.8   &  4.3  & 2.7 & 2    \\
22& I13291-6249	&  13:32:31.23 &	 -63:05:21.8 & -34.7  & 7.61   & 7.1    & 1.73   & 27.2   &  5.2  & 3.7 & 2    \\
23& I13295-6152	&  13:32:53.49 &	 -62:07:49.3 & -44.4  & 3.89   & 6.7    & 0.36   & 19.4   &  3.3  & 2.7 & 2    \\
24& I13471-6120	&  13:50:42.10 &	 -61:35:14.9 & -56.7  & 5.46   & 6.4    & 1.01   & 35.1   &  5.3  & 3.4 & 2    \\
25& I13484-6100	&  13:51:58.64 &	 -61:15:43.3 & -55    & 5.4    & 6.4    & 0.84   & 31.8   &  4.8  & 3.1 & 2    \\
26& I14013-6105	&  14:04:54.56 &	 -61:20:10.7 & -48.1  & 4.12   & 6.4    & 0.80   & 31.4   &  5.0  & 3.2 & 2    \\
27& I14050-6056	&  14:08:42.15 &	 -61:10:43.0 & -47.1  & 3.42   & 6.6    & 0.66   & 32.2   &  4.7  & 2.8 & 2    \\
28& I14164-6028	&  14:20:08.23 &	 -60:42:05.0 & -46.5  & 3.19   & 6.6    & 0.17   & 28.7   &  3.7  & 2.2 & 2    \\
29& I14206-6151	&  14:24:22.81 &	 -62:05:22.7 & -50    & 3.29   & 6.5    & 0.37   & 27.3   &  3.7  & 2.4 & 2    \\
30& I14212-6131	&  14:25:01.08 &	 -61:44:59.4 & -50.5  & 3.44   & 6.5    & 0.57   & 21.8   &  4.0  & 3.0 & 2    \\
31& I14382-6017	&  14:42:02.76 &	 -60:30:35.1 & -60.7  & 7.69   & 6.0    & 1.68   & 28.0   &  5.2  & 3.6 & 2    \\
32& I14453-5912	&  14:49:07.77 &	 -59:24:49.7 & -40.2  & 2.82   & 6.6    & 0.79   & 24.1   &  4.2  & 2.9 & 2    \\
33& I14498-5856	&  14:53:42.81 &	 -59:08:56.5 & -49.3  & 3.16   & 6.4    & 0.74   & 26.7   &  4.4  & 3.0 & 2    \\
34& I15122-5801	&  15:16:06.77 &	 -58:11:40.5 & -60.9  & 9.26   & 5.9    & 2.65   & 27.7   &  5.0  & 3.9 & 2    \\
35& I15254-5621	&  15:29:19.48 &	 -56:31:23.2 & -67.3  & 4      & 5.7    & 0.89   & 33.5   &  5.1  & 3.1 & 2    \\
36& I15290-5546	&  15:32:53.16 &	 -55:56:06.8 & -87.5  & 6.76   & 4.9    & 1.80   & 33.5   &  5.7  & 3.8 & 2    \\
37& I15384-5348	&  15:42:16.57 &	 -53:58:31.3 & -41    & 1.82   & 6.9    & 0.54   & 33.9   &  4.6  & 2.5 & 2    \\
38& I15394-5358	&  15:43:16.48 &	 -54:07:16.9 & -41.6  & 1.82   & 6.9    & 0.46   & 20.7   &  3.7  & 2.9 & 2    \\
39& I15408-5356	&  15:44:43.00 &	 -54:05:44.9 & -39.7  & 1.82   & 6.9    & 0.61   & 34.0   &  4.9  & 2.9 & 2    \\
40& I15411-5352	&  15:44:59.59 &	 -54:02:21.4 & -41.5  & 1.82   & 6.9    & 0.57   & 30.5   &  4.5  & 2.7 & 2    \\
41& I15437-5343	&  15:47:33.11 &	 -53:52:43.9 & -83    & 4.98   & 5.0    & 0.77   & 29.6   &  4.6  & 3.0 & 2    \\
42& I15439-5449	&  15:47:49.82 &	 -54:58:32.1 & -54.6  & 3.29   & 5.9    & 0.69   & 26.8   &  4.4  & 3.0 & 2    \\
43& I15502-5302	&  15:54:06.43 &	 -53:11:38.4 & -91.4  & 5.8    & 4.6    & 1.66   & 35.7   &  5.8  & 3.7 & 2    \\
44& I15520-5234	&  15:55:48.84 &	 -52:43:06.2 & -41.3  & 2.65   & 6.2    & 0.67   & 32.2   &  5.1  & 3.2 & 2    \\
45& I15522-5411	&  15:56:07.74 &	 -54:19:57.8 & -46.7  & 2.73   & 6.2    & 0.64   & 23.3   &  3.8  & 2.9 & 2    \\
46& I15557-5215	&  15:59:40.76 &	 -52:23:27.7 & -67.6  & 4.03   & 5.3    & 0.82   & 20.7   &  3.9  & 3.2 & 2    \\
47& I15567-5236	&  16:00:32.86 &	 -52:44:45.1 & -107.1 & 5.99   & 4.4    & 1.31   & 35.4   &  5.7  & 3.5 & 2    \\
48& I15570-5227	&  16:00:55.56 &	 -52:36:21.2 & -101.5 & 5.99   & 4.4    & 1.66   & 28.7   &  4.8  & 3.4 & 2    \\
49& I15584-5247	&  16:02:19.63 &	 -52:55:22.4 & -76.8  & 4.41   & 5.1    & 0.86   & 23.9   &  4.2  & 3.1 & 2    \\
50& I15596-5301	&  16:03:32.29 &	 -53:09:28.1 & -72.1  & 10.11  & 5.2    & 1.81   & 28.5   &  5.5  & 3.9 & 2    \\
51& I16026-5035	&  16:06:25.43 &	 -50:43:07.2 & -78.3  & 4.53   & 4.9    & 0.81   & 31.8   &  4.9  & 3.0 & 2    \\
52& I16037-5223	&  16:07:38.10 &	 -52:31:00.2 & -80    & 9.84   & 4.9    & 2.15   & 31.4   &  5.6  & 3.8 & 2    \\
53& I16060-5146	&  16:09:52.85 &	 -51:54:54.7 & -91.6  & 5.3    & 4.5    & 1.24   & 32.2   &  5.8  & 3.9 & 2    \\
54& I16065-5158	&  16:10:20.30 &	 -52:06:07.1 & -63.3  & 3.98   & 5.2    & 1.41   & 30.8   &  5.4  & 3.7 & 2    \\
55& I16071-5142	&  16:11:00.01 &	 -51:50:21.6 & -87    & 5.3    & 4.5    & 1.21   & 23.9   &  4.8  & 3.7 & 2    \\
56& I16076-5134	&  16:11:27.12 &	 -51:41:56.9 & -87.7  & 5.3    & 4.5    & 1.57   & 30.1   &  5.3  & 3.6 & 2    \\
57& I16119-5048	&  16:15:45.65 &	 -50:55:53.5 & -48.2  & 3.1    & 5.8    & 0.93   & 24.0   &  4.3  & 3.2 & 2    \\
58& I16132-5039	&  16:17:01.52 &	 -50:46:51.0 & -47.5  & 3.1    & 5.8    & 1.04   & 32.5   &  4.6  & 2.9 & 2    \\
59& I16158-5055	&  16:19:38.63 &	 -51:03:20.0 & -49.2  & 3.57   & 5.4    & 1.35   & 28.3   &  5.2  & 3.5 & 2    \\
60& I16164-5046	&  16:20:10.91 &	 -50:53:15.5 & -57.3  & 3.57   & 5.4    & 1.37   & 31.4   &  5.5  & 3.7 & 2    \\
61& I16172-5028	&  16:21:02.47 &	 -50:35:10.3 & -51.9  & 3.57   & 5.4    & 1.51   & 32.0   &  5.8  & 4.0 & 2    \\
62& I16177-5018	&  16:21:31.49 &	 -50:25:04.5 & -50.2  & 3.57   & 5.4    & 1.25   & 34.0   &  5.5  & 3.6 & 2    \\
\hline
\multicolumn{12}{l}{$^a$ Ref 1: \citet{Faundez2004}; Ref 2: \citet{Urquhart2018}}\\
	\end{tabular}
\end{table*}

\begin{table*}
	\centering
	\contcaption{Basic parameters of the targets in the ATOMS survey.}
	\label{atomsample}
	\begin{tabular}{cccccccccccc} 
		\hline
Number&IRAS  & RA &  DEC & $V_{\rm lsr}$  & Distance &  $R_{\rm GC}$ & Radius  & $T_{\rm dust}$ & log($L_{\rm bol}$)  & log($M_{\rm clump}$) & refs\\
  &&  & & (km~s$^{-1}$) & (kpc)  & (kpc)  & (pc) & (K) & (L$_{\sun}$)  & (M$_{\sun}$) & \\
\hline
63 &I16272-4837	&  16:30:59.08 &	 -48:43:53.3 & -46.6  & 2.92   & 5.8     & 0.84   & 23.1   &  4.3  & 3.2 & 2    \\
64 &I16297-4757	&  16:33:29.45 &	 -48:03:41.2 & -79.6  & 5.03   & 4.2     & 2.00   & 27.1   &  4.9  & 3.4 & 2    \\
65 &I16304-4710	&  16:34:05.12 &	 -47:16:32.6 & -62.8  & 11.32  & 4.9     & 1.37   & 27.5   &  4.9  & 3.5 & 2    \\
66 &I16313-4729	&  16:34:54.98 &	 -47:35:35.0 & -73.7  & 4.71   & 4.4     & 2.06   & 31.0   &  6.7  & 4.7 & 1,2 \\
67 &I16318-4724	&  16:35:34.11 &	 -47:31:11.3 & -119.8 & 7.68   & 3.3     & 1.45   & 29.5   &  5.2  & 3.7 & 2    \\
68 &I16330-4725	&  16:36:42.96 &	 -47:31:29.7 & -75.1  & 10.99  & 4.6     & 3.89   & 33.8   &  6.3  & 4.3 & 2    \\
69 &I16344-4658	&  16:38:09.38 &	 -47:04:58.7 & -49.5  & 12.09  & 5.4     & 2.70   & 25.5   &  5.4  & 4.1 & 2    \\
70 &I16348-4654	&  16:38:29.64 &	 -47:00:41.1 & -46.5  & 12.09  & 5.4     & 2.40   & 23.6   &  5.4  & 4.4 & 2    \\
71 &I16351-4722	&  16:38:50.98 &	 -47:27:57.8 & -41.4  & 3.02   & 5.7     & 0.69   & 30.4   &  4.9  & 3.2 & 2    \\
72 &I16362-4639	&  16:39:57.32 &	 -46:45:06.3 & -38.8  & 3.01   & 5.7     & 0.54   & 24.0   &  3.6  & 2.5 & 2    \\
73 &I16372-4545	&  16:40:54.57 &	 -45:50:49.6 & -57.3  & 4.16   & 4.7     & 0.73   & 26.0   &  4.2  & 2.9 & 2    \\
74 &I16385-4619	&  16:42:14.04 &	 -46:25:25.9 & -117   & 7.11   & 3.1     & 1.34   & 31.9   &  5.1  & 3.2 & 2    \\
75 &I16424-4531	&  16:46:06.61 &	 -45:36:46.6 & -34.2  & 2.63   & 6.0     & 0.54   & 24.6   &  3.9  & 2.7 & 2    \\
76 &I16445-4459	&  16:48:05.18 &	 -45:05:08.6 & -121.3 & 7.95   & 2.8     & 2.54   & 24.6   &  5.0  & 3.9 & 2    \\
77 &I16458-4512	&  16:49:30.41 &	 -45:17:53.6 & -50.4  & 3.56   & 5.1     & 1.42   & 21.4   &  4.5  & 3.6 & 2    \\
78 &I16484-4603	&  16:52:04.29 &	 -46:08:30.1 & -32    & 2.1    & 6.4     & 0.55   & 35.0   &  5.0  & 3.0 & 1,2 \\
79 &I16487-4423	&  16:52:23.67 &	 -44:27:52.3 & -43.4  & 3.26   & 5.4     & 1.01   & 24.6   &  4.4  & 3.0 & 2    \\
80 &I16489-4431	&  16:52:33.50 &	 -44:36:17.7 & -41.3  & 3.26   & 5.4     & 0.71   & 21.8   &  3.8  & 2.9 & 2    \\
81 &I16506-4512	&  16:54:15.77 &	 -45:17:31.8 & -26.2  & 2.42   & 6.1     & 0.86   & 32.4   &  5.0  & 3.1 & 2    \\
82 &I16524-4300	&  16:56:03.06 &	 -43:04:43.3 & -40.8  & 3.43   & 5.2     & 1.06   & 23.4   &  4.4  & 3.4 & 2    \\
83 &I16547-4247	&  16:58:17.26 &	 -42:52:04.5 & -30.4  & 2.74   & 5.8     & 0.69   & 28.9   &  4.8  & 3.2 & 2    \\
84 &I16562-3959	&  16:59:41.42 &	 -40:03:46.6 & -12.6  & 2.38   & 6.1     & 0.72   & 42.3   &  5.7  & 3.2 & 2    \\
85 &I16571-4029	&  17:00:32.21 &	 -40:34:12.7 & -15    & 2.38   & 6.1     & 0.38   & 27.0   &  4.3  & 2.9 & 2    \\
86 &I17006-4215	&  17:04:12.99 &	 -42:19:54.2 & -23.2  & 2.21   & 6.3     & 0.50   & 27.7   &  4.4  & 2.8 & 2    \\
87 &I17008-4040	&  17:04:23.03 &	 -40:44:24.9 & -17    & 2.38   & 6.1     & 1.26   & 31.0   &  4.6  & 3.0 & 1,2 \\
88 &I17016-4124	&  17:05:11.02 &	 -41:29:07.8 & -27.1  & 1.37   & 7.0     & 0.75   & 32.0   &  5.3  & 3.8 & 1,2 \\
89 &I17136-3617	&  17:17:02.04 &	 -36:20:52.5 & -10.6  & 1.34   & 7.0     & 0.45   & 34.8   &  4.6  & 2.5 & 2    \\
90 &I17143-3700	&  17:17:45.65 &	 -37:03:11.8 & -31.1  & 12.67  & 4.7     & 2.95   & 31.0   &  5.6  & 3.8 & 2    \\
91 &I17158-3901	&  17:19:20.34 &	 -39:03:53.3 & -15.2  & 3.38   & 5.1     & 1.18   & 23.3   &  4.8  & 3.4 & 2    \\
92 &I17160-3707	&  17:19:26.81 &	 -37:11:01.4 & -69.5  & 10.53  & 2.7     & 1.69   & 28.5   &  6.0  & 4.1 & 2    \\
93 &I17175-3544	&  17:20:53.10 &	 -35:47:03.0 & -5.7   & 1.34   & 7.0     & 0.35   & 30.6   &  4.8  & 3.1 & 2    \\
94 &I17204-3636	&  17:23:50.32 &	 -36:38:58.1 & -18.2  & 3.32   & 5.1     & 0.60   & 25.8   &  4.2  & 2.9 & 2    \\
95 &I17220-3609	&  17:25:24.99 &	 -36:12:45.1 & -93.7  & 8.01   & 1.3     & 2.41   & 25.4   &  5.7  & 4.3 & 2    \\
96 &I17233-3606	&  17:26:42.73 &	 -36:09:20.8 & -2.7   & 1.34   & 7.0     & 0.37   & 29.9   &  4.6  & 3.0 & 2    \\
97 &I17244-3536	&  17:27:48.71 &	 -35:39:10.8 & -10.2  & 1.36   & 7.0     & 0.30   & 29.8   &  3.7  & 2.1 & 2    \\
98 &I17258-3637	&  17:29:16.99 &	 -36:40:16.5 & -11.9  & 2.59   & 5.8     & 0.92   & 45.5   &  5.9  & 3.3 & 2    \\
99 &I17269-3312	&  17:30:14.77 &	 -33:14:57.5 & -21    & 4.38   & 4.0     & 1.53   & 22.4   &  4.7  & 3.6 & 2    \\
100&I17271-3439	&  17:30:26.21 &	 -34:41:48.9 & -18.2  & 3.1    & 5.3     & 1.34   & 35.0   &  5.6  & 4.0 & 1,2 \\
101&I17278-3541	&  17:31:14.07 &	 -35:44:07.3 & -0.4   & 1.33   & 7.0     & 0.31   & 25.1   &  3.8  & 2.5 & 2    \\
102&I17439-2845	&  17:47:09.20 &	 -28:46:13.9 & 18.7   & 8      & 0.3     & 1.78   & 30.0   &  5.7  & 3.8 & 1    \\
103&I17441-2822	&  17:47:19.79 &	 -28:23:05.7 & 50.8   & 8.1    & 0.2     & 2.51   & 35.0   &  6.6  & 5.4 & 1    \\
104&I17455-2800	&  17:48:41.63 &	 -28:01:44.6 & -15.6  & 10     & 1.7     & 2.18   & 31.0   &  5.9  & 4.1 & 1    \\
105&I17545-2357	&  17:57:34.49 &	 -23:58:04.3 & 7.9    & 2.93   & 5.4     & 0.87   & 23.7   &  4.1  & 3.1 & 2    \\
106&I17589-2312	&  18:01:57.87 &	 -23:12:32.9 & 21.3   & 2.97   & 5.4     & 0.65   & 22.4   &  4.0  & 3.0 & 2    \\
107&I17599-2148	&  18:03:01.83 &	 -21:48:09.0 & 18.6   & 2.99   & 5.4     & 1.15   & 32.0   &  5.2  & 3.4 & 1,2 \\
108&I18032-2032	&  18:06:14.99 &	 -20:31:35.4 & 4.3    & 5.15   & 3.4     & 1.27   & 32.1   &  5.4  & 3.5 & 2    \\
109&I18056-1952	&  18:08:38.18 &	 -19:51:49.0 & 66.7   & 8.55   & 1.6     & 2.32   & 25.1   &  5.7  & 4.4 & 2    \\
110&I18075-2040	&  18:10:34.82 &	 -20:39:16.1 & 31.5   & 3.08   & 5.3     & 0.27   & 23.0   &  3.0  & 2.2 & 2    \\
111&I18079-1756	&  18:10:50.60 &	 -17:55:47.2 & 18     & 1.83   & 6.6     & 0.44   & 25.6   &  3.9  & 2.6 & 2    \\
112&I18089-1732	&  18:11:51.28 &	 -17:31:33.4 & 33.5   & 2.5    & 5.9     & 0.90   & 23.4   &  4.3  & 3.1 & 2    \\
113&I18110-1854	&  18:14:00.83 &	 -18:53:28.2 & 37     & 3.37   & 5.1     & 0.87   & 28.9   &  4.8  & 3.2 & 2    \\
114&I18116-1646	&  18:14:35.76 &	 -16:45:40.8 & 48.5   & 3.94   & 4.6     & 0.99   & 33.8   &  5.1  & 3.1 & 2    \\
115&I18117-1753	&  18:14:39.14 &	 -17:52:01.3 & 36.7   & 2.57   & 5.9     & 1.02   & 23.6   &  4.6  & 3.5 & 2    \\
116&I18134-1942	&  18:16:22.12 &	 -19:41:27.0 & 10.6   & 1.25   & 7.1     & 0.23   & 20.7   &  3.0  & 2.4 & 2    \\
117&I18139-1842	&  18:16:51.72 &	 -18:41:35.2 & 39.8   & 3.02   & 5.4     & 0.45   & 40.4   &  4.9  & 2.5 & 2    \\
118&I18159-1648	&  18:18:54.34 &	 -16:47:51.9 & 22     & 1.48   & 6.9     & 0.48   & 21.6   &  3.8  & 2.8 & 2    \\
119&I18182-1433	&  18:21:09.22 &	 -14:31:46.8 & 59.1   & 4.71   & 4.1     & 0.71   & 24.7   &  4.3  & 3.1 & 2    \\
120&I18223-1243	&  18:25:10.58 &	 -12:42:24.2 & 44.8   & 3.37   & 5.3     & 0.83   & 23.7   &  4.2  & 2.9 & 2    \\
121&I18228-1312	&  18:25:41.81 &	 -13:10:23.3 & 32.3   & 3.21   & 5.4     & 0.81   & 29.8   &  4.7  & 3.0 & 2    \\
122&I18236-1205	&  18:26:25.65 &	 -12:04:01.6 & 25.9   & 2.17   & 6.3     & 0.62   & 19.5   &  3.5  & 2.9 & 2    \\
\hline
	\end{tabular}
\end{table*}

\begin{table*}
	\centering
	\contcaption{Basic parameters of the targets in the ATOMS survey.}
	\label{atomsample}
	\begin{tabular}{cccccccccccc} 
		\hline
Number&IRAS  & RA &  DEC & $V_{\rm lsr}$  & Distance &  $R_{\rm GC}$ & Radius  & $T_{\rm dust}$ & log($L_{\rm bol}$)  & log($M_{\rm clump}$) & refs\\
  &&  & & (km~s$^{-1}$) & (kpc)  & (kpc)  & (pc) & (K) & (L$_{\sun}$)  & (M$_{\sun}$) & \\
\hline
123&I18264-1152	&  18:29:14.28 &	 -11:50:27.0   & 43.2   & 3.33   & 5.3     & 0.81   & 20.3   &  3.9  & 3.2 & 2    \\
124&I18290-0924	&  18:31:43.23 &	 -09:22:28.5   & 84.2   & 5.34   & 4.0     & 1.09   & 22.1   &  4.0  & 3.2 & 2    \\
125&I18308-0503	&  18:33:29.42 &	 -05:00:55.1   & 42.9   & 3.1    & 5.7     & 0.59   & 31.0   &  4.3  & 2.6 & 1    \\
126& I18311-0809	&  18:33:53.43 &	 -08:07:14.4 & 113    & 6.06   & 3.7     & 2.12   & 25.4   &  5.0  & 3.7 & 2    \\
127& I18314-0720	&  18:34:10.25 &	 -07:18:00.1 & 101.5  & 5.82   & 3.9     & 2.54   & 30.4   &  5.7  & 3.8 & 2    \\
128& I18316-0602	&  18:34:20.58 &	 -05:59:41.6 & 42.8   & 2.09   & 6.5     & 0.56   & 23.4   &  4.0  & 2.9 & 2    \\
129& I18317-0513	&  18:34:25.50 &	 -05:10:53.5 & 42.1   & 2.18   & 6.5     & 0.32   & 31.0   &  4.8  & 3.2 & 1,2 \\
130& I18317-0757	&  18:34:24.90 &	 -07:54:50.6 & 80.7   & 4.79   & 4.4     & 1.93   & 30.4   &  5.2  & 3.4 & 2    \\
131& I18341-0727	&  18:36:49.82 &	 -07:24:53.0 & 112.7  & 6.04   & 3.8     & 2.02   & 25.7   &  5.1  & 3.7 & 2    \\
132& I18411-0338	&  18:43:46.07 &	 -03:35:33.0 & 102.8  & 7.41   & 4.0     & 1.98   & 27.6   &  5.1  & 3.7 & 2    \\
133& I18434-0242	&  18:46:03.51 &	 -02:39:26.7 & 97.2   & 5.16   & 4.7     & 1.48   & 35.5   &  5.7  & 3.6 & 2    \\
134& I18440-0148	&  18:46:36.22 &	 -01:45:23.7 & 97.5   & 5.16   & 4.7     & 0.43   & 33.0   &  4.1  & 2.3 & 2    \\
135& I18445-0222	&  18:47:09.76 &	 -02:18:47.6 & 86.9   & 5.16   & 4.7     & 1.30   & 27.0   &  4.6  & 3.4 & 2    \\
136& I18461-0113	&  18:48:41.83 &	 -01:10:01.4 & 96.1   & 5.16   & 4.7     & 0.65   & 27.0   &  4.4  & 3.1 & 2    \\
137& I18469-0132	&  18:49:33.15 &	 -01:29:04.2 & 87     & 5.16   & 4.7     & 0.68   & 32.2   &  4.8  & 3.0 & 2    \\
138& I18479-0005	&  18:50:30.79 &	 -00:01:58.2 & 14.6   & 12.96  & 7.5     & 2.45   & 34.2   &  6.1  & 4.2 & 2    \\
139& I18502+0051	&  18:55:22.63 &	  00:59:16.0 & 53     & 7.1    & 4.7     & 1.24   & 30.0   &  5.3  & 3.9 & 1    \\
140& I18507+0110	&  18:53:18.42 &	  01:15:00.1 & 57.2   & 1.56   & 7.1     & 0.44   & 29.2   &  4.8  & 3.2 & 2    \\
141& I18507+0121	&  18:53:18.12 &	  01:25:22.7 & 57.9   & 1.56   & 7.1     & 0.29   & 22.7   &  3.5  & 2.6 & 2    \\
142& I18517+0437	&  18:54:14.13 &	  04:41:46.2 & 43.9   & 2.36   & 6.6     & 0.23   & 32.0   &  4.6  & 2.9 & 1,2 \\
143& I18530+0215	&  18:55:33.61 &	  02:19:09.0 & 74.1   & 4.67   & 5.3     & 1.06   & 26.1   &  4.7  & 3.3 & 2    \\
144& I19078+0901	&  19:10:13.41 &	  09:06:10.4 & 2.9    & 11.11  & 7.6     & 4.26   & 33.3   &  6.9  & 5.0 & 2    \\
145& I19095+0930	&  19:11:53.90 &	  09:35:45.9 & 43.7   & 6.02   & 5.8     & 0.64   & 34.9   &  5.1  & 3.1 & 2    \\
146& I19097+0847	&  19:12:09.09 &	  08:52:10.8 & 58     & 8.47   & 6.2     & 2.01   & 23.3   &  5.0  & 3.8 & 2    \\
\hline
	\end{tabular}
\end{table*}

\begin{figure*}
\centering
\includegraphics[angle=0,scale=0.9]{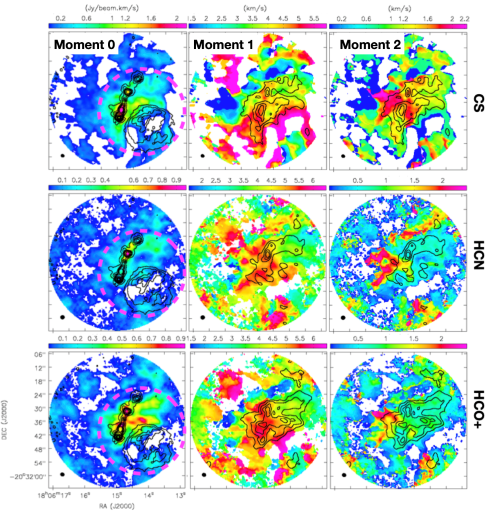}
\caption{The left, middle and right columns present the integrated intensity (Moment 0) maps, intensity-weighted velocity maps (Moment 1) and intensity-weighted velocity dispersion (Moment 2) maps of G9.62+0.19 for the gas tracers indicated on the right of each row. The names of tracers are labeled to the right. The contours on Moment 0 maps show the 3 mm continuum emission. The contour levels are [0.01,0.03,0.05,0.07,0.1,0.2,0.4,0.6,0.8]$\times$74.1 mJy~beam$^{-1}$. The contours on Moment 1 and Moment 2 maps show the corresponding integrated intensity maps of various gas tracers. The contours are [0.2,0.4,0.6,0.8]$\times$F$_{peak}$. The peak integrated intensities ($F_{\rm peak}$) for CS, HCN, HCO$^+$ are 2.16, 0.97, and 0.93 Jy~beam$^{-1}$~km~s$^{-1}$, respectively. The magenta dashed circle on Moment 0 map shows the clump boundary traced by H$^{13}$CN. \label{lineimages1}}
\end{figure*}

\begin{figure*}
\centering
\includegraphics[angle=0,scale=0.9]{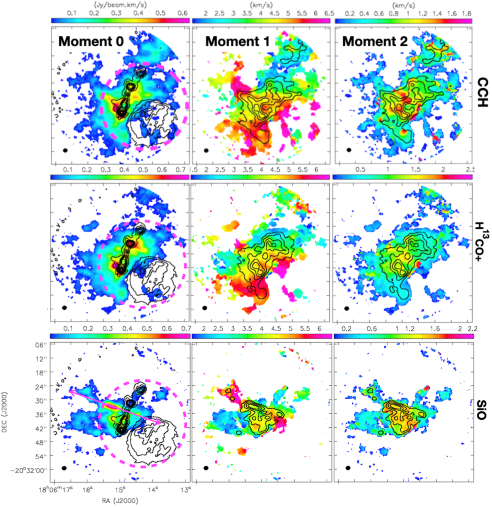}
\caption{The images and contours are same as in Figure \ref{lineimages1}. The peak integrated intensities ($F_{\rm peak}$) for CCH, H$^{13}$CO$^+$ and SiO are 0.63, 0.75, and 0.72 Jy~beam$^{-1}$~km~s$^{-1}$, respectively. The magenta dashed circle on the Moment 0 maps show the clump boundary traced by H$^{13}$CN. The magenta arrow represents the direction of the bipolar outflow driven by MM6. \label{lineimages2}}
\end{figure*}

\begin{figure*}
\centering
\includegraphics[angle=0,scale=0.9]{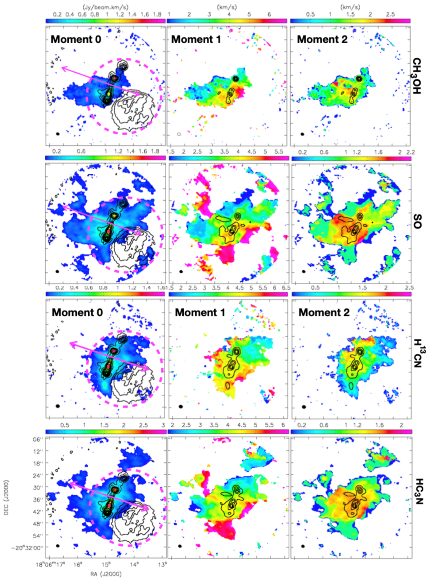}
\caption{The images and contours are same as in Figure \ref{lineimages1}. The peak integrated intensities ($F_{\rm peak}$) for CH$_3$OH, SO, H$^{13}$CN and HC$_3$N are 1.94, 1.98, 1.65, and 3.09 Jy~beam$^{-1}$~km~s$^{-1}$, respectively. The magenta dashed circle on the Moment 0 maps show the clump boundary traced by H$^{13}$CN. The magenta arrow represents the direction of the bipolar outflow driven by MM6.\label{lineimages3}}
\end{figure*}

\clearpage
\onecolumn

\noindent
Author affiliations:\\

\noindent $^{1}$Key Laboratory for Research in Galaxies and Cosmology, Shanghai Astronomical Observatory, Chinese Academy of Sciences, 80 Nandan Road, Shanghai 200030, People’s Republic of China\\
$^{2}$Korea Astronomy and Space Science Institute, 776 Daedeokdaero, Yuseong-gu, Daejeon 34055, Republic of Korea\\
$^{3}$Department of Astronomy, The University of Texas at Austin, 2515 Speedway, Stop C1400, Austin, TX 78712-1205, USA\\
$^{4}$University of Science and Technology, Korea (UST), 217 Gajeong-ro, Yuseong-gu, Daejeon 34113, Republic of Korea\\
$^{5}$Jet Propulsion Laboratory, California Institute of Technology, 4800 Oak Grove Drive, Pasadena, CA 91109, USA\\
$^{6}$Institute of Astronomy and Astrophysics, Academia Sinica. 11F of Astronomy-Mathematics Building,
AS/NTU No. 1, Section 4, Roosevelt Rd., Taipei 10617, Taiwan\\
$^{7}$Center for Astrophysics $|$ Harvard \& Smithsonian, 60 Garden Street, Cambridge, MA 02138, USA\\
$^{8}$National Astronomical Observatory of Japan, National Institutes of Natural Sciences, 2-21-1 Osawa, Mitaka, Tokyo 181-8588, Japan\\
$^{9}$Kavli Institute for Astronomy and Astrophysics, Peking University, 5 Yiheyuan Road, Haidian District,
Beijing 100871, People's Republic of China\\
$^{10}$Department of Physics, P.O. Box 64, FI-00014, University of Helsinki, Finland\\
$^{11}$Departamento de Astronom\'{\i}a, Universidad de Chile, Las Condes, Santiago, Chile\\
$^{12}$School of Physics, University of New South Wales, Sydney, NSW 2052, Australia\\
$^{13}$School of Space Research, Kyung Hee University, Yongin-Si, Gyeonggi-Do 17104, Republic of Korea\\
$^{14}$National Astronomical Observatories, Chinese Academy of Sciences, Beijing 100101, China  \\
$^{15}$University of Chinese Academy of Sciences, Beijing 100049, China\\
$^{16}$Astronomy Department, University of California, Berkeley, CA 94720, USA\\
$^{17}$Department of Astronomy, Yunnan University, and Key Laboratory of Astroparticle Physics of Yunnan Province,
Kunming, 650091, People's Republic of China\\
$^{18}$IRAP, Universit\'{e} de Toulouse, CNRS, UPS, CNES, Toulouse, France\\
$^{19}$Indian Institute of Space Science and Technology, Thiruvananthapuram 695 547, Kerala, India\\
$^{20}$E\"{o}tv\"{o}s Lor\'{a}nd University, Department of Astronomy,
P\'{a}zm\'{a}ny P\'{e}ter s\'{e}t\'{a}ny 1/A, H-1117, Budapest, Hungary\\
$^{21}$Department of Astronomy, Peking University, 100871, Beijing, People's Republic of China\\
$^{22}$Departamento de Astronom\'{\i}a, Universidad de Concepci\'{o}n, Av. Esteban Iturra s/n, Distrito Universitario, 160-C, Chile\\
$^{23}$College of Science, Yunnan Agricultural University, Kunming, 650201, People's Republic of China\\
$^{24}$Shanghai Astronomical Observatory, Chinese Academy of Sciences, 80 Nandan Road, Shanghai 200030 People's Republic of China\\
$^{25}$Key Laboratory of Radio Astronomy, Chinese Academy of Sciences, Nanjing 210008, People's Republic of China\\
$^{26}$School of Physics and Astronomy, Sun Yat-sen University, 2 Daxue Road, Zhuhai, Guangdong, 519082, People's Republic of China\\
$^{27}$NAOC-UKZN Computational Astrophysics Centre, University of KwaZulu-Natal, Durban 4000, South Africa\\
$^{28}$SOFIA Science Centre, USRA, NASA Ames Research Centre, MS-12, N232, Moffett Field, CA 94035, USA

\bsp	
\label{lastpage}
\end{document}